\documentclass[11pt, oneside]{article}   	% use "amsart" instead of "article" for AMSLaTeX format
\usepackage{jheppub}
\usepackage{hyperref}
\usepackage{float}
\usepackage{wrapfig}
\usepackage{comment}

\usepackage{amsmath, amsthm, graphicx,amssymb,subfigure,bbm}
\usepackage[all]{xy}

\newcommand{\be}{\begin{equation}}
	\newcommand{\ee}{\end{equation}}
\newcommand{\bea}{\begin{eqnarray}\displaystyle}
	\newcommand{\eea}{\end{eqnarray}}
\newcommand{\nnm}{\nonumber}

\def\crit{ {\rm crit} }
\def\cZ{\mathcal{Z }} 
\def\one{{\hbox{ 1\kern-.8mm l}}}
\def\zero{{\hbox{ 0\kern-1.5mm 0}}}

\def\mC{ \mathbb{C}} 
 
\def\Phichk{\Phi{{\raisebox{0.02ex}{\Large$\check {}$}} } }

\def\Mult{ { \rm { Mult} } } 
\def\Sym{ {\rm Sym } }
\def\sigC{\sigma_{\scriptscriptstyle C} } 
 
\def\Aut{\mathrm{Aut } } 
\def\clsG{  {\mathrm {Classes}  } ( G ) }

\newcommand{\cA}{\mathcal A}
\newcommand{\cAp}{\mathcal A_+}
\newcommand{\cAm}{\mathcal A_-}

\def\cA{{\cal A}} \def\cB{{\cal B}} \def\cC{{\cal C}}

  \def\cL{{\cal L}}
 \def\cN{{\cal N}} \def\cO{{\cal O}}
  
\def\cS{{\cal S}}  
  
 \def\cZ{{\cal Z}}
\def\Sym{ {\rm{ Sym }} } 
\def\Dim{ {\rm{Dim} } }

\def\Invt{ {\rm Invt} }
\def\crit{ {\rm crit } }

\preprint{DIAS-STP-26-10\\
	\rightline{ QMUL-PH-26-19}}

\author[a]{Denjoe O'Connor,}
\affiliation[a]{School of Theoretical Physics, Dublin Institute for Advanced Studies, \\
       10 Burlington Road, Dublin 4, Ireland.}
\emailAdd{denjoe@stp.dias.ie}
\author[b]{Sanjaye Ramgoolam,}
\affiliation[b]{Centre for Theoretical Physics,  Department of Physics and Astronomy,\\
   Queen Mary University of London, 327 Mile End Road, London E1 4NS, UK.}
\emailAdd{s.ramgoolam@qmul.ac.uk}

\abstract{ We consider the $SO(d)$ and $O(d)$ invariant sectors of the
  bosonic $d$-matrix harmonic oscillator with $U(N)$ gauge
  symmetry. The micro-canonical degeneracy $ \cZ( N , d , k )$ for
  fixed energy $k$ is expressed as a pairing between an $N$-dependent
  vector and a $d$-dependent vector in the space of partitions of the
  integer $k$. This pairing formula is derived by counting invariant
  words in multi-matrix variables $X^i_{j,a}$, using properties of
  Clebsch-Gordan multiplicities (Kronecker coefficients) for the
  symmetric group $S_k$, Schur-Weyl duality and harmonic analysis on
  the homogeneous space $U(d)/SO(d)$.  Analytic formulae for large $N$
  and $k$ with $ k \le N $ are obtained using group integrals over
  $U(N)$ and $SO(d)$ (or $ O(d)$).  The micro-canonical heat capacity
  in this regime is negative and turns positive, at a critical value $
  k_{ \crit }$, due to finite $N$ modifications to the counting, thus
  forming what we denote as a characteristic caloric fold in the $ E $ versus $T$ curve.
  Data from the pairing formula is well fitted by $k_{ \crit } \sim {
    N^2 \over 4 }$ for small values of $d$.  A derivation of this
  large $N$ formula is given using a matrix model approximation and
  semi-classical analysis of the eigenvalue density.  The large $N,d$
  limit of the degeneracies reveals a key role for ribbon graph
  combinatorics. The caloric fold is also notably  a  property of black hole
  thermodynamics in anti-de-Sitter spaces. 
  We propose the spherically symmetric \(SO(d)\) and \(O(d)\) invariant sectors of \(d\)-matrix quantum mechanics as tractable matrix systems for capturing key features  of dual descriptions of black-hole thermodynamics.
  }

  \arxivnumber{}

  \title{Negative heat capacities  in  spherically symmetric  sectors of  $d$-matrix quantum mechanics. }

\begin{document}
\maketitle
%\tableofcontents
\vfill\eject
\section{Introduction}\label{intro}
\begin{wrapfigure}{h}{0.25\textwidth} % {alignment}{width}
  \centering
	\centering
	\includegraphics[width=0.25\textwidth]{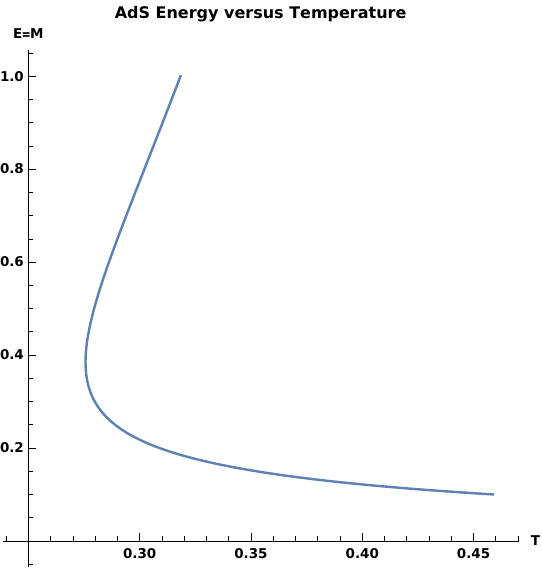} 
	\caption{Caloric fold from black hole in AdS} 
	\label{fig:EnergyVsTemperatureAdSFig1}
 \end{wrapfigure}
In the classic paper of Hawking and Page \cite{HawkPage} the
thermodynamics of black holes in anti-de Sitter (AdS) spacetime was
studied. It was found that AdS black holes have a minimum temperature above which there are, for the same temperature, black holes of two distinct masses.  This energy-temperature curve, 
illustrated in  Figure  \ref{fig:EnergyVsTemperatureAdSFig1},  has a shape which we will refer to as the {\it caloric fold}.
The lower branch describes the small black hole with negative heat capacity, while the upper branch 
 describes a larger black hole with positive heat capacity.  By computing the Euclidean action of the black hole solution, Hawking and Page argued  that there is a first order phase
 transition from the large  AdS black hole to thermal AdS.  
  Witten \cite{WittenThermal} reinterpreted this Hawking-Page  transition
 as the gravitational dual of a  confining-deconfining 
 transition in $ \cN=4$ super-Yang Mills theory with $U(N)$ gauge group at large $N$ in the context of 
 the AdS/CFT correspondence \cite{Malda,GKP,WittenHolog}. The large black hole phase
 is the deconfined phase and the thermal AdS is  the confined phase.
 
 The deconfinement/Hagedorn transition was also found in the zero-coupling super Yang-Mills theory, and in associated simpler gauged matrix models in \cite{Sundborg:1999ue,Aharony:2003sx}. Understanding the small black hole---negative heat capacity--- branch of the black hole caloric fold from dual gauge theory and matrix models has been the subject of investigation in a number of papers \cite{BerSmall,HanMaltz,AspBer2008,Mandal:2009vz,Azuma:2014cfa,Morita:2020liy}.  A standard result in statistical thermodynamics is that heat capacities in the canonical ensemble are 
 always non-negative (see e.g. \cite{PathriaBeale,Reif:Chapter6}). 
 However, negative heat capacities are not prohibited in the microcanonical ensemble. They arise especially in systems with long-range interactions or ensemble inequivalence, notably self-gravitating systems, black holes, and finite systems such as fragmenting nuclei \cite{LyndenBell1968,Hawking1976,HawkingPage1983,Gross1997,Gross2001,Touchette2015}.

 \begin{wrapfigure}{r}{0.25\textwidth} % {alignment}{width}
  \centering
  \includegraphics[width=0.25\textwidth]{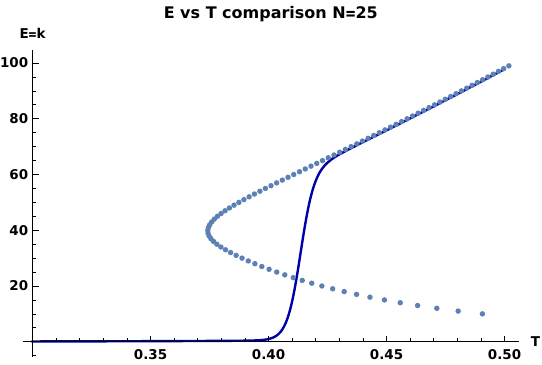}
  \caption{The canonical and microcanonical Energy Temperature plot for a permutation invariant quantum matrix model with $N=25$.}
  \label{fig:MicroAndCanonPIMQMM}
\end{wrapfigure}
Permutation invariant matrix quantum mechanics models have been shown
to admit combinatorial formulae \cite{LMT,PIGMM,BPR21,BPR22,PIMQMPI,PIMQMPF,PIMQTNS} where
both the canonical and micro-canonical thermodynamics is
accessible. This has allowed the identification of negative heat
capacities in the micro-canonical energy temperature, $(E,T)$, curve
for energies $ E \lesssim f ( N ) $ for $ f ( N ) \sim { N \log N
  \over 2 }$ where there is a turn-around to positive heat capacity at
higher energy (see Figure \ref{fig:MicroAndCanonPIMQMM}).  The exact
product formulae for the canonical partition functions, available for
any finite $N$, as a sum over partitions of $N$ allows numerical
comparison of the canonical and microcanonical energy versus
temperature curve. The discussion was generalised to permutation
invariant tensor models \cite{PIMQT} and it was found that the
phenomenon of negative heat capacity at low energy is a generic
feature of permutation invariant matrix and tensor models.  Similar
phenomena have been studied recently in the context of the statistical
thermodynamics of networks \cite{Evnin}.

A further caloric fold was also found in \cite{PIMQTNS} for a zero
charge sector of matrix quantum mechanics with two hermitian matrices $X, Y$  and  simple harmonic oscillator potential. The two matrices $ X , Y $ can be combined into a complex matrix 
$ Z = ( X + i Y )$ which transforms with positive $U(1)$ charge, while the physical states correspond to gauge-invariant 
polynomials in $ Z $ and its hermitian conjugate $ Z^{ \dagger} $, with an equal number of $ Z , Z^{ \dagger}$. 
Equivalently we are gauging this  $U(1)$ in addition to the $U(N)$ and considering physical states in this $U(N)\times U(1)$ gauged model. For the energy range $ 1  \le k \le  N $ the number of states is independent of $N$, and we refer to this as a stable range counting.  The stable $k$-dependent counting function is known in terms of the coefficients in an infinite-product generating function. The asymptotics of these degeneracies for $ k \gg 1 $ was computed in \cite{RWZ} by using mathematical results on multi-variate asymptotics \cite{PemantleWilsonACSV}. 
Representation-theoretic formulae for the microscopic 
degeneracies at finite $N$ \cite{Dolan,BHR1,Collins2008}    were computed explicitly in \cite{PIMQTNS}, using SageMath algorithms, for values of $N$ near $13$ and energies $k$ close to $20$ to demonstrate  that the micro-canonical
heat capacity is negative at low $k$ and then turns around above a criticial $ k_{ \crit } ( N )$. 
In the  high-temperature/high-energy limit, accessible from
the path integral, the heat capacity is positive. Thus the low-energy asymptotics, computational evidence, and the high-energy/high-temperature limit provided evidence that a caloric fold feature holds generally at large finite $N$.

\vskip.2cm

 In this paper we generalise the zero charge sector of the two-matrix quantum mechanics  to $SO(d)$ and $O(d)$ invariant sectors of $d$-matrix quantum mechanics. We consider quantum mechanics with $d$ hermitian matrices, i.e.  matrix variables $ X_{a,j}^i$, with $ a \in\{ 1, \cdots , d \} , i, j \in \{ 1, \cdots , N \}$ obeying 
 \bea 
 X_{a,j}^i =  ( X_{ a , i }^j )^* 
 \eea 
 The simple harmonic oscillator quantum mechanical Lagrangian 
 \bea 
 \cL = \sum_{ a =1}^d \sum_{i , j =1}^{ N }  { 1 \over 2 }  \left (  \partial_t X^i_{ a,j} ( \partial_t X^j_{ a,i})^*  -   X^i_{ a,j} ( X^j_{ a,i})^* \right ) 
 \eea
  is invariant under a $U(N)\times O(d)$ global symmetry where the $d$ matrices transform under $ U \in U(N)$ in the adjoint representation 
 \bea 
 X_a \rightarrow U X_a U^{ \dagger }
 \eea
 while the $a$-index transforms in the vector of $SO(d)$ and $O(d)$. Matrix models  arising in known gauge-string dualities, such as  BFSS matrix model \cite{BFSS}, the IKKT matrix model \cite{IKKT}  and the BMN matrix model  \cite{BMN} contain such orthogonal groups as part of their symmetry groups and $U(N)$ as gauge group. It is natural to wonder about the physical properties of the $SO(d)$ or $O(d)$ invariant sectors of these models and of the physical meaning of the interactions between the invariant and non-invariant sectors. Our work on the $d$-matrix harmonic oscillator can be considered as a stepping stone towards this question. In section \ref{sec2} we consider the path integral for the 
 $U(N) \times SO(d)$ (and $ U(N )\times O(d)$) gauged matrix harmonic oscillator and relate the thermal partition functions to counting functions 
 $ \cZ ( N , d, k) $ (and $ \cZ_O ( N  , d , k ) $ for the $O(d)$ case) defined mathematically for the appropriate invariant polynomials, and amenable to exact finite $N$ algorithms. We also review the path integral formulation for the $U(N)$ gauged $d$-matrix harmonic oscillator which gives the counting function $ Z ( N , d, k) $ for degree $k$ polynomials which are invariant under the $U(N)$ gauge group.

 \vskip.2cm

 The main conceptual result of this paper is to give detailed evidence that the micro-canonical degeneracies $ \cZ( N , d , k )$  and $ \cZ_O ( N , d , k ) $ display the characteristic caloric fold thermodynamics in the regime of $ d \sim 1 $ and large finite  $N$, with a critical energy at the minimum temperature of $ k_{ \crit } \sim { N^2 \over 4 }$. In this case 
 the caloric fold is associated with a Hagedorn transition in the canonical ensemble at $ x_H = d^{-1} $, where $x={\rm e}^{-\beta}$. 
 We also find evidence that when $ d \gg  N$, the regime of large finite  $N$ has the characteristic caloric fold. In this case, the Hagedorn temperature approaches zero as $N$ tends to infinity, and a more detailed study is left for the future. 

\vskip.2cm 

Our main new technical results are: 

\begin{itemize}

\item  For the stable range, i.e.  $ N \ge k $, we derive the  large $k$ asymptotics of $ \cZ( N , d, k )$ for general  $ d $ of order $1$ (see  equation \eqref{LowEnergyCoefficientsSOdGaugedv2}). This large $k$ asymptotics is obtained from a leading factor in the infinite product formula for the counting of $U(N)$ invariants of $d$ matrices, which is related to the counting of $SO(d)$  singlet states in $V_d^{ \otimes k }$. The generating functions for the dimensions of $SO(d)$ invariant subspaces of $ V_d^{ \otimes k }$ for all finite $k$  are also obtained for $ d = 2,3,4,5$,  and involve special functions, such  as Elliptic integrals and Appel hypergeometric functions (equations
 \eqref{SpecFuncsSO} and \eqref{SpecFuncsO}).

\item Exact finite $N$ formulae  for $ \cZ ( N , d, k )$ and $ \cZ_O ( N , d , k )$ are obtained by combining standard properties of  symmetric groups, Schur-Weyl duality and harmonic analysis on  the coset $ U(d)/SO(d)$. They 
 are brought to a simple mathematical expression as a pairing of the form 
 \bea 
\cZ( N , d , k ) =  \langle \Psi ( N , k  ) , \Phi (d , k  ) \rangle ~~ \, , 
\eea
where $ \Psi ( N , k  ) $ and $  \Phi (d , k ) $ are functions of partitions $ p $, of the integer  energy $k$, with integer coefficients given in terms of  symmetric group characters. The precise definitions of these functions are given in the build-up to the equations  \eqref{PairingSO} and \eqref{PairingO}.

\item  The exact finite $N$ pairing formulae and their computational implementations  are used to exhibit the caloric fold in the micro-canonical energy/temperature relation, for the  examples of $ d = 2,3,4,5$ and $6$. The lower energy negative heat capacity branch is governed by stable range asymptotics and there is a critical energy $ k_{ \crit } $ where the heat capacity turns positive.   We give a  large $N$ eigenvalue density calculation which evidences $ k_{ \crit } = { N^2 \over 4 }$. The computed data is consistent with this formula for the turning point $ k_{ \crit }$.

\item  We study the large $N,d$ limit of the partition
functions, finding that $ \cZ(N,d,k)$ and $\cZ_O(N,d,k)$ have the same limit, $\cZ(\infty,\infty,2n)$, which is non-zero for even $k=2n$. This has an interpretation as the number of ribbon graphs, of any genus,  with $n$ edges. 
The study of $ \cZ( \infty , \infty , 2n  )$ shows that it has two distinct expansions over partitions of $2n$ involving the group $ S_{ 2n } $ and its wreath-product subgroup $ S_n [ S_2 ]$. The wreath product $S_n[S_2]$ is the semi-direct product  $S_n \ltimes S_2^{ \times n } $, where \(S_n\) acts on the product group \(S_2^{\times n}\) by permuting its \(n\) factors.This combinatorial duality extends to an analogous duality for any pair consisting of a group $G$ and subgroup $H$.

\end{itemize}

\vskip.2cm 

The paper is organised as follows.  Section \ref{sec2} gives the path
integral formulation of the partition function for $U(N) \times SO(d)$ and $ U(N) \times O(d)$ gauged matrix quantum harmonic oscillators. These 
have physical Hilbert spaces with dimensions at energy $k$
equal to the dimension of the spaces of gauge invariant polynomial functions of degree $k$ in the matrix variables $ X^i_{ a,j}$. The known mathematical Molien-Weyl  group-integration formulae for the counting of the invariants are shown  to be equal to thermal partition function of the gauged quantum mechanical system, following \cite{PIMQMPI}. Thus the central mathematical objects $ \cZ ( N , d , k ) $ and $ \cZ_O ( N ,  d , k ) $ in this paper are given a direct path integral realisation.  

Section \ref{sec:HagInvFull} gives plots of  micro-canonical energy versus temperature curves in the gauged $U(N)
$ multi-matrix model without $SO(d)$ or $O(d)$ gauging to demonstrate the absence of negative heat capacity. 
This sets the stage for our studies of the models with gauged $SO(d)$ or $O(d)$ which have negative heat capacity and a caloric fold curve with the heat capacity turning positive for sufficiently high energy.

Section \ref{sec:LowEnergy} studies the form of $ \cZ ( N , d, k) $ in the regime $ N \ge  k \gg    d  ~~  ( d \sim \cO( 1) )  $. 
The starting point is  the  stable range formulae for the counting of $U(N)$ gauge invariants of $d$ hermitian matrices: the stable range is defined by $ N \ge k $. These take the form of infinite products. Group integration over $SO(d)$ is performed using the explicit form of the Haar measure, for $d=2,3,4,5$ in order  to obtain analytic formulae for the asymptotic forms. 
The $d=2$ case agrees with \cite{RWZ}, and the results take the form 
\bea\label{keyasymps} 
 \cZ ( N , d, k)  \sim { d^k \over k^{d ( d -1) /4  } } 
 \eea 
 for the regime $ N \gg k \gg    d $ (and $ d  \sim \cO( 1) $). The details of the calculations are given in Appendix \ref{LargeNGeneratingFunctions}.

Section \ref{sec:ExactCountingAndAlgorithms}  gives combinatorial derivations of the micro-canonical  degeneracies
$ \cZ( N , d ,  k ) $ and $ \cZ_{O} ( N , d , k )$. Schur-Weyl duality and the Cartan-Helgason theorem for harmonic
analysis on the homogenous space $U(d)/SO(d)$ are the key mathematical tools:  standard references are
\cite{HelgasonGroups,GoodmanWallach}.
 Equations  \eqref{KronHelg1} and \eqref{KronHelgO} give the micro-canonical partition functions for $SO(d)$ and $O(d)$ respectively, as a sum over a restricted set of Kronecker coefficients. We then obtain  character-Pairing formulae. The key formulae in this section have simple  implementations using the computational software for group-theoretic and algebraic algorithms, SageMath. These are described in the Appendix  section \ref{sec:SageMathCode}.

Section \ref{sec:FoldsExact}  displays the caloric folds using the exact formulae and computational implementations. 
We focus on the examples of $ d = 2,3,4,5$ and $6$. The data is consistent with $ k_{ \crit } \sim { N^2 \over 4 }$. 

Section \ref{sec:EigDensAnalytic} discuses the high temperature large
$N$ limit in terms of a continuous eigenvalue distribution and obtains
expressions connecting to the large $N$ low temperature regime. These formulae give a derivation of the large $N$ form 
$ k_{ \crit } \sim { N^2 \over 4 }$ of the critical energy where the micro-canonical temperature is minimum. 
We also give excellent fits to the caloric fold, Figure  \ref{fig:LargeNFold}.
In addition, we find that for $k\ge \frac{N^2}{4}$,  the large $N$ form of $\cZ(N,d,k)$ has an additional exponential suppression factor (see equation \eqref{HighLowLargeNMatched}).

Section \ref{sec:largeNd} discusses the large $N$ large $d$ limit and
its connection to ribbon graphs.  This allows us to return to the
finite $N$ and finite $d$ cases using representation-theoretic Fourier
transforms on the space of ribbon graphs. The section ends with a
discussion of thermodynamics when $d \gg  N, k$; while $k$ varies from
values below to above a fixed large $N$. 

The paper concludes with a discussion of the broader context,
implications for a deeper connection with black hole thermodynamics
and outlook. This includes a conjecture for the dual strongly coupled gauge theory formulation of the difference between the minimum black hole temperature and the transition temperature between large black holes and thermal AdS, a quantity calculated in \cite{HawkPage},  in the context of $ AdS_5 \times S^5$ and its $ \cN=4$ SYM gauge theory dual. 

In addition there are  appendices detailing technical steps and providing useful data. 
We begin with a specialised discussion of the $d=3$ and $d=4$ cases of $ \cZ( N , d, k )$  in Appendix  \ref{sec:Low-dim-isom}, Appendix \ref{sec:SageMathCode} gives the SageMath code that can be used to compute $ \cZ( N , d, k ) $ and $ \cZ_O ( N , d , k )$. A review of black hole thermodynamics including the AdS caloric fold is in  \ref{ReviewOfBlackHoleThermodynamics}. Appendix \ref{LargeNGeneratingFunctions}  gives the derivation of the generating functions in section \ref{sec:LowEnergy} for $SO(d) $ invariants in $ V_d^{ \otimes k }$ with $ d = 2,3,4,5$.  
Appendix \ref{sec:GenTwo} proves an identity relating two ways that a subgroup $H \subset G$ interacts with the conjugacy classes of $G$. On one side of the identity is  a sum over the distributions of elements of a subgroup $H \subset G $ across conjugacy classes of $G$, and on the other side a sum involving orbits of $H$ acting by conjugation on conjugacy classes of $G$. A special case  of this identity with $ G = S_{ 2n} $ and $ H = S_n [ S_2] $ plays a role in  the connection between ribbon graph counting and the counting of $U(N) \times SO(d)$ (or $ U(N) \times O(d)$) invariants in $d$-matrix theory at large $d$, as discussed in section \ref{sec:largeNd}. Appendix \ref{sec:AppIdDimSpec} derives identities relating the generating functions in equations  \eqref{SpecFuncsSO} and \eqref{SpecFuncsO} of section \ref{sec:LowEnergy}  to sums of dimensions of irreducible representations of symmetric groups $S_k$ taken over the sets $ \cA(  k , d) $ and $ \cA_+ ( k , d )$ defined in section \ref{sec:ExactCountingAndAlgorithms}, which we have also checked directly with Mathematica for some  ranges of $k$. 
The Appendix  \ref{Tables} gives tables for  $ \cZ( N , d, k ) $ and $ \cZ_O( N , d, k )$ for different values of $d$. 

\vskip.5cm 

\section{$U(N) \times SO(d )$ gauged multi-matrix harmonic oscillator }
\label{sec2}
We consider a set of $N \times N$  hermitian matrices  $X_a$, with $a=1,\cdots, d$, gauged under the combined adjoint action  of $U(N)$ and vector representation of $SO(d)$ or  $O(d)$.  The transformation property of the matrices is $X_a\rightarrow R_{ab}g X_b g^{-1}$ with $g\in U(N)$ and $R\in SO(d)$ or $ O(d)$. 
The Euclidean action for the gauged harmonic oscillator system is then
\begin{equation}\label{ContinuumGaugeAction}
S[X]=\int_0^\beta d\tau {\bf tr}\left(\frac{1}{2}({\cal D}_\tau X)^2+\frac{1}{2}m^2 X^2\right)\end{equation} 
where ${\cal D}_\tau=\partial_\tau+\omega-i[A,\,\cdot\,]$ is the covariant derivative with
the gauge field $A(\tau)$ being an $N\times N$ hermitian matrix and $\omega$ the $SO(d)$ connection, which is an anti-symmetric $d$-dimensional matrix. For simplicity we have also suppressed both $U(N)$ and $SO(d)$ indices.
By discretising the Euclidean path integral and taking a continuum limit, we show that it gives the Molien-Weyl formula 
for the generating function of gauge-invariant functions of the matrix variables $X^i_{ a , j }$. 

To set up the discrete path integral for this model we need to convert the derivative ${\cal D}_\tau$ to a lattice analogue. This is naturally defined to be
\begin{equation}
  \label{LatticeCovDeriv}
  {\cal D}_\tau X\rightarrow \frac{R_{n,n+1}g_{n,n+1}X_{n+1}g_{n+1,n} -X_n}{a}\ =\frac{{\rm e}^{a{\cal D}_\tau}-1}{a}X_n
\end{equation}
where $g_{n,n+1}$ are the gauge variables on links that parallel transport the unitary group element at $n+1$ back to that at $n$, $g_{n+1,n}=g_{n,n+1}^{-1}$ and $R_{n,n+1}$ parallel transports the vector index from $n+1$ back to $n$.
The lattice Laplacian is then given by 
\begin{equation}
  \label{LatticeCovLap}
  \Delta_{\Lambda,R,g}=\frac{2-e^{a {\cal D}_\tau}-e^{-a {\cal D}_\tau}}{a^2}\
  \end{equation}
and the action is a quadratic form
\begin{eqnarray}
    \label{LatAction1}
    S_\Lambda[X]&=&\sum_{n,n'=1}^\Lambda \frac{a}{2}X_{n'}(\Delta_{\Lambda,R,g}+m^2)_{n',n}X_n\,\nonumber\\
    &=&\sum_{n=1}^{\Lambda}\frac{1}{a}{\bf tr}\left\{-\,X_nR_{n,n+1}g_{n,n+1}X_{n+1}g_{n+1,n}+(1+\frac{\beta^2 m^2}{2\Lambda^2}) X_n^2\right\}.
\end{eqnarray}
and $X_{\Lambda+1}=X_1$ by periodicity of the thermal lattice.
Substituting a path ordered product, denoted  ${\cal P}$, one sees 
\begin{equation}\label{linkU}
g_{n,n+1} ={\cal P}\exp\left[i\int_{na}^{na+a}d\tau\,A(\tau)\right]\quad\hbox{and}\quad R={\cal P}\exp\left[\int_{na}^{na+a}d\tau\,\omega(\tau)\right],
\end{equation}
and taking $a$ to zero one recovers the continuum Laplacian.
The lattice partition function is then
\begin{equation}\label{ZNLambda_with-gs}
  Z_{\Lambda}(N,d)=\int_{{SO(d)}^\Lambda}\int_{{U(N)}^\Lambda} \int_{{\mathbb R}^{dN^2\Lambda}}\prod_{k=1}^\Lambda\mu_{d}(R_{k,k+1})\mu_{N}(g_{k,k+1})\frac{d^{dN^2}X_k}{(2\pi a)^{N^2}}{\rm e}^{-S_{\Lambda,g}}
\end{equation}
where
\begin{equation}\label{GaugeLatAction_v1}
S_{\Lambda,g}=\sum_{n,n'=1}^{\Lambda}\frac{a}{2}{\bf tr}(X_n (\Delta_{\Lambda,g}+m^2)_{n',n}X_n)\, .
\end{equation}
Eqn (\ref{ZNLambda_with-gs}) is a matrix integral over vector valued matrix variables, $X_n$, with Dyson measure $d^{dN^2}{\kern -2.5pt}X_n$,
for each the $\Lambda$ lattice sites, $n=1,\cdots,\Lambda$,
together with $\Lambda$ integrals over the group elements
$g_{n,n+1}$ and $R_{k,k+1}$ associated with each link. These latter integrals
have Haar measure, $\mu_{N}$ and $\mu_{d}$ respectively, both
normalised to unity and as shown in \cite{PIMQMPI} the group
invariance and normalization of the Haar measures ensures that the path
integral can be reduced to a single integration over $R$ and $g$ so
that (\ref{ZNLambda_with-gs}) becomes
\begin{equation}\label{ZLambda_g}
  Z_{\Lambda}(N,d)=\int \mu_{d}(R)\mu_{N}(g)\frac{z_{-}^{\frac{d N^2\Lambda}{2}}}{{\rm\bf det}[{\bf 1}-z_-^\Lambda R\otimes g\otimes g^{-1}]},
  \end{equation}
with ${\rm\bf det}$ denoting the d-fold matrix determinant.  and
$z_{\pm}=1+\frac{\mu^2}{2}\pm \sqrt{\mu^2(1+\frac{\mu^2}{4})}$,
$\mu=\frac{m\beta}{\Lambda}$ and $z_{+}z_{-}=1$.

Again following
\cite{PIMQMPI} one can evaluate (\ref{ZNLambda_with-gs}) and
take the continuum limit,  $\Lambda\rightarrow\infty$ keeping $\beta m$ fixed to obtain the continuum result
\begin{equation}
  Z_{\infty}(N,d)= \int \mu_{d}(R)\mu_{N}(g)\frac{{\rm e}^{-\frac{d N^2\beta m}{2}}}{{\rm\bf det}[{\bf 1}-{\rm e}^{-\beta m} R\otimes g\otimes g^{-1}]}\, .
\end{equation}
The numerator ${\rm e}^{-\frac{d N^2\beta m}{2}}$ comes from the zero
point energy of the quantum system of oscillators and can be removed to yield 
the generating function for the Hilbert-Poincar\'e series as the Molien-Weyl formula
\begin{equation}
\cZ(N,d;x)=\int \mu_{d}(R)\mu_{N}(g)\frac{1}{{\rm\bf det}[{\bf 1}-x R\otimes g\otimes g^{-1}]}\label{MolienWeyl}
\end{equation}

Since the determinant here is a class function, the resulting integration reduces to integration over the maximal torus. For $U(N)$ this amounts to integration over the $N$ diagonal phases in the diagonalized holonomy matrix $g$ with eigenvalues $z_i={\rm e}^{i\varphi_i}$. The measure becomes a product of Vandermonde determinants
\begin{equation}
  \Delta(z)={\displaystyle\prod_{1\leq i < j\leq N}}(z_i-z_j)
\end{equation}
 and we have the explicit expression
\begin{equation}
  \cZ(N,d;x)=\int\mu_{d}(R)\frac{1}{N!}\oint \prod_{i=1}^N\frac{dz_i}{2\pi i z_i}\Delta(z)\Delta(z^{-1}) \prod_{i=1}^N\prod_{j=1}^N\frac{1}{{\bf det}[{\bf 1}-x R z_i z_j^{-1}]}\label{ZnSOd-general}
\end{equation}
and the residual ${\bf det}$ is over the $d$-dimensional representation of the orthogonal group.

For $SO(d)$ one usually treats even and odd $d$ separately as their group theory is associated with the $D_m$ series of Lie algebras, for $d=2m$, and the $B_m$ series for $ d = (2m+1)$.
The maximal torus in both cases is parameterized by \( m \) angles \( \theta_1, \dots, \theta_m \) and reducing (\ref{ZnSOd-general}) to the maximal torus brings $R$ to block-diagonal form:
\begin{eqnarray}
\label{BlockRotationMatrices}
R_{SO(2m)}=\begin{bmatrix}
R_2(\theta_1) & & \\
& \ddots &  \\
& &  R_2(\theta_m)
\end{bmatrix}\qquad R_{SO(2m+1)}=\begin{bmatrix}
R_2(\theta_1) & & & \\
& \ddots & & \\
& & R_2(\theta_m) & \\
& & & 1
\end{bmatrix}
\end{eqnarray}
with 
\[ R_2(\theta) = \begin{pmatrix} \cos\theta & -\sin\theta \\ \sin\theta & \cos\theta \end{pmatrix}.
\]
The resulting measures when pulled back to the maximal torus become
\begin{equation} \mu_{2m}=\prod_{r=1}^m\frac{d\theta_r}{2\pi} \frac{\Delta_{m}^2}{2^{m-1}m!} \quad
\hbox{with}\quad
\Delta_{m}=\prod_{1\leq r<s\leq m}(2\cos(\theta_r)-2\cos(\theta_s))
\label{muSO2mv1}
\end{equation}
while
\begin{equation} \mu_{2m+1}=\prod_{r=1}^{m}\frac{d\theta_r}{2\pi}(2-2\cos(\theta_r))\frac{\Delta_{m}^2}{2^{m}m!}
\label{muSO2mp1v1}
\end{equation}

One can still diagonalize the rotation matrix but in terms of pairs of eigenvalues
\begin{equation}
  ({\rm e}^{i\theta_r},{\rm e}^{-i\theta_r})=(\zeta_r,\zeta_r^{-1})\, .
\label{zetadefined}
\end{equation}
The integration measures are then cast in a form suitable for contour integration:
\begin{eqnarray}
  &&\mu_{2m}(\zeta)=\prod_{r=1}^{m}\frac{d\zeta_r}{2\pi i\zeta_r}\frac{\Delta_m^2(\zeta)}{2^{m-1}m!}\cr
  \hbox{with}\qquad\Delta_m^2(\zeta)&=&\prod_{1\leq r<s\leq m}(1-\frac{\zeta_r}{\zeta_s})(1-\frac{\zeta_s}{\zeta_r})(1-\zeta_r \zeta_s)(1-\frac{1}{\zeta_r\zeta_s})
\label{muSO2mv2}
\end{eqnarray}
and
\begin{eqnarray}
\mu_{2m+1}(\zeta)=\prod_{r=1}^{m} \frac{d\zeta_r}{2\pi i\zeta_r}\prod_{r=1}^m(1-\zeta_r)(1-\frac{1}{\zeta_r})\frac{\Delta_m^2(\zeta)}{2^mm!}
\label{muSO2mp1v2}
\end{eqnarray}
and the $U(N)$ measure is 
\begin{equation}
  \mu_{N}(z)=\prod_{i=1}^N\frac{dz_i}{2\pi i z_i}\frac{\Delta(z)\Delta(z^{-1})}{N!}
  \label{muUN}
  \end{equation}
The classic reference for these formulae is \cite{WeylClassicalGroups}
 while modern presentations are \cite{BumpLieGroups,MeckesClassicalCompactGroups}.

The resulting determinants become polynomials in ratios and products of
the variables $z_i$ and $\zeta_r$ so that the integrand is a ratio of polynomials in these complex variables, which in principle can be evaluated by repeated contour integrations.  For $SO(2m)$, the partition function is then given by 
\begin{equation}\label{ZSO2m}
\cZ(N,2m;x)=\int\mu_{2m}(\zeta)\mu_{N}(z)\prod_{r=1}^{m}\prod_{i,j=1}^N
\frac{1}{(1-x\zeta_r z_iz_j^{-1})(1-x\zeta_r^{-1}z_iz_j^{-1})}
\end{equation}
while for $SO(2m+1)$ we get
\begin{equation}\label{ZSO2m+1}
  \cZ(N,2m+1;x)=\int \mu_{2m+1}(\zeta)\mu_{N}(z)\prod_{i,j=1}^N\frac{1}{(1-x z_iz_j^{-1})}\prod_{r=1}^{m}
  \frac{1}{(1-x\zeta_r z_iz_j^{-1})(1-x\zeta_r^{-1}z_iz_j^{-1})}.
  \end{equation}
The additional $1/(1-xz_iz_j^{-1})$ for the $B_m$ series arises from the additional 1 in the block-diagonal form (\ref{BlockRotationMatrices}).
%with the determinant for $SO(2m)$ becomes
%\[\frac{1}{{\rm\bf det}[{\bf 1}-x R(\zeta)\otimes g(z)\otimes g(z)^{-1}]}=\prod_{r=1}^{m}\prod_{i=1}^N\prod_{j=1}^N\frac{1}{1-x\zeta_r z_i z_j^{-1}}\frac{1}{1-x\zeta_r^{-1} z_i z_j^{-1}}  \]

One can evaluate these expressions exactly for sufficiently small $N$
and small $m$ or expand the expressions in $x$ and evaluate the coefficients
in the series expansion directly by contour integration.

For $N=1$ there is no $g$ integration and we have a system of $d$ oscillators transforming in the vector representation of $SO(d)$.
The resulting partition function is then  
\begin{equation}
  \cZ(1,d;x)=\frac{1}{1-x^2}\quad \hbox{ for all } d
\end{equation}
The next simplest cases that can be evaluated exactly are $N=2$
where we find
\begin{equation}
\cZ(2,2;x)  = \frac{1+x^4}{(1-x^2)^2(1-x^4)^2}
  \end{equation}
\begin{equation}
\cZ(2,3;x)  = \frac{1+x^9}{(1-x^2)^2 (1-x^3) (1-x^4)^2 (1-x^6)}
  \end{equation}
\begin{equation}
\cZ(2,4;x)  = \frac{1+x^{12}}{(1-x^2)^2 (1-x^4)^3 (1-x^6)^2}
  \end{equation}
and for all $d>4$ we have
\begin{equation}
Z(2,d;x)  = \frac{1+x^{12}}{(1-x^2)^2(1-x^4)^2(1-x^6)^2(1-x^8) }
  \end{equation}
The case of $d=2$, i.e. $SO(2)$ with general $N$ is equivalent to the
charge zero sector discussed in \cite{PIMQT}. One can
further understand the general result for $SO(d)$ with $d>4$ as
arising from the equivalence of a system of Hermitian $2\times 2$
matrices to a system of 4-vectors invariant under $SO(d)$ so that when
$d>4$ the additional rotations available in $SO(d)$ are redundant.

\begin{equation}
\cZ(3,3;x)=\frac{P(3,3;x)}{(1-x^2)^2 (1-x^3)^3(1-x^4)^3(1-x^5)^3(1-x^6)^3(1-x^7)(1-x^8)};
\end{equation}
\begin{eqnarray}
P(3,3;x)&=&1-2x^3+x^4-x^5+8x^6+5x^7+19x^8+15 x^9+35x^{10}\nonumber\\
     &&+23x^{11}+63x^{12}+57x^{13}+113x^{14}+116x^{15}+196x^{16}+193x^{17}\nonumber\\
     &&\quad+280x^{18}+262x^{19}+337x^{20}+304x^{21}+367x^{22}+307x^{23}\nonumber\\
     &&\qquad+367x^{24}+304x^{25}+337x^{26}+262x^{27}+280x^{28}+193x^{29}+196x^{30}\nonumber\\
     &&\qquad\quad+116x^{31}+113x^{32}+57x^{33}+63x^{34}+23x^{35}+35x^{36}+15x^{37}\nonumber\\
     &&\qquad\qquad+19x^{38}+5x^{39}+8x^{40}-x^{41}+x^{42}-2x^{43}+x^{46}
  \end{eqnarray}

We further note that the expressions for $SO(d)$ with $U(N)$ and $SU(N)$ have significantly different expressions, e.g. for $d=3$ we have
\begin{equation}
  \cZ(SU(2),3,x) = \frac{1}{(1 - x^2) (1 - x^3) (1 - x^4)};
\end{equation}
and for all $d>3$ we have
\begin{equation}
Z(SU(2),d;x)= \frac{1}{(1 - x^2) (1 - x^4) (1 - x^6)};
\end{equation}

\begin{equation}
  \cZ(SU(3),3;x) = \frac{P(SU(3),3;x)}{(1-x)(1-x^2)(1-x^3)^2(1-x^4)^2(1-x^5)^2(1-x^6)^3(1-x^7)(1-x^8)};
\end{equation}
\begin{eqnarray}
  P(SU(3),3;x)&=&1-x-x^3+x^4-x^5+2x^6+4x^8+x^9+3x^{10}\nonumber\\
  &&-2x^{11}+7x^{12}-3x^{13}+6x^{14}+11x^{16}+x^{17}+13x^{18}\nonumber\\
  &&\quad+x^{19}+11x^{20}+6x^{22}-3x^{23}+7x^{24}-2x^{25}+3x^{26}\nonumber\\
  &&\qquad+x^{27}+4x^{28}+2x^{30}-x^{31}+x^{32}-x^{33}- x^{35}+x^{36}
    \end{eqnarray}

\subsection{Gauging $ U(N) \times O(d)$}
It is straightforward to extend the above discussion from $SO(d)$ to $O(d)$.
For this one needs to include new group elements with determinant ${\bf det}R=-1$. These involve a reflection combined with a rotation $R$ and we denote such elements $R_{-}$   with ${\bf det}R_{-}=-1$. The two terms are to be summed with weight $\frac{1}{2}$ so that the $O(d)$ Molien-Weyl formula becomes 
\begin{eqnarray}
  \cZ_O(N,d;x)&=&\frac{1}{2}\int\mu_{d}(R)\frac{1}{{\bf det}[1-x R\otimes g\otimes g^{-1}]}\nonumber\\&&\qquad\quad+\frac{1}{2}\int\mu^{-}_{d}(R_{-})\frac{1}{{\bf det}[1-x R_{-}\otimes g\otimes g^{-1}]}\, .
  \end{eqnarray}
The net effect of extending to $O(d)$ from $SO(d)$ is that invariants built from the $\delta$'s survive while those involving the $\epsilon$ tensor are removed from the series.  Reflection in an axis changes the orientation and the sign of the $\epsilon$ tensor.

For the $B_m$ series the $R_-$ elements when brought to block diagonal form have a single $-1$ eigenvalue replacing the $+1$ in the
block diagonal form (\ref{BlockRotationMatrices}).  For the $D_m$ series,
since all eigenvalues must be on the unit circle and either come in complex
conjugate pairs or are real, the only options for the eigenvalues in the ${\bf det}R_{-}=-1$ sector are ${\rm e}^{\pm i\theta_r}$ and $\pm1$, and so to get determinant $-1$ we must have a two by two block of eigenvalues 
\begin{equation}
\label{BlockRotationMatrices2x2}
\sigma_3=\begin{bmatrix}
1 & &0  \\
0& & -1
\end{bmatrix}\, .
\end{equation}
So for $O(d)$ reducing (\ref{ZnSOd-general}) to block diagonal matrices yields the block-diagonal forms:
\begin{eqnarray}
\label{BlockRotationMatricesOd}
R_{SO(2m)}=\begin{bmatrix}
R_2(\theta_1) & & & \\
& \ddots & & \\
& &R_2(\theta_{m-1})&   \\
& &  &\sigma_3
\end{bmatrix}\qquad R_{SO(2m+1)}=\begin{bmatrix}
R_2(\theta_1) & & & \\
& \ddots & & \\
& & R_2(\theta_m) & \\
& & & -1
\end{bmatrix}
\end{eqnarray}
The eigenvalues are now $\{\zeta_1,\zeta_1^{-1},\cdots,\zeta_{m},\zeta_{m}^{-1},-1\}$ for $B_m$ and $\{\zeta_1,\zeta_1^{-1},\cdots,\zeta_{m-1},\zeta_{m-1}^{-1},1,-1\}$ for $D_m$.
Since the eigenvalues have changed so too does the measure, but it is still the modulus square of the Vandermonde for all eigenvalues.  We find that the measure for $B_m$ is given by
\begin{equation}
  \mu^{-}_{2m+1}=\prod_{r=1}^{m-1}\frac{d\theta_r}{2\pi}(2+2\cos(\theta_r))\frac{\Delta_{m-1}^2}{2^{m-1}(m-1)!}
  \end{equation}
while that for $D_m$ is given by
\begin{equation}
  \mu^{-}_{2m}=\frac{1}{2^{m-1}(m-1)!}\prod_{r=1}^{m-1}\frac{d\theta_r}{2\pi}(2-2\cos(\theta_r))(2+2\cos(\theta_r))\Delta_{m-1}^2
  \label{muO2m}
  \end{equation}
\begin{equation}
  \Delta_{m-1}=\prod_{1\leq r<s\leq m-1}(2\cos(\theta_r)-2\cos(\theta_s))
\end{equation}
and the integral over the final angle $\theta_m$ is trivial is trivial and performed.

One can again evaluate special cases of $Z_O(N,d)$ for small $N$ and $d$. One finds for example
\begin{equation}
  \cZ_O(2,2;x)=\frac{1}{2}\left(\frac{1+x^4}{(1-x^2)^2(1-x^4)^2}+\frac{1-x^4}{(1-x^2)^2(1-x^4)^2}\right)=\frac{1}{(1-x^2)^2(1-x^4)^2}
\end{equation}
and we see that the power series associated with $x^4$ is associated with the epsilon tensor, which begins with the lowest order invariant ${\bf tr}(\epsilon_{ab}X^aX^bX^cX^c)$.
\begin{eqnarray}
  \cZ_O(2,4,x)&=&\frac{1}{2}\left(\frac{(1-x^{4}+x^{8})(1-x^4)}{(1-x^2)^2(1-x^4)^3(1-x^6)^2}+\frac{(1-x^4+x^{8})(1-x^4)}{(1-x^2)^2(1-x^4)^3(1-x^6)^2}\right)\\
  &=&\frac{(1-x^{4}+x^{8})}{(1-x^2)^2(1-x^4)^3(1-x^6)^2}
  \end{eqnarray}
where in the first expression we have written $1+x^{12}=(1-x^4+x^8)(1-x^4)$ to exhibit the cancellation. For $d\ge6$ and $N=2$ there is no contribution from the $\epsilon$ tensor as there are not enough components to anti-symmetrize and
hen for all $d\ge6$
\begin{equation}
  \cZ_O(2,d;x)=Z(2,d;x)=\frac{1+x^{12}}{(1-x^2)^2(1-x^4)^2(1-x^6)^2(1-x^8) }\, .
\end{equation}
For odd $d$ the ${\bf det}R_{-}=-1$ contribution simply gives $\cZ(N,2m+1,-x)$ so that the $O(d)$ partition function only involves even powers of $x$.

\section{ Positive heat capacity for Hagedorn transition region in $d$-matrix model  }
\label{sec:HagInvFull}

Since the two groups $U(N)$ and $SO(d)$ are independent the
$SO(d)$ invariant case can be obtained by first considering a refined $U(N)$ case where each matrix $X^a$ has its own parameter $x_a$. The
refined $U(N)$ partition function is then
\begin{equation}
Z_{N}(x_1,\cdots,x_d)= \int \mu_{N}\prod_{a=1}^d\prod_{i,j=1}^N\frac{1}{1-x_a z_iz_j^{-1}}.
\label{RefinedUNPartitionFunction}
\end{equation}
By Taylor expanding in each of the $x_a$  arount $0$ one obtains,
at order $k$, a homogenous degree $k$ polynomial in the $d$ variables $x_a$ with
coefficients which depend on the $z_i$.
The $U(N)$ integration can be performed by contour integration the result being a homogeneous degree $k$ polynomial in the $x_a$ with constant coefficients. Further integration over $SO(d)$ gives an alternative computation of the coefficients $\cZ(N,d,k)$.

\subsection{The $U(N)$ gauged system with global $SO(d)$ symmetry.}

It is instructive to first review the gauged $U(N)$ case corresponding to
(\ref{RefinedUNPartitionFunction}) with $x_a=x={\rm e}^{-\beta} , \; \forall\;  a$ so that equation (\ref{RefinedUNPartitionFunction}) becomes
\begin{equation}
Z_{N}(x,\cdots,x)= Z(N,d;x)=\int \mu_{N}\prod_{i,j=1}^N\frac{1}{(1-x z_iz_j^{-1})^d}.
\label{UnRefinedUNPartitionFunction}
\end{equation}
then expanding in $x$ and integrating over the $z_i$ one obtains for small $x$
\begin{equation}
Z(N,d;x)=\sum_{k=0}^\infty Z(N,d,k)x^k
\label{smallxUnrefined}
\end{equation}
Given the efficiently computable expressions for $ Z ( N , d, k )$ as a sum over Young diagrams (see equation (9.24) in \cite{PIMQTNS}), we find it useful to tabulate the differences
%\footnote{Note that 
% $$\cZ(N,d,k)=\sum_{H=1}^N\Delta\cZ(H,d,k)\quad \hbox{with $\Delta\cZ(H,d,k)$ defined in (\ref{KronDim}).}$$}
$$\Delta Z(H,d,k)=Z(H,d,k)-Z(H-1,d,k) $$ which are obtained by summing over diagrams of a fixed height $H$. 
%% \quad\hbox{as }\quad Z(N,d,k)=\sum_{H=1}^N\Delta Z(H,d,k)$$
The analogous efficiently computable formula for the $SO(d)$ invariant micro-canonical state counting function $ \cZ( N , d, k )$,    is  equation \eqref{charformZNdk}  in Section \ref{sec:ExactCountingAndAlgorithms}.  For the $O(d)$ invariant case $ \cZ_O ( N , d , k ) $,  we derive \eqref{ZNdO}. We also tablulate the differences 
$$ \Delta \cZ ( H , d , k ) = 
\cZ( H , d ,  k ) - \cZ( H -1, d , k )$$ 
 which are also  obtained by summing over Young  diagrams of a fixed height $H$. 

Special cases of the canonical ensemble  $U(N)$ gauged partition functions $ Z ( N , d , x )$  ( see the equations \ref{RefinedUNPartitionFunction}) and (\ref{UnRefinedUNPartitionFunction} for example) are known for small $N$,
with the largest unrefined worked example being $N=7$ \cite{KristWil2020}.  Expanding in small $x$, as in (\ref{smallxUnrefined}) provides the microcanonical data necessary to extract energy temperature and specific heat. Tables of $\Delta Z(H,2,k)$ and $\Delta Z(H,4,k)$ are provided in Tables \ref{fig:Z2_delDegens} and \ref{fig:Z4_delDegens} respectively.

The microcanonical temperature is then obtained from the entropy $S(N,d,k)=\ln Z(N,d,k)$ with $k$ the energy via
\begin{equation}
T_{micro}^{-1}=\frac{dS}{dE}\, .
\label{TmicrofromEntropy}
\end{equation}
To approximate the derivative we use the discrete symmetric difference thoroughout this paper.
From our microcanonical data one can extract the microcanonical temperature as
\begin{equation}
T_{micro}(N,d,k)=\frac{2}{\ln \frac{Z(N,d,k+1)}{Z(N,d,k-1)}}\, .
\label{Tmicro}
\end{equation}

\begin{figure}[H] %
\centering
\begin{minipage}{0.3\textwidth}
	\includegraphics[width=\textwidth]{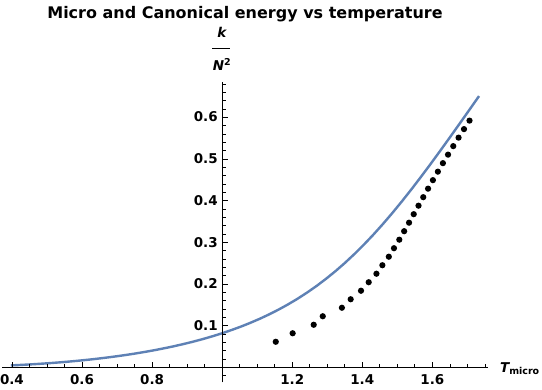} 
	\label{fig:Z2_MicroAndCanonicalEnergyTempZ2fig}
\end{minipage}
\hfill
\begin{minipage}{0.3\textwidth}
	\centering
	\includegraphics[width=\textwidth]{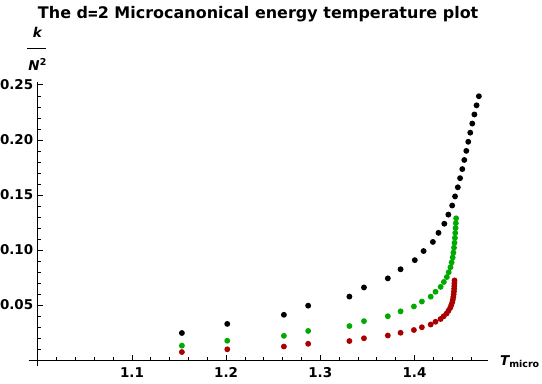} 
	\label{fig:Z2_MicroEvsT}
\end{minipage}
\hfill
\begin{minipage}{0.3\textwidth}
	\centering
	\includegraphics[width=\textwidth]{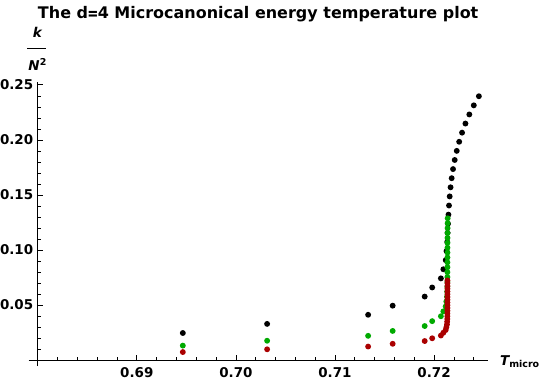}
	\label{fig:Z4_MicroEvsT}
\end{minipage}
\caption{The left figure shows plots of the canonical energy (continuous blue curve) and the microcanonical temperature for the $d=2$ matrix model with black $N=7$ using \cite{KristWil2020}.
	The middle figure shows plots of the energy as a function of the microcanonical temperature for the $d=2$ matrix model with black $N=11$, green $N=15$ and red $N=20$.
	The figure on the right shows plots of the energy as a function of the microcanonical temperature for the $d=4$ matrix model (black dots) $N=11$, green $N=15$ and red $N=20$. These micro-canonical plots are obtained from the SAGE code in Appendix E of \cite{PIMQTNS} implementing equation (9.24) in the paper \cite{PIMQTNS}. 
}
\label{fig:equivalenceofUngaguedEnsembles}
\end{figure}

In Figure \ref{fig:equivalenceofUngaguedEnsembles} the figure on the left provides a comparison of the microcanonical and canonical data for $N=7$. The middle and figure shows how the microcanonical curves change as $N$ is increased, showing how the curve approaches zero with increasing $N$ below the transition, and rises more sharply with increasing $N$ in the immediate neighbourhood of the transition at the Hagedorn temperature.  The monotononic rise of the curve shows that the heat capacity is always positive.
%\vfill\eject

The micro-canonical degeneracies $ Z ( N , d , k )$ are obtained from the formula (9.24) of \cite{PIMQTNS} which is 
\bea\label{ZNdkprev}  
Z ( N  , d , k ) = \sum_{p \vdash k } { d^{C_p}   \over \Sym ~ p }
 \sum_{ \substack { R \vdash k \\  l(R ) \le N  } }  ( \chi^R_p )^2    
\eea
The partition $p$ is presented as $ [ 1^{ p_1 } , 2^{ p_2 }, \cdots , k^{ p_k }]$ and $ \Sym ~ p = \prod_i i^{ p_i} p_i! $ 
while $ C_p = \sum_i p_i$. This was  obtained from the formula in terms of Kronecker coefficients, known from the earlier literature (see refs in \cite{PIMQTNS}) 
\bea 
Z ( N  , d , k ) = \sum_{ \substack { R \vdash k \\ l(R ) \le N }  } \sum_{ \substack { \Lambda \vdash k \\ l(\Lambda  ) \le d  }  }     C ( R , R , \Lambda )  \Dim_{ U(d) } ~ \Lambda 
\eea 
by  observing that the vanishing of  $ \Dim_{ U(d) } ~ \Lambda $ for $ l( \Lambda ) > d $ allows the restriction $ l( \Lambda ) \le d $ to be lifted. This allows a simplication of the Kronecker coefficient formula in terms of characters, using character orthogonality, to obtain \eqref{ZNdkprev}. 

\subsection{The $U(N)\times SO(d)$ gauged system.}
Substituting the $SO(d)$ or $O(d)$  dependence $x_1=\zeta_1 x$, $x_2=\zeta_1^{-1}x$,  $x_3=\zeta_2 x$, ...,  etc.  $\dots$ for $x_a$ as prescribed in Section \ref{sec2} allows one to perform the group integrating as a contour integration over the $\zeta_r$, after expanding in $x$, thus evaluating $\cZ(N,d,k)$ as the coefficients of $x^k$ of the refined partition function (\ref{RefinedUNPartitionFunction}). The result is a series expansion
\begin{equation}
\cZ(N,d;x)=\sum_{k=0}^{\infty}\cZ(N,d,k)x^k\quad\hbox{ and}\quad \cZ_O(N,d;x)=\sum_{k=0}^{\infty}\cZ_O(N,d,k)x^k
\end{equation}
The integers $\cZ(N,d,k)$ and $\cZ_O(N,d,k)$ provide us with the microcanonical
data we seek. These integers are efficiently computed by the methods described in Section 
\ref{sec:ExactCountingAndAlgorithms} using the SageMath code of Appendix \ref{sec:SageMathCode}. 

We present these numbers, as a set of tables, in Appendix \ref{Tables}
for both $SO(d)$ and $O(d)$ for $d=2$ to $d=6$ and up to $k=40$.  For completeness we also include the unrefined
microcanonical micorcanonical data for the $2$ and $4$ matrix models in Tables \ref{fig:Z2_delDegens} and \ref{fig:Z4_delDegens} with up to $k=30$.
The data is presented as $\Delta \cZ(H,d,k)$ so that
$\cZ(N,d,k)=\sum_{H=1}^N\Delta\cZ(H,d,k)$. The tables have a characteristic
upper triangular form with zeros in the lower triangle.
These zeros indicate that the respective $\cZ(N,d,k)$ have become independent of $N$, i.e. become stable, as adding further zeros leaves them unchanged. 
Each column corresponding to a particular value of $k$ has a characteristic
`carrot' shape reaching a maximum at some value $k_{max}(H)$. The entropies $ \log  \cZ(  N , d, k )$ and associated 
micro-canonical curves with negative heat capacity, and  a characteristic caloric fold with a turning point at $ k_{ \crit } \sim { N^2 \over 4 } $ will be studied in detail in the following sections. 
% This maximum appears to grows as $\frac{N^2}{4}$ which is characteristic of
% the approach to the Hagedorn transition.

\begin{figure}[H] %  
\centering
\begin{minipage}{0.3\textwidth}
	\centering
	\includegraphics[width=\textwidth]{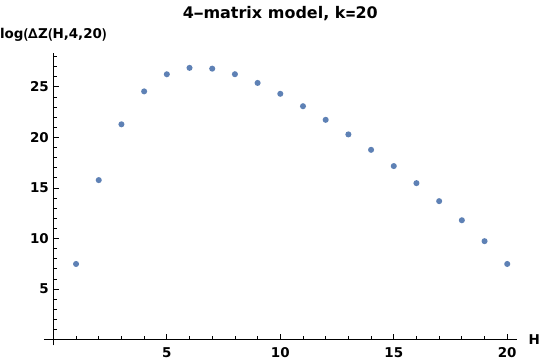} 
	\label{fig:LogDelZd4ofHk20fig}
\end{minipage}
\hfill
\begin{minipage}{0.3\textwidth}
	\centering
	\includegraphics[width=\textwidth]{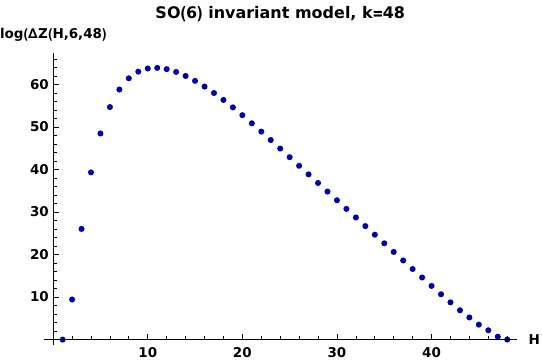} 
	\label{fig:LogDelZofHk24SO6fig}
\end{minipage}
\hfill
\begin{minipage}{0.3\textwidth}%[H] %  
	\centering
	\includegraphics[width=\textwidth]{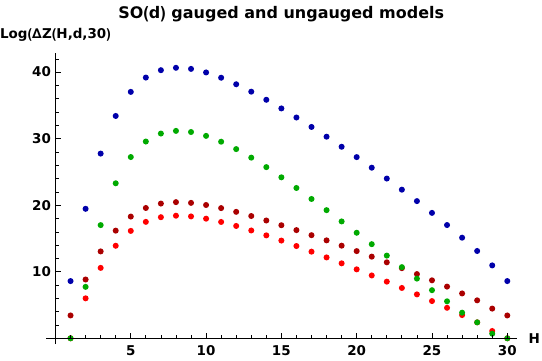} 
	\label{fig:LogDelZofHk30SO6fig}
\end{minipage}
\caption{The three figures show $\ln\Delta Z(H,d,k)$ and $ \ln \Delta \cZ( H , d , k )$  for different models. The figure on the left shows the pure $4$-matrix model and column 20 of Table \ref{fig:Z2_delDegens} and Table \ref{fig:Z4_delDegens} and the maximum occurs for $k_{\max}=5$. The middle figure shows the $SO(4)$ invariant model and the equivalent of column 48 from Table \ref{fig:SO4_delDegens} (not listed in the table), here the maximum occurs at $k_{\max}=11$. The figure on the right compares both $SO(d)$ gauged and ungauged models for $d=2$ and $d=4$ with and $k=30$. The darker red is the two matrix model the brigher red the $SO(2)$ gauged version while the dark blue is the $4$-matrix model and the green the $SO(4)$ invariant model. The maxima all occur at $k_{\max}=8$. Note that the graph for the $SO(d)$ gauged model is always lower than that for the ungauged model as the former has fewer invariants that can contribute. }
%\caption{Two figures}
%\label{Twofigs}
\end{figure}

\section{The low-energy large $N$ Limit: lower branch of the $E$ vs $T$ caloric fold. }
\label{sec:LowEnergy}

In this section, we use the known large $N$ result for the generating function of the $U(N)$ invariants of $d$ matrices   to obtain the asymptotics of the micro-canonical degeneracies 
 $ \cZ( N , d, k )$ and $ \cZ_O ( N , d , k )$ in the regime $ N \ge   k >> 1$, which is valid for $ d $ of order $1$, which is of physical interest here. This is given in section \ref{sec:largekGend}.  The derivation starts from  
 the $n=1$  term 
 \bea\label{faclead}
  { 1 \over ( 1 - x_1 - x_2 \cdots - x_d )} 
 \eea  
  of the  infinite product generating function of $ U(N)$ invariants for $d$ matrices 
   \bea\label{InfProd}
  \prod_{n=1}^{ \infty } { 1 \over ( 1 - x_1^n  - x_2^n  \cdots - x_d^n  )} 
  \eea 
 This formula for $d$-matrix invariants is explained in the context of the finite $N$ construction of orthogonal 
 multi-matrix  bases  in Appendix C of  \cite{BHR1}, and related discussions are in \cite{Willenbring,BDHO,Dolan,Collins2008}. The leading term is known to control the Hagedorn temperature and the leading high energy asymptotics of the partition function as discussed in \cite{Sundborg:1999ue} \cite{Aharony:2003sx}. Our derivation of the asymptotics of  $U(N) \times SO(d)$ invariants  in section \ref{sec:largekGend} uses the  $SO(d)$ projection of \eqref{faclead}. Similarly the $SO(d)$ projection of (\ref{InfProd}) gives the stable coefficients for $N\ge k\ge0$.
  The coefficient of $ x_1^{ m_1 } \cdots x_d^{ m_d }$ from the expansion \eqref{faclead}  has the interpretation in terms of the counting of vectors in 
  $ V_d^{ \otimes ( m_1 + \cdots + m_d ) }$ having charges $ ( m_1 , m_2 , \cdots , m_d )$  for $ U(1)^{ d} \subset U(d)$.  
 In section \ref{sec:GenFuncsFinited} the projection to the $SO(d)$ invariant subspace  is computed using an $SO(d)$ group integration. This has an interpretation as the generating function for the dimensions of  $SO(d)$ invariant subspaces 
 of $ V_d^{ \otimes k }$.   While our main interest, from the point of view of  establishing the caloric fold for the $ U(N) \times SO(d)$ invariants is the asymptotic result in section \ref{sec:largekGend}, we pause  in \ref{sec:GenFuncsFinited} to derive the exact generating functions for $SO(d)$ invariants in $ V_d^{ \otimes k }$ for the cases of $ d = 2,3,4,5$ since they are accessible using the explicit group integration method we use later, and produce explicit results in terms of special functions which have not been discussed in the context of matrix theory previously.

\subsection{ $SO(d)$ invariants in $ V_d^{ \otimes k }$ for low values of $d$  } 
\label{sec:GenFuncsFinited} 

The $U(N)$ partition function in the limit of large $N$  
with $x_a<d^{-1}$, $a=1,\cdots, d$ can be obtained exactly as
\begin{equation}
  Z_{N}(x_1,\cdots,x_d)= \int \mu_{N}\prod_{a=1}^d\prod_{i,j=1}^N\frac{1}{1-x_a z_iz_j^{-1}}\xrightarrow{N\rightarrow \infty} Z(x_1,\cdots,x_d)\, ,
  \label{ZN(x1,..,xd)}
\end{equation}
where
\begin{equation}
Z(x_1,\cdots,x_d)= \prod_{n=1}^\infty\frac{1}{1-\sum_{a=1}^dx_a^n}.
 \label{ZNstable(x1,..,xd)}
\end{equation}
is the generating function for the refined stable sequence. 
When all $x_a=x$ we have
\begin{equation}
  Z(d;x)=\prod_{n=1}^\infty\frac{1}{1-d x^n}
\label{UngaguedLowTempStable}
\end{equation}
which when expanded gives the exact values for $Z(N,d,k)$ with $k\le N$, i.e. it generates the stable sequence for the unrefined $d$-matrix model.
By taking only the $n=1$ term, which governs the nearest singularity to the origin,  one gets the generating function for the large order terms as
\begin{equation}
  Z(d;x) \simeq\frac{1}{\phi(\frac{1}{d})} \frac{1}{1-d x}=\sum_{k=1}^\infty \frac{d^k}{\phi(\frac{1}{d})} x^k
\label{LowEnergyUngagued}
  \end{equation}
where $\phi(q)=(q;q)_{\infty}$ is Euler's function and $(q;x)_{\infty}$ is the q-Pochhammer symbol.
Hence:
\begin{equation}
  Z(N,d,k)\sim \frac{d^k}{\phi(\frac{1}{d})}\quad \hbox{and}\quad N\ge k\gg 1,\quad\hbox{with}\quad N\sim\infty\, .
\label{UngaguedLowTemp}
\end{equation}

By substituting the $\zeta$ dependence of $SO(d)$ into Eqn (\ref{ZNstable(x1,..,xd)}) and projecting to $SO(d)$ invariants, we obtain $Z_{d}(x)$ defined as
\begin{equation}\label{SO2mNinfty}
  \cZ_{2m}(x)=\int\mu_{2m}\prod_{n=1}^\infty\frac{1}{1-
    x^n\sum_{r=1}^m (\zeta_r^n+\zeta_r^{-n})}
\end{equation}
and
\begin{equation}\label{SO2mp1Ninfty}
  \cZ_{2m+1}(x)=\int\mu_{2m+1}\prod_{n=1}^\infty\frac{1}{1-
    x^n-x^n\sum_{r=1}^m (\zeta_r^n+\zeta_r^{-n})}\, .
\end{equation}
By expanding (\ref{SO2mNinfty}) and (\ref{SO2mp1Ninfty}) in $x$ and then
doing the integrations one can obtain the stable coefficients for $SO(d)$
for different $d$. These provide the sum of the columns of the tables
in Appendix \ref{Tables}.

The nearest singularity to the origin arises in the $n=1$ terms of (\ref{SO2mNinfty}) and (\ref{SO2mp1Ninfty}), which diverges at $x=\frac{1}{d}$ 
a value that coincides with the ungauged value, so the Hagedorn temperature
remains $T_H=\frac{1}{\ln d}$. The large order coefficients are then well approximated by 
\[ \cZ_{2m}(x)=\frac{1}{\phi(\frac{1}{2m})}\hat{Z}_{2m}(x)
  \quad\hbox{and}\quad \cZ_{2m+1}(x)=\frac{1}{\phi(\frac{1}{2m+1})}\hat{Z}_{2m+1}(x)
\]
where 
\begin{equation}\hat{Z}_{2m}(x)=\int\mu_{2m}(\zeta)\frac{1}{1-x\sum_{r=1}^m(\zeta_r+\zeta_r^{-1})} \label{Zhat2m}
\end{equation}
and
\begin{equation}\hat{Z}_{2m+1}(x)=\int\mu_{2m+1}(\zeta)\frac{1}{1-x-x\sum_{r=1}^m(\zeta_r+\zeta_r^{-1})} \label{Zhat2m+1}
\end{equation}
which are themselves generating functions for the dimensions of the $k$-fold tensor product of the vector representation of $SO(2m)$ and $SO(2m+1)$ respectively and their expansion gives the integer dimensions of these representations. Explicitly for small $m$ and we find
\begin{eqnarray}\label{SpecFuncsSO} 
  \hat{Z}_{2}(x)&=&\frac{1}{\sqrt{1-4x^2}}\cr
  \hat{Z}_{3}(x)&=&\frac{1}{2x}\left(1-\sqrt{\frac{1-3x}{1+x}}\right)\cr
  \hat{Z}_{4}(x)&=&\frac{1}{x^2} \left(\frac{2 {\bf E}(16 x^2)-(1-16x^2){\bf K}(16 x^2)}{2\pi}-\frac{1}{4}\right)=
%{}_4F_3\!\left(1,1,\tfrac12,\tfrac12;2,2,1;16x^2\right)=
{}_3F_2\!\left(
\begin{matrix}
1,\tfrac12,\tfrac12\\
2,2
\end{matrix}
;16x^2
\right)\cr
  \hat{Z}_{5}(x)&=&\frac{1}{4x^3\sqrt{1+2x-3x^2}}\left((1+3x^2)\sqrt{1+2x-3x^2}-(1-5x)^3 F(2,x)\right.\nonumber\\
  &&\qquad\left.-15(1-5 x)^2xF(3,x)-80(1-5x)x^2F(4,x)-140x^3F(5,x)\right)  \cr 
  && 
  \end{eqnarray}
where $F(n,x)$ is the Appell series
$$F(n,x)=F_1(\frac{1}{2}, \frac{1}{2}, \frac{1}{2}, n, \frac{4 x}{1+3x}, \frac{4 x}{1 - x})$$
and we use the convention
\[
E(m)=\int_0^{\pi/2}\sqrt{1-m\sin^2\theta}\,d\theta 
\quad\hbox{and}\quad
K(m)=\int_0^{\pi/2}{d\theta\over \sqrt{1-m\sin^2\theta}},
\]
so that \(K\) and \(E\) are regarded as functions of the parameter
\(m\).  The formula for \(\widehat Z_5(x)\) involves the Appell
function \(F_1\), a two-variable generalisation of the Gauss
hypergeometric function.  For standard references on elliptic integrals
and Appell functions see, for example, \cite{DLMF,ErdelyiHTF1}.
Then the leading large $N$ behaviour of the low temperature partition functions are given, in thems of these generating functions by
\begin{equation}\cZ(N,d,x)=\cZ_d(x)\simeq\frac{1}{\phi(\frac{1}{d})}\hat{Z}_{d}(x)\, .
  \end{equation}

Inspecting (\ref{Zhat2m}) and (\ref{Zhat2m+1}) we see that for $d=2m$ there are two equi-distant singularities nearest to the origin. They occur at $\zeta_r=\pm 1$ so that $\displaystyle\sum_{r=1}^m(\zeta_r+\zeta_r^{-1})=\pm 2m$ which corresponds to $R={\bf 1}$ or $R=-{\bf 1}$, and reflects the fact that only even powers of $x$ arise.  For $d=2m+1$ the singularity occurs at $\zeta_r=1$ i.e. at $R={\bf 1}$ and both even and odd powers of $x$ are present.

In Appendix \ref{LargeNGeneratingFunctions} we extract the asymptotic form of the coefficients of $\hat{Z}_d(x)$ for small values of $d$.  Defining
\begin{equation}
  \hat{Z}_{d}(x)\simeq\sum_{n=0}^\infty \hat{\cZ}_k(d)x^k
  \end{equation}
we find the coefficients $\hat{\cZ}_k(d)$ are of the form
\begin{equation}
  \hat{\cZ}_k(d)\simeq c(d)\frac{d^k}{k^{\nu(d)}}
\end{equation}
In particular for even $d$ only even powers of $x$ occur. For $d=2$ with $k=2n$
we have 
\begin{eqnarray}\label{asymp2} 
  \hat{\cZ}_{2n}(2)&=&\frac{(2n)!}{(n!)^2}\simeq\frac{2^{2n}}{\sqrt{\pi n}}\quad\hbox{with}\quad \nu(2)=\frac{1}{2}\quad\hbox{and}\quad c(2)=\sqrt{\frac{2}{\pi}}
\end{eqnarray}
\noindent
For $d=3$ the expansion gives $c_k(2)=r_k$ the Riordan numbers \href{https://oeis.org/A005043}{https://oeis.org/A005043}
\begin{equation}
  r_k = \frac{(k - 1)}{k+1}(2r_{k-1} + 3r_{k-2})\quad \hbox{with}\quad r_0 = 1;r_1 = 0;
  \label{RiordanNumbers}
  \end{equation}
For large $n$ we have the asymptotic form
\begin{eqnarray}\label{asymp3} 
  \hat{\cZ}_k(3)&=r_k\simeq&\frac{3\sqrt{3}}{8\sqrt{\pi}}\frac{3^k}{k^{3/2}}\quad\hbox{with}\quad \nu(3)=\frac{3}{2}\quad\hbox{and}\quad c(3)= \frac{3\sqrt{3}}{8\sqrt{\pi}}
  \end{eqnarray}
\noindent
Expanding $\hat{Z}_4(x)$ gives 
\begin{equation}\label{asymp4} 
\hat{Z}_4(x)=\sum_{n=0}^\infty C_n^2x^{2n}%=\sum_{n=0}^\infty\frac{((2 n)!)^2}{(n+1)^2(n!)^4}x^{2n}
\end{equation}
where $C_n = \frac{1}{n+1} \binom{2n}{n} = \frac{(2n)!}{(n+1)!n!}$
are the Catalan numbers. For large $c_k(4)$ we then have the asymptotic
result
\begin{equation}
\hat{\cZ}_{2n}(4)\sim\frac{4^{2n}}{\pi n^3}\quad\hbox{with}\quad \nu(4)=3\quad\hbox{and}\quad c(4)=\frac{8}{3}
\end{equation}
\noindent
For $SO(5)$ the nonanalyticity of $\hat{Z}_5(x)$ occurs at $x=\frac{1}{5}$ where the final argument of the Appell series approaches 1. Using the integral representation of $\hat{Z}_5(x)$ we find the asymptotic behaviour of the coefficients for large n is given by
\begin{equation}\label{asymp5} 
  \hat{\cZ}_{k}(5)\sim \frac{9375}{128 \pi}\frac{5^k}{k^{5}}\quad\hbox{with}\quad \nu(5)=5\quad\hbox{and}\quad c(5)=\frac{9375}{128\pi}
  \simeq{23.3137}\, .
\end{equation}

For $SO(d)$ with $d\ge6$ we have less analytic control over the expressions
however, as we will see below, our best estimate is 
the asymptotic form\footnote{This form is motivated by the asymptotics of $n!$
which yields the more precise estimate for $\hat{\cZ}_{2n}(2)\simeq\frac{2^{2n}}{\sqrt{\pi n}}{\rm e}^{-\frac{1}{8n}}$\,.}
$$ \hat{\cZ}_k(d)\sim A \frac{d^k}{k^\nu}{\rm e}^{-\frac{B}{k}}$$
and fitting requiring $A>0$ and $\nu$ the nearest integer or half integer
fits the data well. A comparison fit for $SO(5)$ gives
$$\hat{\cZ}_k(5)\sim 23.3\frac{5^k}{k^5}{\rm e}^{-\frac{11.64}{n}}$$
with the numerical coefficient $23.3$ and ${\rm e}^{-11.64/n}$ extracted from fitting the first 300 terms in the series expansion. This is excellent agreement
with $c(5)=\frac{9375}{128\pi}\simeq{23.3137}$. Similarly, with this prescription we obtain based on the first 32 coefficients, the fitting estimate 
\begin{equation}
  \hat{\cZ}_{2n}(6)\sim 31.8\frac{6^{2n}}{n^{15/2}}{\rm e}^{-\frac{25.4}{n}} \quad\hbox{and}\quad  \hat{\cZ}_k(7)\sim 0.5\frac{7^{k+8}}{k^{21/2}}{\rm e}^{-\frac{47.}{k}}\, .
\end{equation}
Note from (\ref{c(2m)andc(2m+1)}) below:
$\frac{c(6)}{2^{15/2}}=\frac{6561\sqrt{3}}{64 \pi^{3/2}}\simeq 31.89$ and $\frac{c(7)}{7^8}=\frac{2205 \sqrt{7}}{2048 \pi^{3/2}}\simeq 0.5116$ in excellent agreement with the fits.

%For $d=7,8$ and $9$ we have $c(7)=\frac{2205 \sqrt[7]7^8}{2048 \pi^{3/2}}\simeq 0.5116 7^8$,
%though we have slightly less analytic control. Using numerical fits we obtain
%\begin{equation}
%  \hat{\cZ}_k(7)=\sim 0.5\frac{7^{k+8}}{k^{21/2}}{\rm e}^{-\frac{47.}{k}}\, ,
%\quad  \hat{\cZ}_{2n}(8)\sim \frac{8^{2n}}{n^{14}}{\rm e}^{15.07 -\frac{42.8}{n+1}}
%\quad\hbox{and}\quad
%  \hat{\cZ}_k(9)\sim 4.95 \frac{9^{k+18}}{k^{18}}{\rm e}^{-\frac{31.5}{k}}
%  \end{equation}
%in excellent agreement with the asypotitic estimates below.

\subsection{Large $k$ coefficients for general  sufficiently small  $d$}
\label{sec:largekGend}

The large order coefficients are governed by the singularity nearest the origin in (\ref{Zhat2m}) and (\ref{Zhat2m+1}). For $SO(2m)$ this occurs at the identity $R={\bf 1}$ and $x=1/d$ or $R={\bf -1}$ and $x=-1/d$ respectively. Both are equi-distant from the origin and contribute equally. For $SO(2m+1)$ only the identity and $x=1/d$ contributes as the next nearest singlarity occurs for $x=-1/(d-1)$ and $\zeta_r=-1$.  The singular contribution comes from infinitesimal regions of the group around these singularities, i.e. solid spheres, ${\mathbb B}^m(\delta)\in {\mathbb R}^m$, of radius $\delta$ obtained by expanding $\zeta_r$, in the integrals (\ref{Zhat2m}) and (\ref{Zhat2m+1}),  to quadratic order around $\zeta_r=1$ or $\zeta_r=-1$.

The contribution from the neighbourhood of the identity for $d=2m$ is captured by
 \begin{equation}
  \hat{Z}^{\delta}_{d}(x,{\bf 1})=\int_{{\mathbb B}^m(\delta)}\frac{d^m\theta}{(2\pi)^m}\prod_{1\leq r<s\leq m}\frac{(\theta_r^2-\theta_s^2)^2}{2^{m-1}m!}\frac{1}{1-dx+x\displaystyle\sum_{r=1}^m\theta_r^2}\, .
  \label{B2m}
 \end{equation}
 and that from $R=-{\bf 1}$ is given by
 \begin{equation}
  \hat{Z}^{\delta}_{d}(x,-{\bf 1})=\int_{{\mathbb B}^m(\delta)}\frac{d^m\theta}{(2\pi)^m}\prod_{1\leq r<s\leq m}\frac{(\theta_r^2-\theta_s^2)^2}{2^{m-1}m!}\frac{1}{1+dx-x\displaystyle\sum_{r=1}^m\theta_r^2}\, .
  \label{B2m-}
 \end{equation}
 with the sum resulting in an expression invariant under $x\rightarrow -x$.
 Similarly the expression for $d=2m+1$ is given by 
\begin{equation}
 \hat{Z}^{\delta}_{2m+1}(x,{\bf 1})=\int_{{\mathbb B}^m(\delta)}\frac{d^m\theta}{(2\pi)^m}\prod_{r=1}^m \theta_r^2\prod_{1\leq r<s\leq m}\frac{(\theta_r^2-\theta_s^2)^2}{2^{m}m!}\frac{1}{1-dx+x\displaystyle\sum_{r=1}^m\theta_r^2}\label{B2m+1}\, .
\end{equation}
Going to radial coordinates by taking $\theta_r=r n_r$ with $n_r$ a unit vector in ${\mathbb R}^m$ and defining 
 \begin{equation}
  A_{2m}=\int_{S^{m-1}}\frac{1}{(2\pi)^m}\prod_{1\leq r<s\leq m}\frac{(n_r^2-n_s^2)^2}{2^{m}m!}
  \label{A2m}
\end{equation}
and
\begin{equation}
  A_{2m+1}=\int_{S^{m-1}}\frac{1}{(2\pi)^m}\prod_{r=1}^m n_r^2\prod_{1\leq r<s\leq m}\frac{(n_r^2-n_s^2)^2}{2^{m}m!}\, .\label{A2m+1}
\end{equation}
we can write
\begin{eqnarray}
  \hat{Z}_{2m}^{\delta}(x)&=&\int_0^\delta\frac{dr}{r}\frac{r^{d(d-1)/2}}{2}\{\frac{1}{1-d x+x r^2}+\frac{1}{1+d x-x r^2}\}A_{d}\nonumber\\
  \hbox{and}\\
  \hat{Z}_{2m+1}^{\delta}(x)&=&\int_0^\delta\frac{dr}{r}\frac{r^{d(d-1)/2}}{1-d x+x r^2}\nonumber
\label{Zhatdelta}
\end{eqnarray}
Expanding in $x$ we obtain
\begin{equation}
\cZ_{d}^{\delta}(x)=\sum_{k=0}^\infty \int_0^{\delta}\frac{dr}{r}r^{d(d-1)/2}{(1-\frac{r^2}{d})}^k d^k A_{d} x^k\, .
\end{equation}
and for $d=2m$ only even $k$ contribute. Since the integration over $r$ is concentrated at small $r$, for large $k$ we can replace
$$(1-\frac{r^2}{d})^k\simeq {\rm e}^{-\frac{k}{d}r^2}$$
and send $\delta$ to $\infty$. This gives our asymptotic estimate for the coefficients as
\begin{equation}
  \hat{\cZ}_d(k)=d^k\int_0^{\infty}\frac{dr}{r}r^{d(d-1)/2}{\rm e}^{-\frac{k}{d}r^2}A_{d}
= \frac{1}{2}d^{\frac{d(d-1)}{4}}\Gamma(\frac{d(d-1)}{4})A_d\frac{d^{k}}{k^{\frac{d(d-1)}{4}}}
\end{equation}
thus establishing that
\begin{equation}
  \hat{\cZ}_d(k)=c(d)\frac{d^k}{k^{\frac{d(d-1)}{4}}}
    \label{LowEnergyCoefficientsSOdGauged}
  \end{equation}
with the exponent 
\begin{equation}
  \boxed{\nu(d)=\frac{d(d-1)}{4}}\quad\hbox{and}\quad  \boxed{ c(d)=\frac{1}{2}d^{\frac{d(d-1)}{4}}\Gamma(\frac{d(d-1)}{4})A_d}
  \label{nuAndcd}
\end{equation}
which as we will see in section \ref{sec:EigDensAnalytic} matches continuously with the result from our analysis above the Hagedorn transition energy. We therefore conclude that in the regime $N\ge k\gg 0$ where finite $N$ effects are absent 
\begin{equation}
  \cZ(N,d,k)\simeq\frac{c(d)}{\phi(\frac{1}{d})}\frac{d^k}{k^{\frac{d(d-1)}{4}}}\, .
    \label{LowEnergyCoefficientsSOdGaugedv2}
\end{equation}
with $k=2n$ for $d=2m$ as the odd term $k$ terms calcel in the sum.
Noting the integrals integrals
\begin{equation}
\int_{-\infty}^{\infty}\frac{dx_1}{2\pi} \dots \int_{-\infty}^{\infty}\frac{dx_m}{2\pi}\prod_{1 \le i < j \le m} (x_i^2 - x_j^2)^2 e^{-\scalebox{0.6}{$\frac{1}{2}\displaystyle\sum_{k=1}^m x_k^2$}}=\frac{m!}{(2\pi)^{\frac{m}{2}}} \prod_{j=1}^{m-1}\cdot (2j)!
\end{equation}
\begin{equation}
\int_{-\infty}^{\infty}\frac{dx_1}{2\pi} \dots \int_{-\infty}^{\infty}\frac{dx_m}{2\pi} \prod_{k=1}^m x_k^2  \prod_{1 \le i < j \le m} (x_i^2 - x_j^2)^2 e^{-\scalebox{0.6}{$\frac{1}{2}\displaystyle\sum_{k=1}^m x_k^2$}} \, = \frac{1}{2^m(2\pi)^{\frac{m}{2}}} \prod_{j=1}^{m}(2j)!
\end{equation}
we get 
\begin{equation}
  c(2m)=2\frac{m^{m(2m-1)/2}}{2^{m-1}(2\pi)^{\frac{m}{2}}}
  \prod_{j=1}^{m-1} (2j)! \,\quad\hbox{and}\quad
  c(2m+1)=\frac{(m+\frac{1}{2})^{m(2m+1)/2}}{2^{2m}m!(2\pi)^{\frac{m}{2}}}
  \prod_{j=1}^{m-1} (2j)!
  \label{c(2m)andc(2m+1)}
\end{equation}
and $A_d$ is then given by (\ref{nuAndcd}) and the leading $2$ in $c(2m)$ arises from summing the two terms in (\ref{B2m}).  %$A_d=\frac{c(d)}{\frac{1}{2}d^{\frac{d(d-1)}{4}}\Gamma(\frac{d(d-1)}{4})}$.
As we will see in Section \ref{sec:HighEnergLargeN} the same scaling and coefficients arise in the expansion above the Hagedorn temperature.

As $d$ increases, the leading form of the entropy $ S \sim ( k \log d - { d^2 \over 4}  \log k ) $  obtained here becomes negative for $ d \sim \sqrt { k }$, thus limiting the range of validity of the present derivation.  A treatment of the large $d$ limit, using ribbon graphs and their asymptotics is given in section \ref{sec:calfolLarged}.

\subsection{Large $k$ coefficients for $O(d)$ with fixed finite $d$}

We can perform a similar analysis for the stable coefficients for $O(d)$. These are given by
\begin{equation}
\cZ_{Od}(x)=\frac{1}{2}\cZ_d(x)+\frac{1}{2}\cZ_d^{-}(x)
\end{equation}
where $\cZ_d(x)$ is given by (\ref{SO2mNinfty}) and (\ref{SO2mp1Ninfty}) and
\begin{equation}
  \cZ_{2m}^{-}(x)=\int \mu^{-}_{2m}\prod_{n=1}^\infty\frac{1}{1-x^n\sum_{r=1}^{m-1}(\zeta_r^n+\zeta_r^{-n})}
  \label{cZ^{-}_{2m}}
\end{equation}
and
\begin{equation}
  \cZ_{2m+1}^{-}(x)=\int \mu^{-}_{2m+1}\prod_{n=1}^\infty\frac{1}{1-(-x)^n-x^n\sum_{r=1}^{m-1}(\zeta_r^n+\zeta_r^{-n})}\, .
  \label{cZ^{-}_{2m+1}}
\end{equation}
Expansion in $x$ and performing the contour integrations then gives the coefficients $ \cZ_O( N  , d, k )$  for $O(d)$. However this procedure is only practical for rather small $k$ and a more systematic approach is that of Section \ref{sec:ExactCountingAndAlgorithms} and equation (\ref{PairingO}).

The integrals above are very useful in obtaining the large $k$ asymptotics, for general small $d$, and large $N \ge  k$. 
 The leading singularities near the origin are captured by the $n=1$ term  so that the generating function for these large order terms for $O(d)$ is given by
retaining only these and we have
\begin{equation}
  \hat{Z}_{Od}(x)=\frac{1}{2}\hat{Z}_{d}(x)+\frac{1}{2}\hat{Z}^{-}_{d}(x)
\end{equation}
where $\hat{Z}^{-}_d(x)$ is given by the $n=1$ term in (\ref{cZ^{-}_{2m}}) and (\ref{cZ^{-}_{2m+1}}) for $d=2m$ and $d=2m+1$ respectively.
We can evaluate these again analytically for small values of $d$
finding
\begin{eqnarray}\label{SpecFuncsO} 
  \hat{Z}_{O2}(x)&=&\frac{1}{2}(\frac{1}{\sqrt{1-4x^2}}+1)\cr
  \hat{Z}_{O3}(x)&=&\frac{1}{4x}\left(\sqrt{\frac{1-3x}{1+x}}-\sqrt{\frac{1-3x}{1+x}}\right)\cr
  \hat{Z}_{O4}(x)&=&\frac{1}{2x^2} \left(\frac{2 {\bf E}(16 x^2)-(1-16x^2){\bf K}(16 x^2)}{2\pi}-\frac{1}{4}\right)+\frac{1}{4x^2}(1-\sqrt{1-4x^2})
  \end{eqnarray}
and $\hat{Z}_{O5}(x)=\frac{1}{2}(\hat{Z}_5(x)+\hat{Z}_{5}(-x))$.\\
Expanding $\hat{Z}_{d}(x)$ to obtain the coefficients we get
\begin{eqnarray}
  \hat{Z}_{O2}(x)&=&1+\frac{1}{2}\sum_{n=2}^\infty\frac{(2n)!}{(n!)^2}x^{2n}\nonumber\\
  \hat{Z}_{O3}(x)&=&\sum_{n=0}^\infty r_{2n}x^{2n}\\
    \hat{Z}_{O4}(x)&=&\sum_{n=0}^{\infty}\frac{1}{2}(C_n^2+C_n)x^{2n}\nonumber\, ,
    \end{eqnarray}
where $r_{2n}$ are Riordan numbers and $C_n$ are the Catalan numbers introduced earlier.

Analysis of $\hat{Z}^{-}_{d}(x)$ follows similarly to that of $\hat{Z}_{d}(x)$. For $d=2m+1$ the nearest singularity to the origin now occurs for $R=-{\bf 1}$ and at $x=-\frac{1}{d}$ therefore the contribution from ${\mathbb B}^m(\delta)$ is simply $\hat{Z}^\delta_{2m+1}(-x)$. The net result is that
odd powers of $k$ are eliminated and the asymptotic coefficients
are those of $\hat{Z}_{2m+1}(x)$ restricted to even $k$.

For $d=2m$, $m>1$, we have one less angle contributing and the singularity is again captured by $\zeta_1,\cdots,\zeta_{m-1}$ being $1$ or $-1$. Restricting the integration to the ${\mathbb B}^{m-1}(\delta)$ and expanding the measure (\ref{muO2m}) and integrand to quadratic order around these points gives 
\begin{eqnarray}
  \hat{Z}^{-,\delta}_{d}(x)&=&\sum_{\epsilon=\pm}\int_{{\mathbb B}^m(\delta)}\frac{d^{m-1}\theta}{(2\pi)^{m-1}}\prod_{r=1}^{m-1} \theta_r^2\prod_{1\leq r<s\leq m-1}\frac{(\theta_r^2-\theta_s^2)^2}{2^{m-1}(m-1)!}\frac{1}{1-\epsilon x \left((d-2)-\displaystyle\sum_{r=1}^m\theta_r^2\right)}\nonumber\\
  &=&\int_0^\delta \frac{dr}{r}\frac{r^{(d-1)(d-2)/2}}{2}\{\frac{1}{1- x(d-2)+x r^2}+\frac{1}{1+ x(d-2)-x r^2}\}A^{-}_d\, .
  \label{OB2m-}
 \end{eqnarray}
Then going to radial coordinates, expanding in $x$ as before we find 
\begin{eqnarray}
\hat{\cZ}^{-}_{2m}(k)&\sim&c^{-}_O(d){\left(\frac{d-2}{k}\right)}^{(d-1)(d-2)/4}
\end{eqnarray}
for a constant $ c^{-}_O(d)$, 
which is sub-dominant to $\cZ_{2m}(k)$. We therefore conclude
again
\begin{equation}
\hat{\cZ}_d(k)=\frac{c_O(d)}{\phi(\frac{1}{d})}\frac{d^{k}}{k^{\frac{d(d-1)}{4}}}
\label{LowEnergyCoefficientsSOdGaugedv2O}
\end{equation}
with only $k$ even contributing and $c_O(2m)=\frac{1}{2}c(2m)$
while $c_O(2m+1)=c(2m+1)$.

\section{Multi-matrix words and  symmetry-based counting  of $SO(d) \times U(N)$ invariants in $d$-matrix model  } 
\label{sec:ExactCountingAndAlgorithms}

The Hilbert space of the $d$-matrix harmonic oscillator consists of matrix oscillators $  ( A^{ \dagger} )^i_{ a , j }$ acting on the vacuum. Each oscillator can be taken to have unit energy and the $k$-oscillator states of the form 
\bea 
 ( A^{ \dagger } )^{ i_1 }_{ a_1 , j_1 } ( A^{ \dagger } )^{ i_2 }_{ a_2 , j_2 }  \cdots  ( A^{ \dagger } )^{ i_k }_{ a_k , j_k   }   | 0 \rangle 
\eea
have energy $k$. Imposing $U(N) \times SO(d)$ invariance projects the Hilbert space to  the states spanned by linear combinations of the above states, where all the upper  $U(N)$ indices $ ( i_1 , \cdots , i_k )  $  
are contracted with the lower indices $ ( j_1 , \cdots , j_k )$ with $ U(N)$ invariant Kronecker deltas, while the indices $a_1 , \cdots , a_k $ are contracted with $SO(d)$ invariant deltas and epsilons. Permutations are used to organise the contractions and symmetric group representation theory plays an important role in systematically counting the invariants. 

In section \ref{sec:KronInvDim} we derive the formula \eqref{KronDimOne} for the $ U(N) \times SO(d)$ invariant subspace of the oscillator Hilbert space at energy $k$, denoted $ \cZ ( N , d , k )$, in terms of Kronecker coefficients of symmetric groups and the dimensions of the $SO(d)$ invariant subspace of the $k$-fold tensor power of the $SO(d)$ vector representation $ V_d$. This is adapted straightforwardly to the  $ O(d)$ case.  In section \ref{sec:HomSpace} we use the theory of Fourier analysis on homogeneous spaces, notably the coset space $ U(d)/SO(d)$ to arrive at useful characterisations of these $SO(d)$ and $O(d)$ invariant subspaces of $ V_d^{ \otimes k }$. This is applied in section 
\ref{sec:KronHelg} to obtain the equations \eqref{KronHelg1} and \eqref{KronHelgO} for $ \cZ ( N , d , k )$ and $ \cZ_O ( N , d , k )$ respectively. In section \ref{sec:CharHelg} we use the character expansion of the Kronecker coefficients to re-write these formulae as a pairing of two vectors $ \Psi (N,  k )$ and $ \Phi ( d , k )$ defined on the space of 
partitions of the integer $k$. The key formulae  are in equations \eqref{PairingSO} and \eqref{PairingO}. The expressions 
\eqref{KronHelg1} and \eqref{KronHelgO} in terms of Kronecker coefficients as well  those in terms of characters (equations 
\eqref{PairingSO} and \eqref{PairingO} ) can be coded in SageMath using readily available functions. The codes are briefly mentioned in this section where appropriate, and explained in more detail in Appendix \ref{sec:SageMathCode}.

\subsection{ Counting using  symmetric group Kronecker coefficients and dimensions of $SO(d)$ invariant subspaces of $V_d^{ \otimes k }$.   }
\label{sec:KronInvDim}

Consider the bosonic $d$-matrix  harmonic oscillator quantum mechanics, with matrix variables
\bea 
X^i_{a,j}  : && ~~~     i , j \in \{ 1 , \cdots , N \} \cr 
&& ~~~  a \in \{ 1 , \cdots , d \} 
\eea
which transform in the vector of $SO(d)$  or $O(d)$ and the adjoint of $U(N)$. Taking 
$ V \in SO(d) $ or $ V  \in O(d)$, with $ U \in U(N)$, we have 
\bea 
( U , V ) : X_{ a , j }^{ i } \rightarrow V_{ ba} U^{i}_{~l}   X_{ b , k}^l (U^{ \dagger })^{ k }_{ j}
\eea
Physical states are obtained by applying matrix creation operators to the vacuum. We restrict to 
the $SO(d)$ or $O(d)$ invariant subspace of the $U(N)$ gauge invariant part of the Hilbert space 
as toy models for the microstates of black holes. We will show that these models naturally have a negative heat capacity branch in the energy/temperature curve of the micro-canonical ensemble, which connects smoothly to a positive heat capacity branch. This is similar to the Hawking-Page transition between small and large black holes in semi-classical AdS gravity. The results for 
$SO(d)$ here generalises the $SO(2)$ results from \cite{PIMQT}.   A future direction is to explore the interpretation of states transforming in non-trivial representations of $SO(d)$ or $O(d)$ as toy models for  quantum perturbations or large classical deformations of the black hole geometry. We will start by discussing $SO(d)$ and explain how the results change for $O(d)$. 

The counting of states of energy $k$, transforming in a general representation $ \Lambda $ of 
$SO(d)$ in $U(N)$-gauged harmonic oscillator quantum mechanics is equivalent to the counting of the $SO(d) \times U(N)$ invariant polynomials in $ X_{ a , j }^i$ of degree $k$. 
We will describe a representation theoretic approach to this counting using an approach based on 
\cite{BHR2}, where this approach was used to construct orthogonal bases for these matrix invariants in the free-field inner product, which is also the harmonic oscillator quantum mechanics inner product.  Related results on the counting of gauge invariants for multi-matrix theories are in \cite{BDHO,Dolan,DutGop}. 

We will explain below the derivation, along the lines of \cite{BHR2}, of  the following counting formula 
\begin{equation}
	\boxed{
		\cZ ( N , d , k )
		=
		\sum_{\substack{ R \vdash k \\ l(R) \le N }}
		\sum_{\substack{ T \vdash k \\ l(T) \le d }}
		C(R,R,T)\,\Dim(V_{\Lambda_0,T})
	}
	\label{KronDimOne}
\end{equation}
where $ C(R,R,T)$ is the Kronecker coefficient for the tensor product $R \otimes R \otimes T$ for irreducible representations of $S_k$. This is the multiplicity of the trivial representation in the decomposition of the tensor product
$R \otimes R \otimes T$, where elements $\sigma \in  S_k$ acts diagonally as $ \sigma \otimes \sigma \otimes \sigma $,  into irreducible representations. Equivalently it is the multiplicity of $T$ in the decomposition of $ R \otimes R$, with diagonal action of $ \sigma \in S_k$ as $ \sigma \otimes \sigma$.  $ \Dim(V_{\Lambda_0,T})$ is dimension of the $SO(d)$ invariant subspace of $V_d^{ \otimes k } $ into irreps of $SO(d)$, or equivalently the multiplicity of the trivial representation  of $SO(d)$ in the decomposition of $V_d^{ \otimes k } $  in terms of irreducible representations.

We have an elementary field $X_{ a, j }^i $ where 
$ a$ runs over $ \{ 1 , \cdots , d \} $ and $ i , j $ run over $ \{ 1 , \cdots , N \} $. The invariants of degree $k$ are built from appropriate index contractions of 
\bea\label{Polys}  
X_{ a_1 , j_1}^{ i_1 } \cdots X_{ a_k , j_k}^{ i_k} 
\eea
Equivalently we are counting invariants of both $SO(d)$ and $U(N)$ in $ \Sym^k ( V_N \otimes \bar V_N \otimes V_d ) $. The $k$-fold symmetrisation comes from the fact that the $X$-variables above are commutative. 

Using the standard Schur-Weyl duality, we have  
\begin{align}\label{UNSW}  
	V_N^{ \otimes k } & = \bigoplus_{ \substack{  R \vdash k \\ l(R) \le N }  }  V_R^{ U(N)} \otimes V_{ R }^{ S_k } \cr 
	\bar V_N^{ \otimes k } & =    \bigoplus_{ \substack{  S \vdash k \\ l(S) \le N }  }  \bar V_S^{ U(N)} \otimes V_{ S }^{ S_k } \end{align}
We have a direct sum over Young diagrams with $ k$ boxes and no more than $N$ rows. 
Projecting to $U(N) $ invariants sets $ R = S $ in the expansions \eqref{UNSW}, and for each such pair there is a unique invariant. 
We therefore have 
\bea 
\Invt_{ U(N) } (  V_{ N }^{ \otimes k } \otimes  \bar V_N^{ \otimes k }  )  = \bigoplus_{ \substack{  R \vdash k \\ l(R) \le N }  } 
\bigoplus_{ \substack{  R \vdash k \\ l(R) \le N }  }  V_R^{S_k } \otimes V_R^{ S_k} 
\eea

The $k$'th  tensor power of the  $d$-dimensional representation of $SO(d)$ can be decomposed into irreps of $SO(d)$ in two steps. 
\begin{align}\label{kfoldSOd} 
	V_{ d}^{ \otimes k } & =  \bigoplus_{\substack { T \vdash k \\ l(T) \le d  } }  V_T^{ S_k} \otimes V_d^{ U(d) }   \cr 
	& =     \bigoplus_{ \Lambda \in {\rm Rep } (SO(d)) } V_{ T }^{ S_k }  \otimes  V_{ \Lambda }^{ SO(d) }  \otimes V_{  T . \Lambda }^{U(d) \rightarrow SO(d)  } 
\end{align} 
In the first step we consider it as a representation of $U(d)$ and decompose into irreps of $U(d)$ labelled by Young diagrams with $k$ boxes and no more than $d$ rows.
In the second step we decompose the irreps $V_{ T}^{U(d)}$ into irreps of $ V_{ \Lambda}^{ SO(d)} $ with a  multiplicity space 
$V_ {\Lambda , T } $ 
\bea\label{UdOdred}  
\Dim ( V_{ \Lambda , T }^{U(d) \rightarrow SO(d)  }  ) = \Mult ( V_{T}^{ U(d)}  \rightarrow   V_{ \Lambda }^{ SO(d) } ) 
\eea
The irreps $\Lambda $ of $SO(d)$ appearing in this decomposition are the tensor representations, which are indexed by Young diagrams with $k$ or fewer  boxes,  and no more than $\lfloor \frac{d}{2} \rfloor $ rows. In the case of even $d$, there is the additional subtlety that for Young diagrams with exactly $ \frac{d}{2} $ rows, there is a pair of distinct representations with plus and minus chiralities \cite{FultonHarris}. The general rules for the reduction multiplicities in \eqref{UdOdred} are available in the mathematical literature 
\cite{Littlewood,King,KoikeTerada,HoweTanWillenbring}.

Thus 
\bea 
&& \Invt_{ U(N)} ( V_N^{ \otimes  k } \otimes V_N^{ \otimes k } \otimes V_d^{ \otimes k } ) =  \bigoplus_{ \substack{  R \vdash k \\ l(R) \le N }  } \bigoplus_{ \Lambda \in {\rm Rep } (SO(d)) }  V_R^{ S_k} \otimes V_R^{ S_k}\otimes V_T^{ S_k  } \otimes   V_{ \Lambda }^{ SO(d) }  \otimes V_{  T . \Lambda }^{U(d) \rightarrow SO(d)  } \cr 
&& 
\eea
Projecting to $U(N)$ invariants in the $k$-fold symmetrised tensor power is equivalent to projecting to $ U(N) \times S_k$ invariant subspace  in the tensor product 
\bea 
\Invt_{ U(N)} ( \Sym^k (  V_N \otimes V_N \otimes V_d )) 
= \Invt_{ U(N) \times S_k }  (  V_N^{ \otimes  k } \otimes V_N^{ \otimes k } \otimes V_d^{ \otimes k } )  
\eea
Hence 
\bea\label{UNSkInvtVVV} 
&& \Invt_{ U(N) \times S_k }  (  V_N^{ \otimes  k } \otimes V_N^{ \otimes k } \otimes V_d^{ \otimes k } )  
= \bigoplus_{ \substack{  R \vdash k \\ l(R) \le N }  } \bigoplus_{ \Lambda \in {\rm Rep } (SO(d)) }  V_{RRT}^{ S_k}  \otimes  V_{ \Lambda }^{ SO(d) }  \otimes V_{  T . \Lambda }^{U(d) \rightarrow SO(d)  } \cr 
&& 
\eea
where  $ V_{RRT}^{ S_k} $ is the $S_k$ invariant subspace of  $V_R^{S_k}\otimes V_R^{S_k}\otimes V_T^{S_k}$: 
\bea 
 V_{RRT}^{ S_k} = \Invt_{  S_k }  ( V_R^{S_k}\otimes V_R^{S_k}\otimes V_T^{S_k} ) 
\eea 
The dimension of the invariant subspace is given by the Kronecker coefficient $ C ( R , R , T )$ 
for the triple $ ( R , R , T )$, which is also the Clebsch-Gordan multiplicity for $ V_T{ (S_k) } $ in the 
decomposition of the tensor product representation $ V_R^{ S_k } \otimes V_R^{ S_k } $ into irreducibles 
\bea 
\Dim ( V^{S_k}_{ RRT } ) = C ( R , R , T )
\eea 
The Kronecker coefficient can be expressed in terms of characters of $S_k$. 
The result is that the number of states in the irrep 
$\Lambda $ , denoting it as  $  \cZ ( N , d , k  ; \Lambda ) $ is given by 
\bea 
\cZ ( N , d , k  ; \Lambda )   = \sum_{ \substack { R \vdash k \\ l(R) \le N } } 
\sum_{ \substack { T \vdash k \\ l(T) \le d  } } C ( R , R , T ) \Dim ( V^{U(d) \rightarrow SO(d)  }_{ \Lambda , T }  )
\eea
Since we are after trivial irreps of $SO(d)$, we  specialise  $ \Lambda $ on the right hand side of   \eqref{UNSkInvtVVV}  to $ \Lambda_0$ and obtain 
\bea\label{InvtSubspUSkSO}  
&& \Invt_{ U(N) \times SO(d) \times  S_k }  (  V_N^{ \otimes  k } \otimes V_N^{ \otimes k } \otimes V_d^{ \otimes k } )  
= \bigoplus_{ \substack{  R \vdash k \\ l(R) \le N }  } \bigoplus_{ \Lambda \in {\rm Rep } (SO(d)) }  V_{RRT}^{ S_k}  \otimes  V_{ \Lambda_0  }^{ SO(d) }  \otimes V_{   \Lambda_0 , T  }^{U(d) \rightarrow SO(d)  } \cr 
&& 
\eea 
and 
\bea\label{KronDimTwo}  
\cZ ( N , d , k   ) && = \cZ ( N , d , k  ; \Lambda_0  )  \cr 
&& = \sum_{ \substack { R \vdash k \\ l(R) \le N } } 
\sum_{ \substack { T \vdash k \\ l(T) \le d  } } C ( R , R , T ) \Dim ( V^{U(d) \rightarrow SO(d)  }_{ \Lambda_0  , T }  )
\eea

Changing the invariance group from $ SO(d)$ modifies the derivation above in a straightforward way, leading to 
\bea\label{ZOrth} 
\cZ_{O} ( N , d , k  ; \Lambda )   = \sum_{ \substack { R \vdash k \\ l(R) \le N } } 
\sum_{ \substack { T \vdash k \\ l(T) \le d  } } C ( R , R , T ) \Dim ( V^{U(d) \rightarrow O(d)  }_{ \Lambda , T }  )
\eea 
and 
\bea\label{KronDimO}  
\cZ_{O} ( N , d , k   ) && = \cZ ( N , d , k  ; \Lambda_0  )  \cr 
&& = \sum_{ \substack { R \vdash k \\ l(R) \le N } } 
\sum_{ \substack { T \vdash k \\ l(T) \le d  } } C ( R , R , T ) \Dim ( V^{U(d) \rightarrow O(d)  }_{ \Lambda_0  , T }  )
\eea
While these formulae are useful for general representations $ \Lambda $ and the algorithms for general $ \Lambda  $ can be specialised to $ \Lambda_0$, it is possible to obtain more efficient algorithms for \eqref{KronDimOne} and \eqref{KronDimO} by taking advantage of results on harmonic analysis on  spaces 
$U(d)/SO(d)$ and $ U(d)/O(d)$.

\subsection{ $SO(d)$ and $O(d)$ invariant sectors for general $d$ 
	using homogeneous spaces  } 
	\label{sec:HomSpace} 

In this section we use results from harmonic analysis on the homogeneous space $U(d)/SO(d)$ to derive
\begin{equation}\label{SOAkdresult}
	\boxed{
		\begin{aligned}
			\Dim ( V^{U(d) \rightarrow SO(d)  }_{ \Lambda_0  , T }  )
			& = 1 ~~  \hbox{if} ~~  T \in \cA( k , d )  \cr 
			& = 0   ~~  \hbox{if} ~~ T \notin \cA( k , d ) 
		\end{aligned}
	}
\end{equation}
where 
\begin{align}\label{defAkd1}
	\cA( k , d ) = \cA_+ ( k , d )  \cup  \cA_- ( k , d )
\end{align}
with 
\bea\label{defAkd2} 
\cA_{ + } ( k , d )  = \hbox{Set of Young diagrams with $k$ boxes, even row lengths and up to $d$ rows. }   \cr 
\cA_{ - } ( k , d ) = \hbox{Set of Young diagrams with $k$ boxes, odd row lengths and exactly $d$ rows.  }   \cr 
& 
\eea
The result for $ O(d) $ invariant sector is 
\begin{equation}
	\boxed{
		\begin{aligned}
			\Dim ( V^{U(d) \rightarrow O(d) }_{ \Lambda_0 , T } )
			& = 1 ~~ \hbox{if} ~~ T \in \cA_+ ( k , d )  \cr
			& = 0 ~~ \hbox{if} ~~ T \notin \cA_+ ( k , d )
		\end{aligned}
	}
\end{equation}

\noindent 
{\bf Comparing $O(d)$ and $SO(d)$  }

For $d$ odd, diagrams in $ \cA_- ( k  , d) $ have an odd number of boxes, since they have, by definition, exactly $ d$ rows of odd lengths. This means that  $ \cA_- ( k  , d) $  is empty for even $k$. 
Furthermore,  $ \cA_+ ( k , d )$ for odd $d$ and odd $k$ is empty, since diagrams in this set must have even row lengths. 
This means that 
\begin{align}
\hskip1cm  \hbox{ Odd  $d$  case } \cr 
~~\hbox{For }  \hbox{ even energies,  }  k = 2n ~ : ~  \cA ( 2 n , d ) & =   \cA_{+} (  2 n  , d ) \cr 
\cZ_{SO}  ( N , d , 2 n ) & =   \cZ_{O}  ( N , d ,2  n ) \cr 
~~\hbox{For } \hbox{ odd energies,    } k = (2n+1) ~  :  ~ \cA (2 n +1  , d ) & =   \cA_{- } ( 2 n +1  , d ) \cr 
\cZ_{O}  ( N , d , 2 n +1  ) & =   0  
\end{align}

Now consider even $d$,  $ \cA_- ( k , d ) $ as well as $ \cA_{ + } ( k , d )$  are  empty for odd $k$. 
Hence in this case we have 
\begin{align}
 ~~~~~\hskip2.8cm  & \hbox{ ~~~~~~~~~~~~~~~~~~~~~ Even  $d$ case }  \cr 
 \hbox{Odd energies,  } k &= (2n+1) ~:  ~ \cZ_{ S O } (  N , d , 2 n +1 ) = \cZ_{ O } (  N , d , 2n  +1 ) = 0 \cr 
&
\end{align}
In this case, for even $k$, generically $ \cZ_{ S O } (  N , d , 2n ) $ and $ \cZ_{ O } (  N , d , 2n ) $ are both non-zero and different. 

\subsubsection{  Cartan-Helgason theorem and $SO(d)$ invariant subspace of $ V_d^{ \otimes k }$  }  
\label{sec:Cartan-Helgason} 

In \eqref{kfoldSOd} we described the $SO(d)$ invariant subspace of the 
$k$-fold tensor product $V_d^{\otimes k}$ by first decomposing it into 
$U(d)$ irreducible representations using Schur--Weyl duality, and then 
reducing these $U(d)$ irreps to $SO(d)$ irreps. Specialising this 
reduction to the trivial $SO(d)$ representation $\Lambda_0$ leads to 
\eqref{InvtSubspUSkSO}.

A general result from harmonic analysis on homogeneous spaces gives the 
decomposition of the space of functions on a coset as 
\bea 
{\rm Functions}(G/H) 
= \bigoplus_{ T \in {\rm Rep}(G) }  
V^{G}_{T} \otimes V^{ G \rightarrow H}_{ T , \Lambda_0  }
\eea
in terms of the multiplicity spaces $ V^{ G \rightarrow H}_{ T , \Lambda_0  } $, 
which describe the decomposition of the irreducible representation $T$ of $G$ 
into representations of $H$, with $\Lambda_0$ denoting the trivial representation of $H$. 

 In the present context we restrict attention to polynomial 
(regular) functions on $G/H$. These provide an algebraic basis for the 
representation-theoretic decomposition of functions on $G/H$, modulo the 
analytic completion appearing in the Peter--Weyl theorem. The 
decomposition above is a standard consequence of harmonic analysis on 
compact homogeneous spaces. See for example Helgason \cite{Helgason} for 
a systematic treatment of functions on $G/H$, and Fulton--Harris 
\cite{FultonHarris} for background on polynomial representations of 
$U(d)$.

Interesting instances of these homogeneous spaces are ones where the multiplicity spaces have dimension  zero or one. 
We will find the Cartan--Helgason result for the coset $SU(d)/SO(d)$ to be useful:
\bea 
{\rm Functions }  ( SU(d)/SO(d) )  =
\bigoplus_{ k }  \bigoplus_{ T \in \cB ( k ,  d )  }  V_T^{ SU(d) } 
\eea
where 
\bea 
\cB ( k ,  d )  = \hbox{ Set of Young diagrams with $k$ boxes, up $ d -1$ rows, all  of even length.  } 
\eea 
Equivalently 
\begin{align}
	\Dim ( V^{SU(d) \rightarrow SO(d)  }_{ \Lambda_0  , T }  )
	& = 1 ~~  \hbox{if} ~~  T \in \cB( k , d )  \cr 
	& = 0   ~~  \hbox{if} ~~ T \notin \cB( k , d ) 
\end{align}

Irreducible representations of $U(d)$ can be obtained from those of $SU(d)$ 
by tensoring with powers of the determinant representation.

We can therefore relate the $SU(d)$ result above to the $U(d)$ result we are after. 
The Young diagrams labelling irreducible representations of $U(d)$ appearing 
in the decomposition of the tensor space $V_d^{\otimes k}$ can be obtained 
from Young diagrams in $\cB(k' = k - md , d)$ by attaching $m$ columns of 
height $d$. This corresponds to tensoring with $m$ copies of $\Lambda^d(V_d)$, 
which is the determinant representation and is $SO(d)$ invariant. 
The set $\cA(k,d)$ is precisely the set of Young diagrams obtained from 
$\cB(k',d)$ with $k' = k - md$ by attaching $m$ columns of length $d$.

Attaching a column of height $d$ increases each row length by one and 
introduces a $d$-th row if it was previously absent. Starting from Young 
diagrams in $\cB(k',d)$ with even row lengths and at most $d-1$ rows, 
attaching $m$ such columns produces diagrams whose row lengths have 
parity determined by $m$. If $m$ is even the resulting diagrams have 
even row lengths and at most $d$ rows, giving the family $\cA_+(k,d)$. 
If $m$ is odd the resulting diagrams have odd row lengths and exactly 
$d$ rows, giving the family $\cA_-(k,d)$.

\subsection{Partition function using  Kronecker coefficients and homogeneous spaces }
\label{sec:KronHelg} 

At this stage, we can combine \eqref{KronDimOne} and \eqref{SOAkdresult} to write for the $ U(N ) \times SO(d)$ invariants 
\bea\label{KronHelg1}  
\cZ( N , k , d )  
&& = \sum_{ \substack { R \vdash k \\ l(R) \le N } } 
\sum_{ \substack { T \in  \cA( k , d )  } } C ( R , R , T ) 
\eea

If we change the invariance imposed from $SO(d)$ to $O(d)$, the totally 
antisymmetric tensor $\epsilon_{a_1 \dots a_d}$ is no longer invariant. 
As a result, contractions involving the $\epsilon$-tensor are not allowed. 
For even $k$ the invariants constructed using only Kronecker deltas are those which remain from the $SO(d)$ case. For odd $k$ no invariants exist, consequently the microcanonical $E$–$T$ curves in the $O(d)$ invariant 
sector will be simpler.
\bea\label{KronHelgO}  
\cZ_O ( N , k , d )  
&& = \sum_{ \substack { R \vdash k \\ l(R) \le N } } 
\sum_{ \substack { T \in  \cA_+ ( k , d )  } } C ( R , R , T ) 
\eea

Both formulae \eqref{KronHelg1} and \eqref{KronHelgO}  are implemented in the ancillary SageMath notebook
\texttt{OrthogKroneckerHelgason.ipynb}. The routines
\texttt{Z\_Kronecker\_SO(N,d,k)} and \texttt{Z\_Kronecker\_O(N,d,k)}
compute respectively the $SO(d)$ and $O(d)$ invariant counting functions by evaluating
Kronecker products of Schur functions and summing over the Cartan--Helgason sets
$A(k,d)$ and $A_+(k,d)$.

\subsection{ Simplification from Kronecker coefficients to  character sums  } 
\label{sec:CharHelg}

For the regime, $ k \le N $, it is useful to use the character expansion of the Kronecker coefficient 
\bea\label{charformZNdk0} 
\cZ ( N , d , k ) = \sum_{ \substack { R \vdash k \\ l(R ) \le N }}   
\sum_{ p \vdash k  } { 1 \over \Sym ~ p } ( \chi^R_p )^2 \sum_{ S \in \cA ( k , d ) } \chi^S_p 
\eea
and change the order of sums, to do the $R$-sum first 
\bea\label{charformZNdk} 
\cZ ( N , d , k ) =   \sum_{ p \vdash k  } { 1 \over \Sym ~ p } \sum_{ \substack { R \vdash k \\ l(R ) \le N }} 
( \chi^R_p )^2 \left (   \sum_{ S \in \cA ( k , d ) } \chi^S_p  \right )\, . 
\eea 
It is useful to define two functions on partitions $ p \vdash k $, which carry information about finite $N $ and $ d$ cut-offs respectively 
\bea\label{defPhi}  
&& \Psi (N, k ;  p ) =\sum_{ \substack { R \vdash k \\ l(R) \le  N } }  ( \chi^R_p )^2 \cr 
&& \Phi ( d , k ;  p ) = \sum_{ S \in \cA ( k , d ) } \chi^S_p   
\eea 
This allows us to write 
\bea\label{charformZNd1} 
\cZ ( N , d , k ) = \sum_{ p \vdash k  } { \Psi  (N,  k ;  p )    \Phi ( d ,k ;  p ) \over \Sym ~ p } 
\eea
We may think of $ \Psi ( N , k ;  p )$ as the coefficients of a vector  $ \Psi ( N  , k )$ in the vector space of  partitions $p $ of the integer $k$, and likewise 
$ \Phi(  d , k ; p )$ as the coefficients of a  vector $ \Phi ( d , k )$ : 
\bea 
&& \Psi ( N , k  ) = \sum_{ p \vdash k  } \Psi  (N,  k ;  p ) ~  p  \cr 
&& \Phi ( d  , k ) = \sum_{ p  \vdash k } \Phi  (d, k ;   p )  ~  p 
\eea
and 
\begin{align}\label{PairingSO}
\boxed{
\cZ ( N , d , k )  = \sum_{ p \vdash k  } { \Psi  (N, k  ;  p )    \Phi ( d ,k ;  p ) \over \Sym ~ p }
 =  \langle \Psi ( N, k  ) , \Phi ( d  , k )    \rangle
} 
\end{align}

For the $O(d)$ case, we have 
\begin{equation}\label{ZNdO} 
\cZ_O ( N , d , k ) = \sum_{ p \vdash k  } { 1 \over \Sym ~ p } \sum_{ \substack { R \vdash k \\ l(R ) \le N }}  ( \chi^R_p )^2  \sum_{ S \in \cA_+ ( k , d ) } \chi^S_p 
\end{equation}
and defining 
\bea 
\Phi_+  ( d , k ;  p ) = \sum_{ S \in \cA_+  ( k , d ) } \chi^S_p   
\eea 
we have 
\begin{align}\label{PairingO}  
\boxed{
\cZ_O  ( N , d , k ) = \sum_{ p \vdash k  } { \Psi  (N, k ;  p )    \Phi_+  ( d , k ; p ) \over \Sym ~ p }  = \langle \Psi ( N  , k ) , \Phi_+ ( d , k  )    \rangle
}
\end{align}

The equations \eqref{PairingSO} and \eqref{PairingO} give our most efficient formulae for the computation of the micro-canonical partition functions. The sage implementation is described in Appendix \ref{sec:SageMathCode}.

\section{ Caloric fold curves from exact finite $N$ micro-canonical degeneracies.  } 
\label{sec:FoldsExact} 

\begin{figure}[H] %
	\centering
	\begin{minipage}{0.3\textwidth}
		\includegraphics[width=\textwidth]{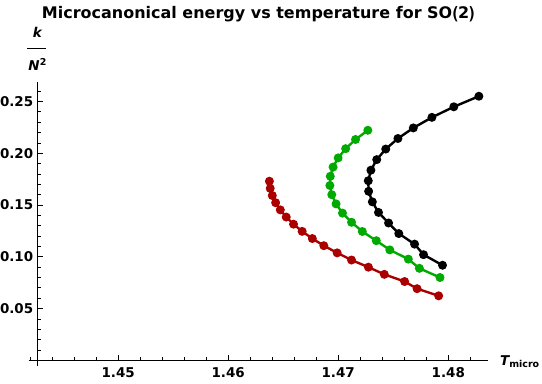} 
		\label{fig:SO2_MicroEvsT}
	\end{minipage}
	\begin{minipage}{0.3\textwidth}
		\centering
		\includegraphics[width=\textwidth]{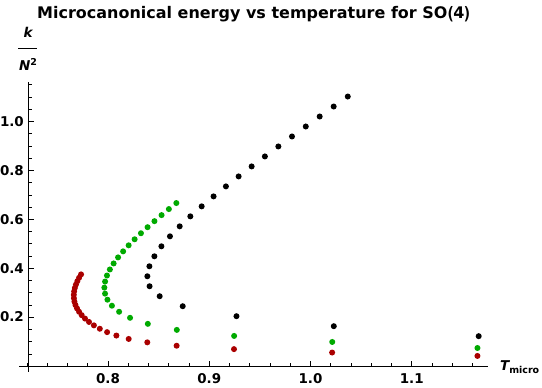} 
		\label{fig:SO4_MicroEvsT}
	\end{minipage}
	\begin{minipage}{0.3\textwidth}
		\centering
		\includegraphics[width=\textwidth]{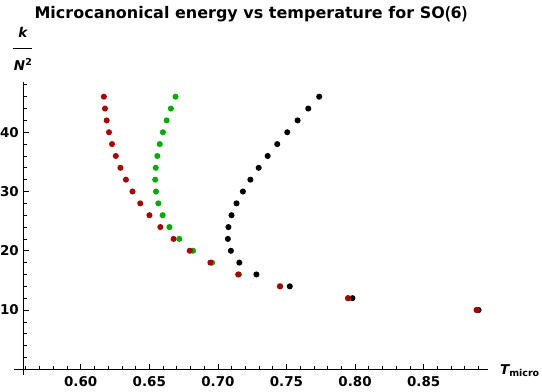} 
		\label{fig:SO6_MicroEvsT}
	\end{minipage}
	\caption{The figures shows plots of the energy as a function of the microcanonical temperature different $N$ with with black $N=14$, green $N=15$ and red $N=17$ for the $SO(2)$ invariant model on the left.  The middle figure is the $SO(4)$ invariant model with $SO(6)$ on the right both with black $N=7$, green $N=9$ and red $N=12$. }
	\label{SO2SO4SO6_MicroEvsT}
\end{figure}

By focusing on the microcanonical temperature energy relation (\ref{Tmicro})
we can understand the micro-canonical  thermodynamics of these $SO(d)$ invariant systems.
From the tables in Appendix \ref{Tables} we can extract different plots
which all have a similar caloric fold. Below we present plots for $SO(2)$, $SO(4)$ and $SO(6)$ in Figure \ref{SO2SO4SO6_MicroEvsT}. One sees that there is a
minimum temperature and characteristic energy $k_{\crit}(N)$ where the curve folds back from a lower energy negative heat capacity branch to an higher energy
positive heat capacity, i.e. the temperature decreases with increasing energy in the lower branch but increases with increasing energy in the upper branch.

\begin{figure}[H] %
	\centering
	\begin{minipage}{0.45\textwidth}
		\includegraphics[width=\textwidth]{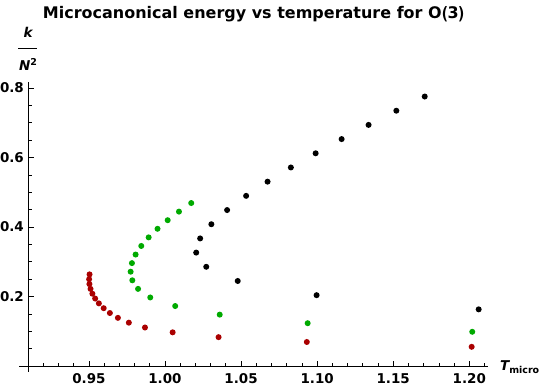} 
		\label{fig:O3_MicroEvsT}
	\end{minipage}
	\begin{minipage}{0.45\textwidth}
		\centering
		\includegraphics[width=\textwidth]{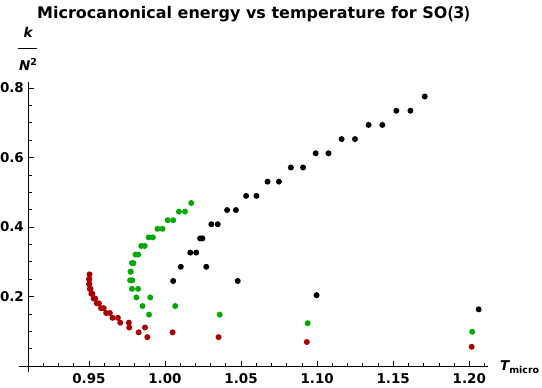}
		\label{fig:SO3_MicroEvsT}
	\end{minipage}
	\caption{In these figures we shows plots of the microcanonical temperature for $O(3)$ on the left and $SO(3)$ on the right with black $N=7$, green $N=9$ and red $N=12$. The $SO(3)$ figure on the right shows shows that as the transition is approached the odd and even sequences for $SO(3)$ merge into a single curve. The plots are for the symmetric derivative with odd and even $k$ treated separately. The behaviour is similar the derivative is between odd and even.}
	\label{fig:SO3O3_MicroEvsT}
\end{figure}

\begin{figure}[H] %
	\centering
	\begin{minipage}{0.45\textwidth}
		\includegraphics[width=\textwidth]{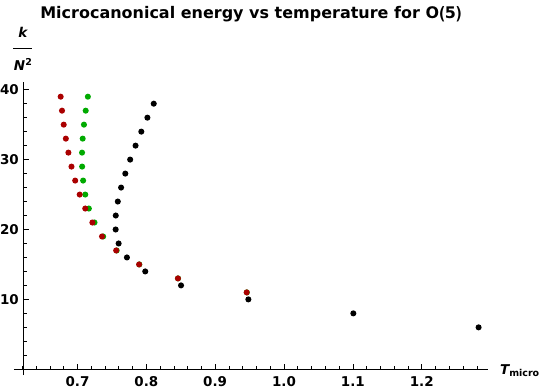} 
		\label{fig:O5_MicroEvsT}
	\end{minipage}
	\begin{minipage}{0.45\textwidth}
		\centering
		\includegraphics[width=\textwidth]{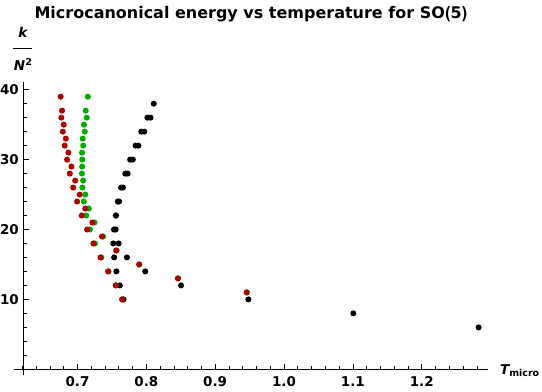} 
		\label{fig:SO5_MicroEvsT}
	\end{minipage}
	\caption{This figure shows the caloric fold for $O(5)$ on the left contrasted with $SO(5)$ on the right, with black $N=7$, green $N=9$ and red $N=12$. The behaviour is similar to $O(3)$ and $SO(3)$. }
	\label{fig:SO5O5_MicroEvsT}
\end{figure}

Then in Figure \ref{fig:SO3O3_MicroEvsT} we contrast $O(3)$ with $SO(3)$ showing
the more complicated low energy structure for $SO(3)$ where there are two low energy branches that merge at higher energy. These are small $N$ effects and there is little difference as the transition is approached at large $N$. We then show in Figure \ref{fig:SO5O5_MicroEvsT} that $O(5)$ and $SO(5)$ are qualitatively similar to
the $d=3$ case.

As observed in the figures there is a critical energy at which the energy temperature curves fold back on themselves which we label $k_{crit}(N)$.

\begin{figure}[H] %  
	\centering
	\begin{minipage}{0.45\textwidth}
		\includegraphics[width=\textwidth]{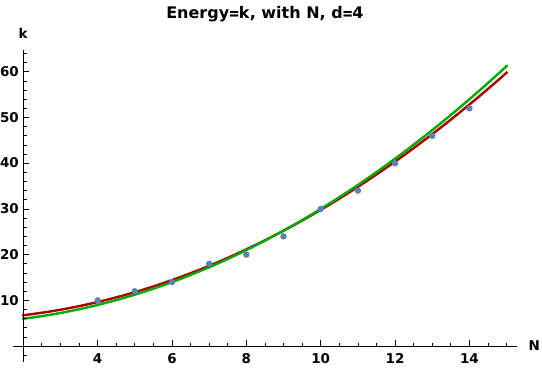} 
		\label{fig:SO4_betavsN}
	\end{minipage}
	\hfill
	\begin{minipage}{0.45\textwidth}  
		\centering
		\includegraphics[width=\textwidth]{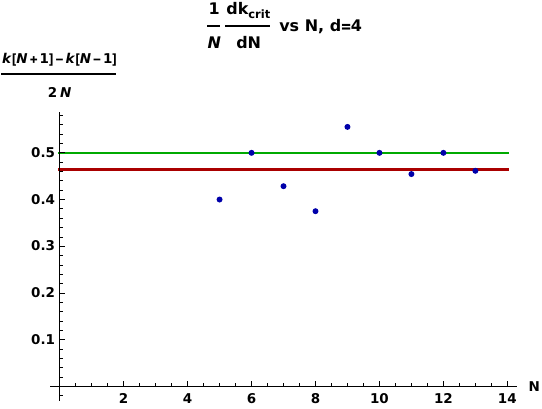} 
		\label{fig:SO4extractnNsqfig}
	\end{minipage}
	\caption{The figure on the left shows the $SO(4)$ critical
		energy $E=k_{\crit}$ at which the microcanonical temperature is minimum.
		The two fitting curves are, $k_{crit}=5.8+0.24N^2$ for the red curve which
		was a two parameter fit and $4.5+\frac{N^2}{4}$ for the green curve, a
		one parameter fit where the coefficient of $N^2$ is fixed to
		be $\frac{1}{4}$. The figure on the right shows $1/N$ times the discrete
		symmetric derivative, $\frac{k(N+1)-k(N-1)}{2}$, plotted against $N$ with
		the numerical constant fit of $0.464$ (red line) and the expected value
		of $0.5$ (green line).}
	\label{fig:SO4kcritfit}
\end{figure}

\begin{figure}[H] %  
	\centering
	\begin{minipage}{0.3\textwidth}  
		\includegraphics[width=\textwidth]{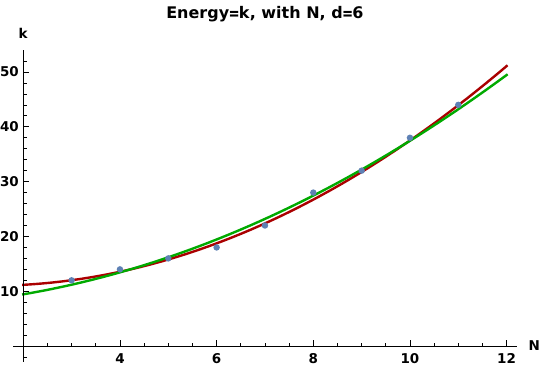} 
		\label{fig:SO6kcritofNfig}
	\end{minipage}
	\hfill
	\begin{minipage}{0.3\textwidth}  
		\centering
		\includegraphics[width=\textwidth]{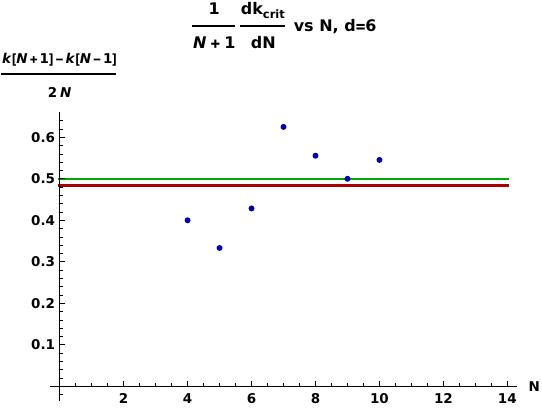}
		\label{fig:SO6extractnNsqfig}
	\end{minipage}
	\hfill
	\begin{minipage}{0.3\textwidth}  
		\centering
		\includegraphics[width=\textwidth]{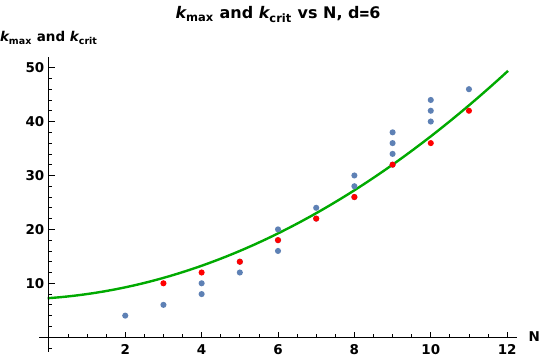} 
		\label{fig:SO6kmaxAndkcritfig}
	\end{minipage}
	\caption{These figures are for $SO(6)$ with $k_{\crit}$ as a function of $N$ on the left. The two fitting plots are, a 3 parameter fit, $k_{\crit}=11.6 - 0.91 N + 0.35 N^2$ (red curve), and a one parameter fit $7.2+\frac{(N+1)^2}{4}$ (green curve). A two parameter fit with a linear term and the coefficient of $N^2$ fixed to $\frac{1}{4}$ gave a coefficient very close to $0.5$ which could be naturally absorbed in a shift of $N$.   
		The middle plot shows the discrete symmetric derivative $\frac{k(N+1)-k(N-1)}{2}$ divided by $N+1$ (suggested by the fit of $k_{\crit}(N)$ for $SO(6)$) plotted against $N$ with the numerical constant fit of $0.484$ (red line) and the expected value of $0.5$ (green line).  The figure on the right shows the value of $k$ for which $\Delta \cZ(N,6,k)$ is maximum, denoted $k_{\max}(N)$, and $k_{\crit}(N)$ for $SO(6)$, with the green curve $7.2+\frac{(N+1)^2}{4}$. Note that $k_{\max}$ is constant for a range of $N$ due to the integer nature of both $k$ and $N$.}
	\label{fig:SO6kcritfit}
\end{figure}

We find that for $SO(4)$
\begin{eqnarray}
	\{N,k_{\crit}(N)\}&=&\{(4, 10), (5, 12), (6, 14), (7, 18), (8, 20), (9, 24),
	(10, 30),\nonumber \\
	&&\qquad (11,34), (12, 40), (13, 46), (14, 52)\}
\end{eqnarray}
while for $SO(6)$ we get
\begin{eqnarray}
	\{N,k_{\crit}(N)\} &=& \{(3, 12), (4, 14), (5, 16), (6, 18), (7, 22), (8, 
	28), (9, 32),\nonumber\\
	&&\qquad\quad (10, 38), (11, 44)\};
\end{eqnarray}
we plot this data and fit it to quadratic function. The fit matches well with
the expected behaviour that the turning point of the caloric fold approaches
the Hagedorn temperature as $N$ grows. The best fit for $SO(4)$ shown in Figure \ref{fig:SO4kcritfit} is
\begin{equation}
	k_{\crit}(N)=4.5+\frac{N^2}{4}
\end{equation}
while that for $SO(6)$ given in Figure \ref{fig:SO6kcritfit} is
\begin{equation}
	k_{\crit}(N)=7.2+\frac{(N+1)^2}{4}
\end{equation}

The main focus of our  investigations of the micro-canonical $E$ vs. $T$ curve has been the turning point at $ k_{ \crit } \sim  { N^2 \over 4 } $. We will approach this by considering the high energy regime of the eigenvalue distribution in the next section, deriving an entropy/energy relation for the high energy regime which is then matched to the low-energy entropy obtained by computing the asymptotics in section \ref{sec:LowEnergy}. A closely related phenomenon, in the combonatorial data  $ \cZ( N  , d, k )$ is the maximum in the  difference $ \Delta \cZ( H , d , k ) = (\cZ( H , d, k ) - \cZ( H-1  , d, k ) )$ which is obtained by summing over Young diagrams of fixed height.  This maximum occurs at $k_{ \max } $ which also closely fits formulae compatible with $ k_{ \max } \sim  { N^2 \over 4 } $ at large $N$, as shown in Figure \ref{fig:SO6kcritfit}. 

The numerical investigations here and the fits in the next section should be viewed as  evidence for mathematical conjectures $ k_{ \crit } \sim { N^2 \over 4 } $ and $ k_{ \max } \sim { N^2 \over 4 } $ at large $N$, which are conjectures concerning the purely combinatorial quantities $ \cZ( N , d, k ) $ (and the $O(d)$ analog $ \cZ_O( N , d, k)$). 
Proving these conjectures at a mathematical level of rigour is an interesting problem for the future.

\section{ Large $N$ eigenvalue density approximation and analytic fits for the caloric folds}
\label{sec:EigDensAnalytic} 

In this section we switch to an analysis of the partition functions in
the high energy regime just above the Hagedorn temperature so that we
are interested in $x\ge x_H=\frac{1}{d}$ and $N$ large but not
infinite. In this regime the eigenvalues $z_i={\rm e}^{i\varphi_i}$ of
$U(N)$ are treated as a continuous distribution and the partition
function is dominated by the saddle point of this distribution.  Our
treatment is close to that of \cite{Sundborg:1999ue} and
\cite{Aharony:2003sx} for the $d$ matrix model.  The large $N$
eigenvalue distribution can be obtained as a function of $R$, the
$SO(d)$ rotation matrix entering equations
(\ref{ZnSOd-general}). Because the resulting effective action is
proportional to $N^2$ we can again evaluate the integrals over $SO(d)$
by a saddle point approximation to obtain a free energy. This free
energy gives the canonical entropy by a Legendre transform yielding
the entropy as a function of the energy Eqn (\ref{EntropyHigh}).
We find that the transition occurs precisely at energy $E=\frac{1}{4}$.
From the underlying quantum theory (a system of oscillators) we have
$E=\frac{k}{N^2}$ and the
transition occurs at $k=\frac{N^2}{4}$. The stable counting is therefore
valid subject to irrelevant corrections at large $N$ up to $k=\frac{N^2}{4}$.

Given the entropy as a function of $k$ can make contact with the
analysis from $x\leq\frac{1}{d}$. For this we need to include the one
loop prefactor though it is suppressed by $N^2$. The result connects
continuously with the microcanonical expressions of Section
\ref{sec:LowEnergy} and we recover
(\ref{LowEnergyCoefficientsSOdGauged}) with an exponential decay.

\subsection{The high energy  Large $N$ Limit}
\label{sec:HighEnergLargeN}

The expression (\ref{MolienWeyl}) or equivalently (\ref{ZSO2m+1}) and (\ref{ZSO2m+1}) can be evaluated in the large $N$ limit above the Hagedorn 
transition, i.e. for $x\ge\frac{1}{d}$ by going to an eigenvalue distribution
and evaluating for large $N$ by using the saddle point approximation.
Rewriting (\ref{MolienWeyl}) in angular variables with $z_i={\rm e}^{i\theta_i}$ we have
\begin{equation}
  \cZ(N,d;x)=\int\mu_{d}(R)\frac{1}{N!}\prod_{i=1}^{N}\int_0^{2\pi}\frac{d\theta_i}{2\pi} {\rm e}^{-S_{eff}(x_a(R),\theta_i)}
  \end{equation}
\begin{eqnarray}
  S_{eff}(x_a,\theta_i)&=&\frac{1}{2}\sum_{a=1}
    ^d\sum_{i, j= 1}^N\ln\vert 1-x_a{\rm e}^{i(\theta_i-\theta_j)}\vert^2-\frac{1}{2}\sum_{i\neq j=1}^N\ln\vert1-{\rm e}^{i(\theta_i-\theta_j)}\vert^2
  \label{Seff}
\end{eqnarray}
and $\{x_a(R)\}=\{x\zeta_r,x\zeta_r^{-1}\}$, with $r=1,\dots,m$ for even $d=2m$ and $\{x_a(R)\}=\{x,x\zeta_r,x\zeta_r^{-1}\}$ with $r=1,\dots,m$ for odd $d=2m+1$ and the last term comes from the Vandermonde of the $U(N)$ measure. 
In the large $N$ limit the $\theta_i$ become continuous and 
the effective action (\ref{Seff}) scales with $N^2$ giving  
\begin{eqnarray}
  \frac{S_{eff}}{N^2}&=&\sum_{a=1}^d\frac{1}{2}\int_{-\pi}^\pi \rho(\alpha)\int_{-\pi}^\pi\rho(\beta)\ln\vert 1-x_a{\rm e}^{i(\alpha-\beta)}\vert^2d\alpha d\beta\nonumber\\
  &&\qquad\qquad  -\frac{1}{2} P\int_{-\pi}^{\pi}\rho(\alpha)\int_{-\pi}^\pi\rho(\beta)\ln\vert1-{\rm e}^{i(\alpha-\beta)}\vert^2 d\alpha d\beta\, .
  \label{Sofrho}
\end{eqnarray}
with $\int_{-\pi}^{\pi}\rho(\theta)d\theta=1$. Then knowledge of $\rho(\theta)$ determining the partition function in this regime. 

Following \cite{Skagerstam:1983gv}, \cite{Sundborg:1999ue} and \cite{Aharony:2003sx} one can solve for the eigenvalue distribution in the neighbourhood of the transition temperature by expanding in $x_a$ and only retaining the leading term, which corresponds to the
large $N$ approximation used in the previous section, i.e. the low temperature regime.  The resulting model is then an $a_1$ type model \cite{Dutta:2015noa,Liu:2004vy}, but with a further integration over $SO(d)$,  and the partition function (\ref{MolienWeyl}) in this approximation is given by
\begin{equation}
\cZ(N,d;x)\simeq\int \mu_{d}(R)\mu_N(g)]{\rm e}^{a_1(R){\rm Tr}(g){\rm Tr}(g^{-1})}\, .
\end{equation}
Any overall constants can also be neglected for most purposes but will play a r\^ole in connecting to the low temperature phase and the microcanonical ensemble.
In the current case $a_1(R)=\sum_{a=1}^dx_a(R)=x\; {\bf tr}(R)$ and (\ref{Sofrho}) becomes
\begin{equation}
  S_{eff}[\rho]=-a_1(R)\vert \rho_1\vert^2 -\frac{1}{2} P\int\rho(\alpha)\rho(\beta)\ln\vert1-{\rm e}^{i(\alpha-\beta)}\vert^2 d\alpha d\beta\, .
\label{Za1}
\end{equation}
where $\rho_1$ is the first moment of the distribution $\rho(\theta)$ i.e.
\begin{equation}
  \rho_1=\int_{-\pi}^\pi d\theta\rho(\theta){\rm e}^{i\theta} =\int_{-\pi}^\pi\rho(\theta)\cos(\theta)\, .
\end{equation}
Minimising the functional $S_{eff}[\rho]$ for the distribution $\rho$ yields
the integral equation
\begin{equation}
\rho_n=\int_{-\pi}^\pi d\theta \rho(\theta)\cos(n\theta)
\end{equation}

As shown originally in \cite{Sundborg:1999ue} (see also \cite{Aharony:2003sx})
the eigenvalue distribution associated with the model (\ref{Za1})
is given by  below the transition is the uniform distribution on the interval $[-\pi,\pi]$ while in the high temperature phase it is restricted to $[-\theta_0,\theta_0]$ and given by
\begin{eqnarray}
  \rho(\theta)=\left\{\begin{matrix}\frac{1}{2\pi}\hfill&\qquad\hbox{for}\qquad  a_1 \le 1\\
  \frac{1}{\pi \sin^2(\frac{\theta_0}{2})}\sqrt{\sin^2(\frac{\theta_0}{2})-\sin^2(\frac{\theta}{2})}\cos(\frac{\theta}{2})&\qquad\hbox{for}\qquad a_1 \ge 1
  \end{matrix}\right.
    \end{eqnarray}
and  $\theta_0$ is specified by  
\begin{equation}
  s^2\equiv\sin^2(\frac{\theta_0}{2})=1-\sqrt{1-\frac{1}{a_1}}\, .
\label{sofa1}
\end{equation}
The Hagedorn transition occurs at $a_1=1$. Then noting that 
\begin{equation}
 P\int_{-\theta_0}^{\theta_0}d\theta\int_{-\theta_0}^{\theta_0}d\theta'\rho(\theta)\ln\vert\sin\frac{\theta-\theta'}{2}\vert\rho(\theta')=\frac{1}{2}\ln \frac{s^2}{4} -\frac{1}{4}
  \end{equation}
Using $\cos(\theta)=2\sin(\frac{\theta}{2})-1$ we can evaluate
the moments of the distribution as
\begin{equation}
  \rho_n=\int_{-\theta_0}^{\theta_0} \rho(\theta)\cos(n\theta) d\theta
  \end{equation}
to obtain
\begin{equation}
  \rho_1=1-\frac{s^2}{2},\quad \rho_2=(1-s^2)^2,\quad \rho_3=(1-s^2)^2(1-\frac{5}{2}s^2), \quad \rho_4=(1-s^2)^2(1-6s^2+7s^4)\; .
  \end{equation}
Substituting into (\ref{Za1}) one obtains the large $N$ free energy
as
\begin{eqnarray}
 S_{eff}(a_1) = \left\{\begin{matrix}0\hfill&\qquad\hbox{for}\qquad  a_1 \le 1\\
\frac{1}{2}-\frac{1}{2s^2}-\frac{1}{2}\ln s^2&\qquad\hbox{for}\qquad a_1\ge  1\label{Seffofa1}
  \end{matrix}\right.
\end{eqnarray}
and
\begin{equation}
 \cZ(N,d;x)\simeq\int\mu_{d}(R){\rm e}^{-N^2S_{eff}(a_1)}=\int\mu_{d}(R){\rm e}^{-N^2\left(\frac{1}{2}- \frac{1}{2s^2}-\frac{1}{2}\ln s^2\right)}.
  \end{equation}
Due to the exponential fall off with $N^2$, the subsequent integration over $R$ can equally be performed by the saddle point method. From (\ref{sofa1}) and (\ref{Seffofa1})  one can
check that both $s$ and $S_{eff}(a_1)$ are a monotonically decreasing functions of $a_1$ and
since $a_1(R)=x\, {\bf tr}(R)$ with $R$ given in (\ref{BlockRotationMatrices}) minimised by $R={\bf 1}$ i.e. $\theta_r=0$, hence in the high temperature regime the effect of restricting to $SO(d)$ singlets is that $a_1(R)= x d$ at this large $N$ saddle. 

The energy can similarly be found noting that
\begin{equation}
  E=-a_1\frac{d S_{eff}(a_1)}{da_1}
  \label{EfromSeff}
\end{equation}
so that for $a_1= d x$ we have
\begin{eqnarray}
  E=\left\{\begin{matrix}&0\hfill&\qquad\hbox{for}\qquad x \le \frac{1}{d}\,\\
   &\frac{1}{4}\left(\frac{2}{s^2}-1\right)\simeq \frac{1}{4}+\frac{\sqrt{d}}{2}\sqrt{x-\frac{1}{d}}+\cdots&\qquad\hbox{for}\qquad x \ge \frac{1}{d}\, .
  \end{matrix}\right.
  \label{Enegryofs}
  \end{eqnarray}
which near the Hagedorn temperature gives
\begin{eqnarray}
  E=\left\{\begin{matrix}0\hfill&\qquad\hbox{for}\qquad  x \le \frac{1}{d}\\
  \frac{1}{4}+\frac{\sqrt{d}}{2}\sqrt{x-\frac{1}{d}}+\cdots&\qquad\hbox{for}\qquad x \ge \frac{1}{d}\, .
  \end{matrix}\right.
  \label{Energyofx}
\end{eqnarray}
and hence the energy is therefore predicted to be discontinuous across
the transition, undergoing a jump discontinuity of $\frac{1}{4}$. Also
we see that the for large $N$ in terms of the microcanonical variables
the transition occurs at $k_H=\frac{N^2}{4}$. To make contact with the
microcanonical data we need to extract the entropy as a function of
the energy. We can do this by noting the standard thermodynamic
relation\footnote{The infinitisimal form of this relation is $T
dS=dE$, i.e. $\frac{dS}{dE}=\beta(E)$}, $F=E-TS$, with
$\beta F=S_{eff}$ becomes
\begin{equation}
  S[E]=\beta(E) E-S_{eff}
  \label{ZEntropyEnergy}
\end{equation}
with $\beta(E)$ obtained from 
(\ref{Enegryofs}) to get
\begin{equation}
  \beta(s)=\ln d+\ln(1-(1-s^2)^2)\quad\hbox{with}\quad s^2=\frac{1}{2(E+\frac{1}{4})}
\end{equation}
  So that
  \begin{equation}
    \beta(E)=\ln d
  +\ln\left(1-(1-\frac{1}{2}\frac{1}{E+\frac{1}{4}})^2\right)\simeq\ln d -4{(E-\frac{1}{4})}^2+\cdots
\end{equation}
and hence substituting into (\ref{ZEntropyEnergy}) yields
\begin{equation}
  S(E)=E \ln(E d)-\frac{1}{2}\ln 2+(E-\frac{1}{4})-2(E+\frac{1}{4})\ln(E+\frac{1}{4})% \simeq\frac{\ln d}{4}-\frac{4}{3}{(E-\frac{1}{4})}^3+\cdots
\label{EntropyHigh}
\end{equation}
Setting $E=\frac{k}{N^2}$ and expanding near $E=\frac{1}{4}$ gives 
\begin{equation}
  S(\frac{k}{N^2})\simeq \frac{k}{N^2}\ln d-\frac{4}{3}{\left(\frac{k}{N^2}-\frac{1}{4}\right)}^3+\cdots\, .
\label{HighEnergyZUngauged}
\end{equation}

Equation (\ref{HighEnergyZUngauged}) is the leading large $N$ contribution 
but to get agreement with the low temperature regime we need to include the $\frac{1}{N^2}$ corrections. For this we perform the steepest decent more carefully including the 1-loop correction or saddle point Hessian prefactor.
Then expanding around $S_{eff}(a_1)$ with $a_1=x( d-\displaystyle\sum_{r=1}^m\theta_r^2+\cdots)$ and using (\ref{EfromSeff}) gives
\begin{equation}
S_{eff}(a_1)=S_{eff}(x d)+\frac{E}{d}\sum_{r=1}^m\theta_r^2+\cdots
\end{equation}
So that the Hessian contributes diagonal entries
$\frac{E}{d}=\frac{k}{N^2d}$ and, in a ball, of radius $\delta$ in the neighbourhood of the identity of $SO(d)$ and we are left with
the prefactor being given by
\begin{equation}
  I_{d}=\int_{{\mathbb B}^m(\delta)}\mu_d {\rm e}^{-\frac{k}{d} r^2}
  \end{equation}
and we recover exactly the integrals of Section \ref{sec:LowEnergy}.  

Including these saddle point prefactors we find that the entropy $S=\ln \cZ$ is, as expected, continuous across the transition\footnote{There is also no evidence for a transition in the entropy from our finite $N$ analysis in agreement with our analysis on either side of the transition.} and adjusting, by a subleading constant in large $N$, the argument of $\ln k$ in both the low and high energy to be $\ln\frac{4k}{N^2}$, this translates to the entropy, modulo an overall $N$ dependent constant,  across the transition being:
\begin{equation}
S(\frac{k}{N^2})=\left\{\begin{alignedat}{3}
  &\frac{k}{N^2} \ln d- \frac{\nu(d)}{N^2}\ln\frac{4k}{N^2}+\cdots\quad&\hbox{for}&\quad k\le \frac{N^{2}}{4} \\
   &\frac{k}{N^2} \ln d- \frac{\nu(d)}{N^2}\ln\frac{4k}{N^2} -\frac{3}{4}\left(\frac{k}{N^2}-\frac{1}{4}\right)^3+\cdots\quad&\hbox{for}&\quad k\ge \frac{N^2}{4}
\end{alignedat}\right.
\label{entropyTransition}
\end{equation}
Then 
\begin{equation}
  \cZ(N,d,k)=c(d){\rm e}^{N^2S(\frac{k}{N^2})}
\label{HighEnergygauged}
\end{equation}
we have
\begin{equation}
 \cZ(N,d,\frac{N^2}{4})=c(d)d^{N^2/4}\, .
\label{EndOfHagedorn}
\end{equation}
with the low and high energy constants matching\footnote{The scale  $\frac{N^2}{4}$, as a characteristic of the onset of  high-energy branch above the Hagedorn transition in the case without projection to the rotationally invariant sector,  was found previously and independently in \cite{Berenstein:2023srv} and \cite{O'connor}, with the latter using the methods described here.} at $k=\frac{N^2}{4}$. 

Differentiating (\ref{entropyTransition}) one can see that, at large $N$, the heat capacity is negative for $k<\frac{N^2}{4}$ and positive for $k>\frac{N^2}{4}$ implying that the large $N$ limit of $k_{\crit}$ is precisely $k_{\crit}=\frac{N^2}{4}$. Furthermore Eqn (\ref{EndOfHagedorn}) gives the asymptotic large $N$ form of $\cZ(N,d,k_{\crit})$.

The bound $ d < \sqrt {k } $ for the validity of the low-temperature formula discussed at the end of \eqref{sec:largeNZNdk} also applies to the matching formula obtained here  from a  high temperature analysis.
We will turn to the large $d$ limit in section \ref{sec:largeNd}, where we find that ribbon graphs play an important role.

\subsection{Approximate analytic fits of the caloric fold }
\label{sec:ApproxAnalyticCal}
First let us consider the $d$ matrix model, without a global ungauged $SO(d)$ symmetry, in the vicinity of the transition and at large $N$. Combining the high energy (\ref{HighEnergyZUngauged}) and low energy (\ref{LowEnergyUngagued}) results one obtains:
\begin{equation}
  Z(N,d,k)=\left\{\begin{alignedat}{3}&c(d)d^k\quad&\hbox{for}\quad&k \le \frac{N^2}{4},\\
   &c(d) d^k{\rm e}^{-N^2\frac{4}{3}(\frac{k}{N^2}-\frac{1}{4})^3+\cdots}\quad&\hbox{for}\quad&k \ge \frac{N^2}{4}\, . 
  \end{alignedat}\right.
\label{HighLowLargeNMatchedUngauged}
\end{equation}
At the Hagedron energy, $E=\frac{k}{N^2}=\frac{1}{4}$ we have
\begin{equation}
  Z(N,d,\frac{N^2}{4})\sim c(d)d^{N^2/4}\, .
  \end{equation}

For the $SO(d)$ gauged model at low energies we have found that
$\cZ(N,d,k)$ is given by (\ref{LowEnergyCoefficientsSOdGaugedv2}) while at high
energies it is given by (\ref{HighEnergygauged}) i.e. 
\begin{equation}
  \cZ(N,d,k)=\left\{\begin{alignedat}{3}&d^k\frac{c(d)}{k^{d(d-1)/4}}\quad&\hbox{for}\, &k \le \frac{N^2}{4},\\
   &d^k\frac{c(d)}{k^{d(d-1)/4}}{\rm e}^{-N^2\frac{4}{3}(\frac{k}{N^2}-\frac{1}{4})^3+\cdots}\quad&\hbox{for}\, &k \ge \frac{N^2}{4}\, .
  \end{alignedat}\right.
\label{HighLowLargeNMatched}
\end{equation}
The modification due to the $SO(d)$ singlet constraint is then the same $1/k^{d(d-1)/4}$ factor to both the high and low energy domains. 

\begin{figure}
    \begin{minipage}{0.45\textwidth}
    \includegraphics[width=\textwidth]{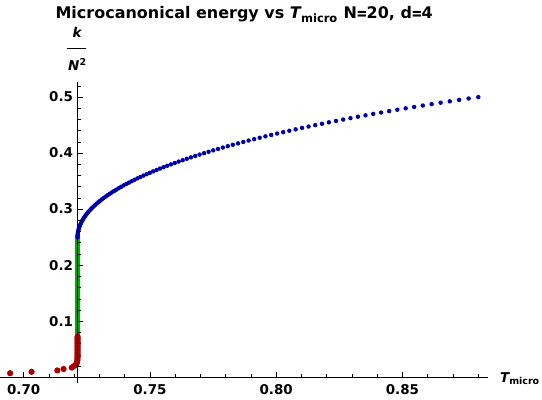}
    \end{minipage}
    \hfill
  \begin{minipage}{0.45\textwidth}
    \includegraphics[width=\textwidth]{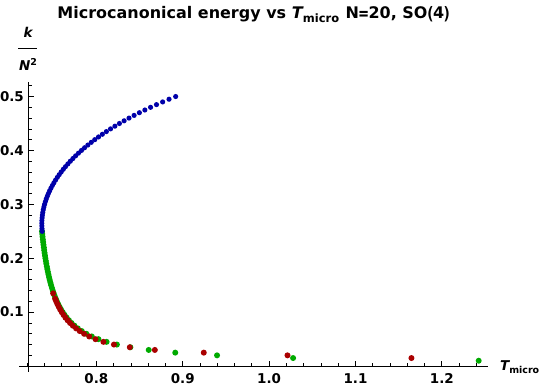}
    \end{minipage}
   \label{fig:LargeNFold}
   \caption{Plot of the energy versus microcanonical temperature for the
     ungauged $d=4$ model on the left and the $SO(4)$ gauged model on the right. We used
     Enq (\ref{HighLowLargeNMatchedUngauged}) for the ungauged model with $Z(20,4,k)$ in red and Eqn (\ref{HighLowLargeNMatched}) with $N=20$ for the gauged model with the red data obtained from the exact degeneracies $\cZ(20,4,k)$. The curves merge surprisingly well and match at $N^2/4$. The green and blue data correspone to $k\le \frac{N^2}{4}$ and $k\ge \frac{N^2}{4}$ respectively. The axes origin is set at $(\frac{1}{\ln d},0)$ in both graphs. The discrete symmetric derivative as in (\ref{Tmicro}) was used throughout.}
\end{figure}

We use (\ref{HighLowLargeNMatched}) to evaluate the caloric fold and perform comparisons with our exact $\cZ(N,d,k)$ degeneracies, finding surprisingly good agreement even for relatively small $N$, see Figure \ref{fig:LargeNFold} which shows: on the left a plot of the $4$-matrix model with $N=20$ and on the right we show the $SO(4)$ gauged model again with $N=20$. Both
graphs match smoothly across the transition and are in excellent
agreement with our overall expectaions.

The eigenvalue-density analysis in this section gives a natural large-$N$
physics explanation for the scale $k_{\rm crit}\sim N^2/4$, in agreement with
the exact finite-$N$ data obtained from the microcanonical degeneracies.  It
would be valuable to sharpen the status of this result in two complementary
directions.  One direction is analytic: to put the eigenvalue-density treatment
on a more rigorous asymptotic footing, including more systematic control of the
saddle point, subleading corrections, and the matching to the microcanonical
degeneracies near the turn of the caloric fold.  A second, more directly
combinatorial, direction is to seek a derivation of the $N^2/4$ scale directly
from the exact finite-$N$ formulae of Section~5, perhaps by analysing the onset
of finite-$N$ trace relations or the representation-theoretic cutoffs in the
sums over Young diagrams along the lines of \cite{Berenstein:2023srv}.

\section{ Ribbon graphs from large $ N , d $ limit.  }
\label{sec:largeNd} 

In this section, we study $ \cZ ( N , d, k )$ in the regime of parameters where $ N \ge k , d >  k $. In this regime, 
$ \cZ( N , d , k )$ is independent of $N,d$ and we denote these stable values as $ \cZ ( \infty , \infty , k )$. 
These are non-zero for even $k$ and we have non-trivial functions $ \cZ ( \infty , \infty , 2n  )$. After large $N$ and large $d$ simplifications in \ref{sec:largeNZNdk} and \ref{sec:largeNDfromCompForm}, we derive 
two expansions of $ \cZ(  N , d , k )$ as sums over partitions of $k$. These two expansions are \eqref{Zpcharsum0} and  \eqref{ZpPhiq}.  The first expansion gives an efficient computational approach based on cycle indices for a wreath-product  sub-group $S_n [ S_2 ] $ of $S_{2n} $. The second expansion has a geometrical interpretation in terms of a counting of ribbon graphs.  Finite $N,d$ counting functions can be recovered  geometrically from the $ \cZ ( \infty , \infty , 2n )$ by doing a Fourier transformation on the space of ribbon graphs (section \ref{subsec-finiteNd}).
In section \ref{sec:calfolLarged} we consider thermodynamics in the regime where $ d >> N$, but where $ k $ values are considered below and above $N$.  In this case, we explain  that there is again the phenomenon of the caloric fold, with the difference from the small $d$ cases, that the Hagedorn temperature is now zero in the large $N$ limit. 

 As an interesting  side-remark, developed in Appendix \ref{sec:GenTwo}, the existence of two distinct expansions for $ \cZ( \infty , \infty , k ) $ has a group-theoretic generalisation where $S_k$ is replaced by any finite group $G$ and $ S_n[S_2]$ by a any subgroup $H \subset G$. 

\subsection{ Large $N$ simplification of  counting }
\label{sec:largeNZNdk} 
	
In the regime $N \ge k $, we can simplify  \eqref{charformZNdk0} and use the definition \eqref{defPhi}  to write 
\bea 
\cZ ( N , d , k ) \rightarrow \cZ ( N\ge k , d , k ) = \sum_{ p \vdash k  } { 1 \over \Sym ~ p } \sum_{ \substack { R \vdash k }}  ( \chi^R_p )^2  \Phi ( d , k ;  p ) 
\eea 
since none of the Young diagrams $R$ exceed $N$ in height. Now use character orthogonality 
\bea 
\sum_{ \substack { R \vdash k }}  ( \chi^R_p )^2 = \Sym ~ p 
\eea 
to obtain 
\bea\label{largeNZNdk} 
 \cZ ( \infty  , d , k )   \equiv  \cZ ( N \ge k , d , k ) = \sum_{ p \vdash k  } \Phi ( d , k ;  p )  
\eea

\subsection{ Large $d$ simplification of $ \cA ( k , d )$ and $ \Phi ( d , k ;  p )$ }
\label{sec:largeNDfromCompForm}

We give the large $N,d$ limit of the character-pairing formulae we derived in earlier sections. 
In the large $d$ limit, $ d >   k $, the epsilon contractions are absent and the counting simplifies 
\begin{itemize}
	
	\item Since $ \cA_- ( k , d ) $ only includes Young diagrams with $k$ boxes and length exactly equal to $d$, it becomes empty when $ d >  k $. An immediate consequence is 
	\bea 
	\cZ_{ SO } ( N , d >   k  , k  ) = \cZ_{ O } (  N , d >  k , k )
	\eea

	\item It follows that $ \cA( k , d ) = \cA_+( k , d ) $. The Young diagrams in this set have even length rows. Therefore $ k $ must be even, and we write it as $ k = 2n$ for positive integer $n$  
	\bea 
	\cZ_{ SO } ( N , d > 2n+1  , 2n +1   ) = \cZ_{ O } (  N , d > 2n+1 , 2n +1  ) =0 
	\eea 
	
	\item The set  $ \cA( k = 2 n  , d ) $ becomes independent of $ d$ in the range $ d \ge  2 n $. We denote the set in this range as 
	$ \cA( 2n   , d = \infty )$. There is an interesting representation theoretic property of $ \cA( 2 n  , \infty ) $ 
	\bea\label{RepThyFct}  
	\cA ( 2 n , \infty ) && = \hbox{Set of Young diagrams with even row lengths  only } \cr 
	&& = \hbox{Set of irreps of $S_{ 2n }$ which contain the trivial irrep of $S_n [ S_2 ] $ with multiplicity one }  \cr 
	&& 
	\eea
	All irreps of $S_{ 2n }$ not belonging to $\cA( 2n , \infty )$ do not contain the trivial irrep of $S_n [ S_2 ]$ 
	
	\item There is an interpretation of $ \cZ( N = \infty , d = \infty , 2k )$ in terms of counting ribbon graphs. The 
	connection to $S_{n } [ S_2 ]$ plays an important role in this connection, as we will elaborate further along in this section.

\end{itemize}

\subsection{ Two partition expansions of the large $N,d$ limit } 
\label{sec:twoPartExps}

Using \eqref{RepThyFct}, we have 
\bea\label{Zpcharsum}
\Phi ( \infty  , 2n  ;  p )  \equiv \sum_{ S \in \cA( 2n  , \infty  ) } \chi^S_p  
\eea 
Combining with the large $N$ formula  \eqref{largeNZNdk},  
\bea\label{Zpcharsum0} 
\cZ ( \infty  , \infty  , 2n ) \equiv 
\cZ ( N \ge 2n , d > 2n  , 2n ) &&= \sum_{ p \vdash 2n   } \Phi ( \infty , 2n ;   p )  
\eea
As we explain  in section \ref{sec:IntegZInfp}, $ \Phi  ( \infty , p ) $ are non-negative integers and can be zero (see Tables \ref{tab2}). We will shortly write another equation for $\cZ ( \infty  , \infty  , 2n ) $ as a sum over partitions where each summand is strictly positive.

Using the representation-theoretic fact (equation \eqref{RepThyFct}) above, along with character orthogonality, we have 
\bea\label{charsumSRedct} 
&&  \Phi  (  \infty , 2n  ; p   )    
=  { 1 \over 2^n n! }  \sum_{S \vdash 2n}  \chi^S_p \sum_{ \gamma \in S_n [S_2] } \chi^S ( \gamma ) \cr 
&& = { 1 \over 2^n n! } \sum_{\mu \in S_{ 2n } } \sum_{\gamma \in S_{ n } [S_2 ]}  \delta ( \gamma \mu \sigma_p  \mu^{-1} )
\eea 
The delta function $ \delta ( \sigma )$  on  group elements  $ \sigma \in S_{2n}$ is defined to be $ 1$ if $ \sigma = \sigma_0  $ is the identity and zero otherwise. More generally, for group algebra elements $ A = \sum_{\sigma \in S_{ 2n} } A_{ \sigma } \sigma $, it is defined to be $A_{ \sigma_0 }$. In the above equation, it ensures that $ \gamma = \mu \sigma_p^{-1} \mu^{-1}$, so that $ \gamma \in S_n [ S_2] $ is in the class $ \cC_p$. It is also useful to observe that the number of permutations $ \mu \in S_{ 2n }$ which obey $ \mu \sigma_p \mu^{-1} = \sigma_p $ is $ \Sym ~ p = \prod_{i=1 }^{ 2n } ( i^{ p_i } p_i! ) $. Let us denote by $  \cZ ( S_n [ S_2 ] , p )  $ the  number of permutations in $ S_{ n } [ S_2 ] \subset S_{ 2n }$ which belong to the conjugacy class $\cC_p$,  divided by the order of group $ S_n [ S_2 ]$. We may therefore write  
\bea\label{Zlowerp} 
\Phi  ( \infty  , 2n  ;  p  ) = { 1 \over 2^n n! } \sum_{\mu \in S_{ 2n} } \sum_{\gamma \in S_{ n } [S_2 ]}  \delta ( \gamma \mu \sigma_p  \mu^{-1} )=  \cZ ( S_n [ S_2 ] , p ) ~ \Sym ~ p 
\eea
and 
\bea\label{dNinf-from-Cycle} 
\cZ ( \infty  , \infty  , 2 n ) && = \sum_{p \vdash 2n }  \Phi  ( \infty  , 2n  ;  p  ) =  \sum_{p \vdash 2n } \cZ ( S_n [ S_2 ] , p ) ~ \Sym ~ p 
\eea 
$\cZ ( S_n [ S_2 ] , p ) $  is related to the {\it cycle index  
	polynomial }  $ \cZ( S_n [ S_2 ] , x_1 , \cdots , x_{2n} )   ) $ of $ S_n [ S_2]$ : 
\bea 
\cZ ( S_n [ S_2 ] , p ) = {\rm Coeff } \left ( \prod_{i =1} x_i^{ p_i} , \cZ( S_n [ S_2 ] , x_1 , \cdots , x_{2n} )   \right ) 
\eea 
Thus we may write 
\bea\label{CycIndForm}  
\cZ ( \infty  , \infty  , 2 n ) = \sum_{p \vdash 2n }   ~{\rm Coeff } \left ( \prod_{i =1} x_i^{ p_i} , \cZ( S_n [ S_2 ] , x_1 , \cdots , x_{2n} )   \right )  ~ \Sym ~p  
\eea 
The cycle index polynomial of the wreath product \(S_n[S_2]\) may be conveniently packaged in the following exponential generating function \cite{BeLaLe}:
\begin{equation}
	\label{CycIndWreathEGF}
	\sum_{n\geq 0}
	t^n\,
	\cZ\!\left(S_n[S_2];x_1,x_2,\ldots\right)
	=
	\exp\left[
	\sum_{j\geq 1}
	\frac{t^j}{j}
	\left(
	\frac{x_j^2+x_{2j}}{2}
	\right)
	\right].
\end{equation}
Equivalently, extracting the coefficient of \(t^n\) gives the explicit formula
\begin{equation}
	\label{CycIndWreathExplicit}
	\cZ\!\left(S_n[S_2];x_1,\ldots,x_{2n}\right)
	=
	\sum_{r\vdash n}
	\frac{1}{\Sym(r)}
	\prod_{j\geq 1}
	\left(
	\frac{x_j^2+x_{2j}}{2}
	\right)^{r_j}.
\end{equation}
Here \(r=(1^{r_1}2^{r_2}\cdots)\) and
\[
\Sym(r)=\prod_{j\geq 1}j^{r_j}r_j! .
\]

Using these cycle polynomials \eqref{dNinf-from-Cycle} reproduces the OEIS sequence A268558 \cite{OEIS-A268558}. As we will briefly describe in the following, the formulae  \eqref{charsumSRedct} and \eqref{dNinf-from-Cycle} are related to the counting of ribbon graphs as previously explained in \cite{FeynString} (there was an error in converting  \eqref{dNinf-from-Cycle},  which is the correct equation (8.24) in \cite{FeynString}, into the numerical sequence written in equation (D.9) there. The error also propagated to the numerical sequence equation (D.10)  of that paper).

It is useful to rewrite \eqref{charsumSRedct} by replacing the fixed $ \sigma_p \in \cC_p$ with a sum over $ \sigma \in \cC_p$, while dividing by the number of elements $ |\cC_p|$ in the conjugacy class. 
\bea\label{csr1} 
\cZ ( \infty  , \infty  , 2 n ) && =   { 1 \over 2^n n! }  \sum_{p \vdash 2n } \sum_{\beta  \in S_{ 2n} } \sum_{\gamma \in S_{ n } [S_2 ]}  \sum_{\sigma \in \cC_p } { 1 \over |\cC_p| }  \delta ( \gamma \beta  \sigma  \beta^{-1} )  \cr 
&& =   { 1 \over 2^n n! }  { 1 \over (2n)! } \sum_{p \vdash 2n } \sum_{\beta \in S_{ 2n} } \sum_{\gamma \in S_{ n } [S_2 ]}  \sum_{\sigma \in \cC_p } \delta ( \gamma \beta \sigma  \beta^{-1} )  \Sym ~ p 
\eea 
We have also made use of the fact that 
\begin{equation}
	|\cC_p| = { ( 2n )! \over \Sym ~ p } \nnm 
\end{equation}
The $ \Sym ~p $  factor in \eqref{csr1} can be replaced by a sum over $ \alpha $ in $ S_{2n}$ and constrained to leave $ \sigma $ fixed under conjugation.  We also   combine the sums over $ p , \sigma \in \cC_p$ into a sum over all permutations in $S_{ 2n}$. 
\bea\label{csr2}  
\cZ ( \infty  , \infty  , 2 n ) &&= { 1 \over 2^n n! }  { 1 \over (2n)! } \sum_{p \vdash 2n } \sum_{\beta \in S_{ 2n} } \sum_{\gamma \in S_{ n } [S_2 ]}  \sum_{\sigma \in \cC_p }  \sum_{\alpha \in S_{ 2n} } \delta ( \gamma \beta \sigma  \beta^{-1} )  \delta ( \alpha \sigma \alpha^{-1} \sigma^{-1} )\cr 
&& = { 1 \over 2^n n! }  { 1 \over (2n)! }  \sum_{\beta \in S_{ 2n} } \sum_{\gamma \in S_{ n } [S_2 ]}  \sum_{\sigma \in S_{ 2n }  }  \sum_{\alpha \in S_{ 2n} } \delta ( \gamma \beta \sigma  \beta^{-1} )  \delta ( \alpha \sigma \alpha^{-1} \sigma^{-1} )
\eea
We can now use the first delta function to solve for $ \sigma $  and find $ \sigma  = \beta^{-1} \gamma^{-1} \beta $. 
Substituting into the second delta function 
\bea 
\cZ ( \infty  , \infty  ,2 n ) && = { 1 \over 2^n n! }  { 1 \over (2n)! }  \sum_{\beta \in S_{ 2n} } \sum_{\gamma \in S_{ n } [S_2 ]}   \sum_{\alpha \in S_{ 2n} } \delta  ( \alpha ~  \beta^{-1} \gamma^{-1} \beta ~  \alpha^{-1} ~  \beta^{-1} \gamma \beta ) 
\eea
We can now use the invariance of the $ \alpha $ sum under conjugation by $ \beta $ to obtain 
\bea\label{WrOrbCount}  
\cZ ( \infty  , \infty  , 2n ) && = { 1 \over 2^n n! } \sum_{\gamma \in S_{ n } [S_2 ]}   \sum_{\alpha \in S_{ 2n} }   \delta  ( \alpha ~   \gamma^{-1} ~  \alpha^{-1} ~   \gamma  ) 
\eea
It is instructive to define, for partitions $ q \vdash 2n$,  
\bea\label{WrOrbCountq} 
 \Phichk ~~( \infty ,   2n ; q )  = { 1 \over 2^n n! } \sum_{\gamma \in S_{ n } [S_2 ]}   \sum_{\alpha \in \cC_q \in  S_{ 2n} }   \delta  ( \alpha ~   \gamma^{-1} ~  \alpha^{-1} ~   \gamma  ) 
\eea
so that 
\begin{equation}\label{ZpPhiq} 
	\boxed{~~ \cZ ( \infty  , \infty  , 2n )  = \sum_{q \vdash 2n }  \Phichk ~( \infty ,   2n ; q ) ~~ } 
\end{equation}
The equation \eqref{WrOrbCount} makes the connection between $ \cZ ( \infty  , \infty  , 2n ) $  and  ribbon graph counting easy to see, as we discuss shortly.  This interpretation makes it manifest that  $ \Phichk ~~( \infty ,   2n ; q )   $ are {\it positive integers}. It turns out that $ \Phi  ( \infty , 2n ; p  )$ are in general  {\it non-negative integers}, as shown in Section \ref{sec:IntegZInfp}. By the way, the integrality of $ \Phi  (  \infty , 2n  ; p )$ also follows from the character sum formula \eqref{Zpcharsum} since characters of symmetric groups are integers (possibly negative), but section \ref{sec:IntegZInfp} also shows non-negativity. 

\begin{table}[ht]
	\centering
	\begin{tabular}{c|cc}
		$p,q$ & $ \Phichk (\infty,4 ; q )$ & $\Phi (\infty,4 ; p )$ \\ \hline
		$[1^4]$   & $1$ & $3$ \\
		$[2,1^2]$ & $2$ & $1$ \\
		$[2^2]$   & $2$ & $3$ \\
		$[3,1]$   & $1$ & $0$ \\
		$[4]$     & $2$ & $1$
	\end{tabular}
	\caption{For $n=2$, the functions $ \Phichk (\infty,4 ; q ) $ and $ \Phi (\infty,4 ; p ) $ are distinct as functions of the partition $q,p\vdash 2n$, although their sums over all partitions of $2n$ are equal. }
	\label{tab2} 
\end{table}

{\bf Remark:}
It is important to note that $ \Phi ( \infty , 2n ; p )  $ and $ \Phichk ( \infty , 2n ; q )$ are distinct functions of partitions of $(2n)$ which allow partition expansions of the large $N,d$ limit $ \cZ ( \infty , \infty , 2n )$.  
$p$ is the original partition which organises character sums in \eqref{charformZNdk} and  \eqref{Zpcharsum0}.
$q$ is the vertex structure of the ribbon graphs. Both formulae have to do with $ S_n [ S_2 ]$. But 
$ \Phi ( \infty , 2n ; p )  $ is related to cycle indices of permutations in $ S_n [ S_2]$ when they are embedded in $ S_{ 2n}$. $ \Phichk ( \infty , 2n ; q )   $ is related to orbits of $ S_n [ S_2 ] $ acting by permutations on conjugacy class $ \cC_q$ in $ S_{ 2n }$. The delta function formulae are both simple, but different - they involve $ \gamma $ and its conjugate in $ \Phichk ( \infty , 2n ; q )   $  ( equation \eqref{WrOrbCountq}   )but only $ \gamma $ in $ \Phi ( \infty , 2n ; p )   $ (equation \eqref{charsumSRedct}).

\subsection{ Ribbon graphs and $ \cZ( \infty , \infty , 2n )$ }
\label{sec:Zinfribb}

This formula \eqref{WrOrbCount} has a neat interpretation in terms of counting orbits of $ S_n [ S_2]$ action by conjugation on permutations in $ S_{ 2n}$. The counting of such orbits is equal, by Burnside's Lemma, to the average over  $ \gamma \in S_n [ S_2 ]$ of the number of fixed points of the action, i.e. the number of solutions to 
\bea 
\gamma \sigma \gamma^{-1}  = \sigma 
\eea
as $ \sigma $ runs over $S_{ 2n }$. This average is  computed by \eqref{WrOrbCount}

We can arrive at the same expression \eqref{WrOrbCount} using a different route, with a starting point that connects directly with the counting of ribbon graphs on surfaces.  Ribbon graphs on surfaces have a cyclic orientation at each vertex. A convenient way to  describe  a ribbon graph is to 
treat the existing vertices of the graph as black nodes  of a new embedded graph. The new graph is also endowed with white vertices in the middle of the edges of the original graph, thus dividing each edge of the original graph into two half-edges. We can assign labels $ \{ 1 , 2 , \cdots , 2n \}$ to the half-edges. Traversing each of the black vertices in the direction of the orientation of the surface and collecting the cycles produces a permutation $ \sigma_1 \in S_{ 2n }$ of some cycle structure we may call $ q \vdash 2n $, which is specified by $\{ q_1 , q_2 , \cdots , q_{ 2n} \}$ which give the number of vertices of valency $ 1,2, \cdots 2n $. Traversing the edges around the white vertices gives a permutation $ \sigma_2 \in [2^n] \in S_{ 2n }$, i.e. a permutation in the conjugacy class of $n$ cycles of length $2$.  A relabelling of the edges 
with $ i \in \{ 1 , \cdots , 2n \}$ being renamed as $ \alpha ( i ) $, using a permutation $ \alpha \in S_{ 2n}$, leads to :
\bea\label{permequivs} 
&& \sigma_1 \rightarrow \alpha \sigma_1 \alpha^{-1} \cr  
&& \sigma_2 \rightarrow \alpha \sigma_2 \alpha^{-1}
\eea
If we generalise the conjugacy class of $\sigma_2 $ we get general bi-partite embedded graphs, which are related to Belyi maps \cite{LandoZvonkin} (for a physics discussion linking these to matrix model correlators  see \cite{DMSRBElyi}\cite{TomComplex} and earlier references therein).

Using the description in \eqref{permequivs}, the counting of ribbon graphs with vertex structure $ q$ is  given, using Burnside's Lemma,  by 
\bea\label{Ribbcount} 
\cZ_{\mathrm { ribb}  ; q } ( 2n ) = {  1 \over  ( 2n)!  }\sum_{\alpha \in S_{ 2n } } \sum_{\sigma_1 \in \cC_q} \sum_{\sigma_2 \in [ 2^n ] \in S_{ 2n } } 
\delta ( \alpha \sigma_1 \alpha^{-1} \sigma_1^{-1 }) \delta ( \alpha \sigma_2 \alpha^{-1} \sigma_2^{-1 } )
\eea 
If we  choose $ \sigma_2 = \sigma_2^{*} = ( 12) ( 34 ) \cdots ( 2n-1, 2n ) $, which can be viewed as a partial  gauge-fixing of the $S_{ 2n }$ symmetry. The residual gauge symmetry is the group of permutations $ \alpha \in S_{ n } [ S_2 ]  \subset S_{ 2n} $, which preserve $ \sigma_2^{*}$. The sum $ \sigma \in [ 2^n ]$ in \eqref{Ribbcount} contains $ { (2n)! \over 2^n n! }$ such choices of gauge-fixed versions of $ \sigma_2$, which means we can write 
\bea 
\cZ_{\mathrm { ribb}  ; q } ( 2n ) = {  1 \over 2^n n ! } \sum_{ \alpha \in S_{ n } [ S_2 ]}  \sum_{\sigma_1 \in \cC_q } \delta ( \alpha \sigma_1 \alpha^{-1} \sigma_1^{-1})
\eea 
Recalling \eqref{WrOrbCountq}, we can identify 
\bea
\cZ_{\mathrm { ribb}  ; q } ( 2n ) = \Phichk ( \infty  , 2n ; q )
\eea 
This identification immediately implies that  $  \Phichk ( \infty  , 2n ; q ) $ are positive integers 

\subsection{ $ \Phi ( \infty , 2n ; p )$ : weighted orbit counting and  non-negative integers  }
\label{sec:IntegZInfp} 

We now give an interpretation of $ \cZ_p ( \infty , \infty , 2n )$, which comes directly from our pairing formula (following equations \eqref{charsumSRedct} and \eqref{Zlowerp} ), as a counting of orbits. This proof comes from an expedient contribution from ChatGPT, when prompted to prove that $ Z_p ( \infty , \infty , 2n  )$ is integer.  

Starting from \eqref{Zlowerp}, reproduced here for convenience,  
\bea
 \Phi ( \infty , 2n ; p ) = { 1 \over 2^n n! } \sum_{\mu \in S_{ 2n} } \sum_{\gamma \in S_{ n } [S_2 ]}  \delta ( \gamma \mu \sigma_p  \mu^{-1} )=  \cZ ( S_n [ S_2 ] , p ) ~ \Sym ~ p 
\eea
we note that 
\bea 
 \Phi ( \infty , 2n ; p ) = 	\frac{ \Sym ~ p   }{|S_n[S_2] |} \, |\cC_{ p } \cap S_n[S_2]  |,
\eea 
and $ \Sym ~ p  $ is the order of the subgroup of $ S_{ 2n}$ which fixes a permutation $ \sigma_p$ under conjugation, expressed as 
\bea 
\Sym ~ p = |{\rm Centralizer }_{S_{ 2n} }(\sigma_p)|.
\eea
Decomposing $|\cC_{ p } \cap S_n[S_2]  |$ as a union of orbits $\cO_s$ under conjugation by $S_n [ S_2]$, we have 
\bea 
\cC_p \cap H = \bigsqcup_{s } \cO_s , 
\eea 
where $s$ runs over the complete set of orbits. By the orbit-counting lemma, the cardinality of the orbit is  
\bea 
|\mathcal O| =
\frac{|S_n [ S_2]|}{|{ \rm Centralizer}_{ S_n [ S_2 ]}(\sigma_{s })|}
\quad (\sigma_{ s } \in \mathcal O_s ),
\eea 
Hence, 
\bea 
  \Phi ( \infty , 2n ; p ) && =   (\Sym ~ p   ) 	\sum_{ s }
\frac{1}{|{\rm Centralizer}_{ S_{ n }[S_2] }(\sigma_{s })|}  \cr 
&& = 	\sum_{ s } { |{\rm Centralizer }_{S_{ 2n} }(\sigma_{s })| \over |{\rm Centralizer}_{ S_{ n }[S_2] }(\sigma_{s})| } 
\eea
Since $  {\rm Centralizer}_{ S_{ n }[S_2] }(\sigma_{s}) $ is a subgroup of ${\rm Centralizer }_{S_{ 2n} }(\sigma_{s })$, the above equation shows that $\cZ_p  ( \infty  , \infty  , 2 n ) $ is non-negative integer. It can be zero when 
$\cC_p \cap H$ is empty. For $n=2$, this occurs for $ p = [ 3,1]$ as we see from Table \ref{tab2}.

\subsection{ Finite $N , d $ from Fourier transform on Ribbon graphs }
\label{subsec-finiteNd} 

The ribbon-graph interpretation of $ \cZ ( \infty , \infty , 2n )$ based on its expansion in  terms of $ Z^{(q)}( \infty, \infty , 2n )$ has two other closely related descriptions. It was proved in \cite{FeynString},  using the permutation-based counting of Feynman graphs,  that the counting of vacuum graphs of Yukawa theory (Dirac fermion coupled to a scalar), or quantum electrodynamics (without the Furry constraint from gamma matrix combinatorics which sets the fermion loops with an odd number of vertices to zero), coincide with the ribbon graph counting. Generalisations to include graphs with external legs have been given in \cite{KLS2018a,KLS2018b}.   A more direct derivation  of $ Z^{(q)}( \infty, \infty , 2n )$  by using permutations $ \alpha \in S_{ 2n }$ to contract indices of the matrices to form mesonic composites which are $O(d ) \times U(N)$ invariant polynomials of matrix variables $X^{ i}_{ a , j }$ was given in \cite{KRS}. This paper also explains how we can go from the ribbon-graphs/index-contractions to a Fourier basis using representation theory labels which can be  truncated to finite $N,d$ for the $U(N) \times O(d)$ theory and matches the finite $N$ counting for this theory.  

As explained above, ribbon  graphs are in 1-1 correspondence with orbits in $ S_{ 2n}$ generated by conjugation with permutations in the subgroup $S_n [ S_2]$. For $ \gamma \in S_{ n } [ S_2 ] , \sigma \in S_{ 2n }$, the action is : 
\bea 
\gamma : \sigma \rightarrow \gamma \sigma \gamma^{-1} 
\eea
There is an associative algebra associated with this equivalence, which is a sub-algebra of the group algebra 
$ \mC( S_{ 2n })$. A basis is labelled by the orbits. For each orbit, the basis element is the sum
of elements in the orbit, each with unit coefficient. These basis elements commute with $ S_{n }[S_2]$. This defines a permutation centraliser algebra \cite{PCA}, a structure which appears in the construction of orthogonal bases for multi-matrix and tensor invariants \cite{KR1,BHR1,BHR2,BCD2007,BDS2008,KR2,PCA,PCAKron}. 

A Fourier basis on the algebra \cite{KRS} is given by  
\bea\label{FouBasRib} 
Q^{ R , S , \tau } = \sum_{ \sigma \in S_{ 2n } }  \cC^{ R , R , S ; \tau  }_{I , J , K  }   D^R_{ IJ  } ( \sigma  ) 
\cB^{ S \rightarrow \mathbf { 1 }  }_{ K }  \sigma 
\eea 
$D^R_{IJ}(\sigma ) =  \langle e^R_I|  D^R( \sigma ) |e^R_J\rangle $ are the matrix elements of the permutation $ \sigma \in S_{ 2n }$ in an orthonormal basis  for the irreducible representation $R$ of $S_{ 2n}$. $\cC^{ R , R , S ; \tau  }_{I , J , K  } $ are the Clebsch-Gordan coefficients for the irrep $S$ in the tensor product $ R \otimes R$. $\tau $ is a multiplicity index which runs over a basis of the multiplicity space, which has dimension equal to the Kronecker coefficient $ C ( R , R , S )$. The label $S$ runs over Young diagrams with $2n$ boxes and even row lengths. 
These are the only irreps of $S_{2n}$ which admit an $S_{n}[S_2]$ invariant subspace. This subspace is one-dimensional and 
$\cB^{ S \rightarrow \mathbf { 1 }  }_{ K }$ is  the coefficient of $e^R_K$ of the unit-normalised invariant vector.  
The invariance under conjugation by $ \gamma \in S_n [ S_2 ] $ of the Fourier basis elements \eqref{FouBasRib}
follows by using the equivariance property of the Clebsch-Gordan coefficients and the invariance property of the branching coefficients. The number of these Fourier basis elements agrees with the $O(d)$ version of the finite $N,d$ formula \eqref{ZNdO}. 

It is shown \cite{KRS} that this  basis diagonalises the inner product at finite $N,d$ coming from matrix theory. 
Setting the constraints $ l( R ) \le N , l( S ) \le d $ gives the finite $N,d$ Hilbert space. These constraints can be implemented using central elements in the group algebra of $S_{ 2n} $ acting on the space of ribbon graphs, along the lines of \cite{PRS}.

\subsection{ Caloric fold  at large $d$ }
\label{sec:calfolLarged}

We now consider the partition function $ \cZ( N , d , k )$ in a regime where $ d \gg  N , k $,  while $k$ is allowed to vary from values smaller than a large finite  $ N$ to values much greater than $N$. It is useful to write $ \cZ( N , \infty , k )$ for the counting function in this regime.  

From the discussion above, $ \cZ( N , \infty , k =2n )$ counts ribbon graphs with $ n $ edges when $ 2n  \le N$. 
The large $k$ asymptotic formula for connected ribbon graphs, $\cZ^{{\rm conn}}(\infty,\infty,k=2n)$ is known  \cite{DrmotaNedela} to be given by 
\[
 \cZ^{\rm conn}( \infty , \infty , k =2n )
 \sim \sqrt{2}\,(2n)^n e^{-n }.
\]

The  corresponding disconnected count $  \cZ(\infty,\infty,k=2n) $ is  obtained by allowing
arbitrary finite collections of connected components. 
A disconnected object with total size \(n\) is assembled from connected
components whose sizes sum to \(n\).  Thus the possible component-size
decompositions include
\[
n,\qquad (n-1)+1,\qquad (n-2)+2,\qquad
(n-2)+1+1,\qquad (n/2)+(n/2),\ldots .
\]
The leading term comes from the single connected component of size \( n \),
namely \( \cZ^{ { \rm conn } }( \infty , \infty , 2n )   \).  This dominance follows from the super-exponential growth
\[
\cZ^{ { \rm conn } }( \infty , \infty , 2n )  \sim \sqrt{2}\left(\frac{2 n }{e}\right)^n 
\]
For example, the contribution from two components of comparable size,
schematically \(n/2+n/2\), is of order
\[
 ( \cZ^{ { \rm conn } }( \infty , \infty , n  )  )^2 
\sim
2\left(\frac{n}{e}\right)^n,
\]
which is  exponentially suppressed relative to the
single-component term.

The entropy at large $k$ is thus 
\bea 
S = \log (  \cZ^{ { \rm conn } }( \infty , \infty , 2n )  ) = n \log n - ( 1 - \log 2 ) n 
\eea
and 
\bea  
 { \partial^2  S \over \partial n^2  } \sim { 1 \over n } 
\eea
which is positive and gives negative heat capacity. 

For finite $N$,  $ d = \infty $, we get 
\bea\label{OrderdlargNlargk} 
\cZ( N , \infty , 2k ) && = \sum_{ \substack { R\vdash 2k \\ l(R ) \le N } }  \sum_{p \vdash 2k }  { ( \chi^R_p )^2 \Phi ( \infty , p ) \over \Sym ~ p } \cr 
&& =  \sum_{ \substack { R\vdash 2k \\ l(R ) \le N } }  \sum_{p \vdash 2k }   ( \chi^R_p )^2 \cZ ( S_k [ S_2 ] , p ) 
\eea 
where we used 
\bea 
\Phi( \infty , p ) = \cZ ( S_k [ S_2 ] , p )  \Sym ~ p 
\eea
derived earlier. Computations with $ N =4$ for example  show the caloric fold.  A more systematic study of $ k_{ \crit} ( N )$ in this large $d$ regime using  \eqref{OrderdlargNlargk} is left for the future. It would aim to tabulate the values for a range of $N$, find plausible functional fits and derivations of these analogous to the   discussion we have given for small $d$ in this paper.

The $ n \log n $ growth of the entropy means that the Hagedorn temperature is vanishing at large $N$, since $ \sum_n  n^n e^{ - \beta n }$ diverges for all finite $ \beta $. The super-exponential growth is similar to the permutation invariant matrix model and the complex 3-index tensor model studied in \cite{PIMQTNS}.  The thermodynamics and zero-temperature Hagedorn temperature of tensor model cases was also discussed in \cite{Klebanov,Tseytlin}.

\section{ Summary, Discussion  and Outlook }\label{secConc}

We start with a summary of the results of the paper accompanied by a discussion of  natural generalisations and related literature. This  includes a conjecture motivated by AdS/CFT and the results of the present paper.

\subsection{ Summary, discussion  and related literature  } 
In this paper we have studied the states of quantum mechanical
matrices $X_a$ with $a=1,\cdots,d$ invariant under both the adjoint
action of $U(N)$ and the vector action of $SO(d)$.  These state form
subsectors of the full Hilbert space of multi-matrix models where only
$U(N)$ and $SO(d)$ (or $O(d)$) singlets are permitted.  Our purpose here
was to study these subsystem from a microcanonical point of view. For
this we count the number of states at each level, $k$, of the Hilbert
space where $k$ counts the number of matrix creation operators acting
on the vacuum.  The generating function for such counting is the
thermal canonical ensemble partition function of an $ U(N) \times SO(d)$
(or $ U(N) \times O(d)$) gauged gaussian matrix model.
Gauging the orthogonal group imposes the
further constraint that only rotationally invariant
combinations of the matrices are permitted. In the $U(N) \times O(d)$
case we further demand invariance under reflection of any individual
matrix, thus excluding contractions with the $\epsilon$-tensor of $SO(d)$.

We found that, in the micro-canonical ensemble,  these systems naturally have negative heat capacity at
low energies which turns over in what we term a ``caloric fold'' to
positive heat capacity at high energies. At large $N$ they have a
Hagedorn transition at $T_H=\frac{1}{\ln d}$, i.e. at the same
temerature as the $U(N)$ gauged models. There is equivalence of
ensembles in the high energy regime but not for low energies. The
characteristic thermal behaviour of these systems is very reminiscent
of black holes in anti De Sitter spacetime and we conjecture that the
gauge invariant states that are in addition rotationally invariant
should be identified with black hole states in the dual gravity
picture.  At low energies the subsector of rotationally invariant
states has negative heat capacity and is the natural candidate for the
small black hole configurations which are unstable and decay. But in
the larger system, below the Hawking-Page/confining-deconfining
transition, they are in thermal equilibrium with the thermal radiation
and fluctuate in and out of existence as one would expect in a finite
system undergoing a first order phase transition. In the terminology of
\cite{Nagle1973,BhattacharjeeNagleHuseFisher} the transition is
`a $3/2$-order transition' i.e. it appears first order as seen from the
low temperature side as the internal energy in the canonical ensemble
jumps from zero to $\frac{1}{4}$ but on cooling from the high
temperature side it appears to be 2nd order (or continuous) with
a divergent heat capacity see Eqn (\ref{Energyofx}).

This work is the starting point for a larger  research programme. A 
natural next step is  to consider fermionic matrices. Here there are
options for how the fermions transform under the orthogonal
group. They might transform as either a vector or as a spinor, both are natural for matrix models, e.g. fermions transform under the spinor representation of $SO(9)$ for the BFSS model. When transforming
under the vector representation the analysis would be very similar to
that performed here and one can anticipate based on the large $N$
connection with the $a_1$ model that the features would be largely
unchanged. However it is less obvious what would happen to fermions in
the spinor representation.

We believe that the broad features of the caloric fold uncovered in this paper are shared by a 
wide class of  multi-matrix model systems and defines a universality class including black holes in AdS space. Adding appropriate small perturbations to the
systems studied here will result in small changes to how the energy
depends on the level $k$ and one should be able to invert this to get
$k(E)$ and then use the counting formulae as the counting of
invariants is model independent. Thus as long as $\frac{dk}{dE}$ remains
monotonic the general shape of the caloric curve is expected to remain unchanged. Furthermore since we find $k_{\crit} \sim \frac{N^2}{4}$, which corresponds to a transition in the number of invariants, one would expect that
$k (E_{ \crit} ) \sim \frac{N^2}{4}$ should more generally specify the transition energy at large $N$. 

The turn of the micro-canonical curve from negative to positive heat capacity
at energies $k_{\crit}\sim{N^2\over4}$ can be attributed  to finite $N$ trace relations being sufficiently significant at the thermodynamic level. 
Trace relations and the role of the $N^2$ scale has also been discussed in connection with the Hironaka forms of the rational generating functions of finite $N$ counting of $ U(N)$ invariants in matrix models
\cite{deMelloKoch:2025ngs}, and the gauging of orthogonal symmetry groups have also been considered in this context  \cite{Gaikwad:2025ugk}. Implications of the counting of invariants using the Molien-Weyl integral representation have also been used recently in the context of BFSS and BMN models in \cite{Badis1,Badis2}. The interplay of the invariant theory of orthogonal groups, which has played an important role in this paper, with matrix or tensor  theories and gauge-string duality  has been a fruitful avenue over the years, 
see e.g. \cite{NRS93,SRSO93,AABF02,CDD13,BNT21,ATV23,GK22,KKMT23,ABT24}, and promises to hold future lessons and surprises.

One sees in the Tables of Appendix  \ref{Tables} that in addition to $k_{\crit}$ each table for $\Delta\cZ(N,d,k)$ and indeed $\Delta Z(N,d,k)$ has a maximum. For large $N$ this corresponds to
the second derivative of the micro-canonical partition function having a zero. Our preliminary evidence suggests that this $k_{\max}$ also scales as $\frac{N^2}{4}$ and we
conjecture that this is the signal for the end of the Hagedorn transition when approached from the low temperature side.

Much effort has gone into Hybrid-Monte Carlo studies of matrix models
see
e.g. \cite{Catterall:2022wjq,Filev:2015hia,Berkowitz:2016jlq,Pateloudis:2022oos,Asano:2018nol}. Such
studies are based on accessing the canonical distribution and have proved useful for comparison with gravitational dual predictions.  However, access to
the microcanonical data is very difficult in such numerical studies.
Our work is complementary in that we access the microcanonical distribution directly,
although in a simplified setting. 

It would  be interesting to study the large-$k$  asymptotics of the
stable-range $SO(d)$-invariant counting functions $\cZ(\infty,d,k)$ defined in \eqref{largeNZNdk}  as $d$ is
varied from small  to large values.  The closely related stable-range problem of counting $U(N)$-invariant
operators in $d$-matrix models, without imposing $SO(d)$ invariance, was studied
in \cite{YangLeiRamgoolam}, where a qualitative change in the convergence
properties of the large-$k$ asymptotic expansion occurs at $d=13$.  The formulae
developed here introduce the additional projection to $SO(d)$ singlets, making
it natural to ask whether analogous $d$-dependent changes occur in this
invariant sector.
 
As shown in \cite{PIMQTNS} the caloric fold feature, and the detailed comparison of canonical and micro-canonical ensemble in the region of the cross-over for large finite $N$  (which becomes the Hagedorn transition at $ N \rightarrow \infty $)
is possible in systems with the $U(N)$ gauge symmetry is replaced by a discrete symmetry. 
For matrix or tensor systems with continuous symmetry,
representation-theoretic summation formulae are known for the
microcanonical degeneracies, but explicit closed forms for the
canonical partition functions for adequately large $N$ are not known. There are available formulae in terms of group integrals,  but these become unwieldy as $N$ increases. An interesting challenge is to find new ways to calculate the  characteristics of the comparison between micro-canonical and canonical ensembles in the transition region for large finite $N$ in the case of continuous symmetries.

In studying diverse thermodynamic aspects of the micro-canonical degeneracy functions $ \cZ ( N , d , k )$ and $\cZ_O ( N , d, k ) $, representation theoretic structures such as Young diagrams and combinatorial structures of stringy geometrical nature, notably ribbon graphs, have played an important role.  The threshold of $N^2/4$, supported here by a large-N eigenvalue-density analysis, has also been discussed  in the context of multi-matrix Hagedorn transitions  from the point of view of Young diagram distributions \cite{VershikKerov,LoganShepp,Berenstein:2023srv,O'connor}.
Young diagrams play an important role in mapping the combinatorics of quantum states in gauge theory to dual giant-graviton and bulk gravitational states in highly supersymmetric sectors \cite{MST,CJR,LLM}. The present work is giving strong indications that representation theory structures and the appropriate invariant theory problems arising in gauge-string duality also provide a promising approach to a microscopic understanding of black hole thermodynamics.

\subsection{ A strong coupling AdS/CFT conjecture for the minimum black hole temperature in AdS   } 

Our results establishing the caloric fold as a generic feature of $SO(d)$ and $O(d)$  invariant subsectors in matrix models suggests an interesting conjecture.   Following the Hawking-Page discussion of black holes in AdS, we have a transition temperature $T_1$ (in their notation) between large black hole and thermal AdS and a minimum temperature $T_0$, below which the black hole does not exist. The $AdS $ spacetime, as part of $AdS_5 \times S^5$  is dual to large $N$ maximally supersymmetric Yang-Mills theory with $U(N)$ gauge group. The dual gauge theory should have some realisation of this minimum temperature. The point of view suggested by the results of  this paper is that an appropriate $SO(4)$ invariant sector of the large $N$ SYM at strong coupling  captures the microscopic degrees of freedom of the black hole in $AdS_5$ in the region near the turning point of the black hole caloric fold. As a result we would expect that the counting of states in the micro-canonical ensemble for the $SO(4) \subset SO(4,2) $ invariant sector of the  strongly coupled (large 't Hooft coupling, $ g_{ YM}^2 N $) SYM  gauge theory will yield a caloric fold of the kind we are finding here for matrix quantum mechanics. In the strongly coupled SYM set-up, we would  then expect that  the  minimum temperature remains distinct from the Hagedorn temperature at large $N$. The Hagedorn temperature is expected, as usual, to be  calculable in principle  from the canonical partition function for the full  SYM theory on $ S^3 \times S^1$. This discussion offers a conjectural, purely SYM gauge theory large $N$ dual  formulation of the non-zero temperature difference 	$(T_1 - T_0 ) $ of the Hawking-Page discussion of black holes in AdS.

\vskip2cm 

\centerline{\bf{Acknowledgments}}
\vskip.2cm
We thank Joseph Ben Geloun, Robert de Mello Koch, Brian Dolan, Yang Lei, Adrian Padellaro, Ryo Suzuki   for helpful conversations.
S.R. is supported by the
Science and Technology Facilities Council (STFC) Consolidated Grant ST/X00063X/1
“Amplitudes, strings and duality,”  a Royal Society International Exchanges grant IEC\textbackslash NSFC\textbackslash 242376 and a Visiting Professorship (Summer 2026) at the Dublin Institute for Advanced Studies. 

\vskip2cm 

% \vfill\eject

\begin{appendix}

\section{$U(d)/SO(d)$ harmonic analysis  and low-dimensional isomorphisms    }
\label{sec:Low-dim-isom}

We have established the general framework  for reasonably efficient computation of the partition functions $ \cZ ( N , d, k )$ at any $ d $. This was achieved with the help of harmonic analysis on $U(d)/SO(d)$. Specialising to $d=3,4$ connects 
this harmonic analysis to special classes of Kronecker coefficients and Plethysm coefficients, which admit independent computational formulations.  We collect these observations here. 

\subsection{ The case  $d=4$ and a special class of 2-row Kronecker coefficients from $U(4)/SO(4)$ harmonic analysis  }

Here we are considering the $ SO(4) \times U(N)$ invariants in 
$ \Sym^k ( V_N \otimes \bar V_N \otimes V_4)$ where $ V_4$ is the four-dimensional vector  representation of $SO(4)$. As in the general $d$ case, we consider 
\bea 
\Invt_{ U(N) \times S_k } ( V_N^{ \otimes k } \otimes \bar V_N^{ \otimes k } \otimes V_4^{ \otimes k }   )
\eea

We can use the isomorphism $ SO(4 ) = SU(2) \times SU(2)$. The vector is isomorphic to $ V_{ 2 } \otimes V_{ 2 } $, the tensor product of two-dimensional irreps of  $ SU(2) \times SU(2) $. 
\bea 
V_4^{ \otimes k } = ( V_2 \otimes V_2)^{ \otimes k  } = V_2^{ \otimes k }  \otimes V_2^{ \otimes k }
\eea
and 
\bea 
\Invt_{ SO(4)  } ( V_4^{ \otimes k })  = \Invt_{ SU(2 ) } ( V_2^{ \otimes k }   ) \otimes \Invt_{ SU(2 ) } ( V_2^{ \otimes k }   ) 
\eea
The two-dimensional representation $V_2$ of $SU(2)$ lifts to the two-dimensional representation of $U(2)$, which allows easy application of Schur-Weyl duality 
\bea 
V_2^{ \otimes k }  = \bigoplus_{ \substack { T\vdash k  \\ l(R) \le 2 } } V_T^{ U(2) }  \otimes V_T^{ S_k }
\eea
For $ k $ even and $ T = [ { k \over 2 } ,  { k \over 2 } ]  \equiv T^* $, $V^{U(2)}_{ T^*}$ is the invariant under the action of the $SU(2)$ subgroup. All other $T \ne T^*$ are not $SU(2)$ invariant. Thus, we conclude that 
\bea 
\Invt_{  SO(4)  } ( V_4^{ \otimes k }  )  =  V_{T^*}^{ (S_k) } \otimes V_{T^*  }^{ (S_k ) } = \bigoplus_{ S \vdash k }  V_S \otimes V_{T^* T^* S  }^{ (S_k)} 
\eea 
where $ V_{T^* T^* S  }^{ (S_k)}$ is the $S_k$ invariant subspace of $ V_{ T^*}^{ (S_k) } \otimes V_{T^*}^{ (S_k)}  \otimes V_S^{(S_k)}  $. The dimension of this invariant space is given by the Kronecker coefficient $C ( T^* , T^* , S ) $ for this triple 
\bea 
\Dim ( V_{T^* T^* S  }^{ (S_k)} ) 
= C ( T^*  , T^*  , S  )
\eea
It follows that 
\bea\label{DimInvtSO4} 
\Dim ( \Invt_{  SO(4)  } ( V_4^{ \otimes k }  )  )  =  
\sum_{ S \vdash k } 
C ( T^*  , T^*  , S  )
\eea

By following the arguments which lead to \eqref{InvtSubspUSkSO} we find
\bea 
\Invt_{ U(N) \times SO(4) \times S_k } (V_N^{ \otimes k }  \otimes \bar V_N^{ \otimes k } \otimes V_4^{ \otimes k }  )  = 
\bigoplus_{ \substack {  R\vdash k \\ l(R) \le N }}  V_{ R R T^* T^* }^{ (S_k) }
\eea
where 
\bea 
V_{ R R T^* T^* }^{ (S_k) } 
\eea
is the $S_k$ invariant subspace of $ V_R \otimes V_R \otimes V_{ T^* } \otimes V_{T^* }$. By using the character expansion formula for this invariant, it follows that 
\bea 
\Dim ( V_{ R R T^* T^* }^{ (S_k)} ) =  
\sum_{ S \vdash k } C ( R , R , S )  C ( T^* , T^* , S  )
\eea

Thus the partition function of $SO(4) \times U(N) $ invariants is now 
\bea 
\cZ ( N , k ) = \sum_{ \substack{ R \vdash k \\ l(R) \le N } }
\sum_{ S \vdash  k } C ( R , R ,S ) C (  T_*  , T_*  , S)
\eea

Comparing the equation 
\eqref{SOAkdresult} above with 
\eqref{SOAkdresult} specialised to $ d=4$ 
we deduce that 
\bea 
C ( T^* , T^* , S )  &=&  1  ~\hbox { if }  S \in \cA( k , 4 ) \cr 
& =&  0  ~  \hbox{if  }   S \notin \cA( k , 4 )
\eea 
The set of Young diagrams $ \cA( k , 4)
$ is given by specialising equations  \eqref{defAkd1},\eqref{defAkd2} to $ d=4$.  General formulae for Kronecker coefficients of 2-row Young diagrams are given in \cite{Rosas} and this result should be a special case of the formulae there. We have also checked the equation by computations in SAGE.

\subsection{ The case $ d=3$ and  $SU(2)$  Plethysm coefficients from $U(3)/SO(3)$      }  

In the case $d=3$, we can take advantage of the isomorphism of Lie algebras between  $SO(3)$ 
and  $SU(2)$. The 3-dimensional vector representation of $SO(3)$ is  the spin 1 irrep of $SU(2)$. 
We are taking the $k$-fold tensor power of the spin one irrep, and we want to resolve the $SU(2)$ invariant subspace in terms of irreps of $S_k$. 

There are some known formulae for $SU(2)$ plethysms which are relevant \cite{King1985}. 
For the $(m+1)$ dimensional irrep of $SU(2)$, and any $ \Lambda $ (specified as an irrep of dimension $(l+1)$), we have 
\bea 
\Dim ( V_{ \Lambda (l )   , T }   ) 
= \hbox { Constant term in } \left [ 
q^{ n ( T  ) - { m \over 2 } k + { l \over 2 } } ( 1 - q ) \prod_{ (i,j) \in T  } 
{ ( 1 - q^{ m +1 - i + j } )    \over ( 1 - q^{ h ( i , j )   }  ) }  \right ]
\eea 
where  $i$ labels the rows of the Young diagram $ T $, $ j $ labels the columns, and $(i,j) $ specifies a box, 
\bea 
n ( T ) && = \sum_{ ( i,j ) \in T }  ( i -1 )  \cr 
%c ( T)  && = \sum_{ ( i, j ) \in T} c ( i,j ) \equiv \sum_{ ( i, j ) \in T } 
%( j - i ) \cr 
h ( i , j ) && = \hbox{ hook length of the box at } ( i, j ) 
\eea
We specialise to $ \Lambda = \Lambda_0$, i.e. $ l = 0$, and $m=2$ to get 
\bea\label{dim3}  
\Dim ( V_{ \Lambda_0  , T } )  
= \hbox { Constant term in } \left [ 
q^{ n ( T  ) -  k } ( 1 - q ) \prod_{ (i,j) \in T  } 
{ ( 1 - q^{ 3 - i + j } )    \over ( 1 - q^{ h ( i , j )   }  ) }  \right ]
\eea 
$T$ is a Young diagram with $k$ boxes and no more than $3$ rows. 
By describing $T$ as a sequence of column  lengths $ [ c_1  ,c_2 , ... ] $  we can also write
\bea 
n(T) = \sum_{ j } \sum_{ i =1 }^{ c_j  } ( i  -1 ) = \sum_{ j } { c_j ( c_j -1 ) \over 2 } 
\eea

We have verified that the formulae for the partition function $ \cZ( N  , d=3, k ) $  obtained by the general method based on harmonic analysis on $ U(d) /SO(d)$, notably \eqref{SOAkdresult} and \eqref{KronHelg1} agree with the formulae obtained by using these $SU(2)$-plethysm formulae. 

\subsection{Low-dimensional group  isomorphisms and counting equivalences.  } 

The general $SO(d)$ discussion, when specialised to $SO(3)$, was compared above to $SU(2)$. The precise isomorphism is  $ SU(2)/Z_2 = SO(3)$, but the $Z_2$ acts trivially in the $3D$ irrep. Therefore the  counting  $ SO(3)$-invariants  in $V_3^{ \otimes k }$ is the same as the counting of $ SU(2)$ invariants. This ensures that the $SU(2)$ plethysm calculations in this appendix give the same result as the $SO(d)$ formulae specialised to $d=2$. 

Likewise for $ SO(4)$ the precise isomorphism is $ SO(4) = SU(2) \times SU(2) / Z_2 $. Again  the $Z_2 $ acts trivially on $ V_4$. This allows the matching of the $SO(d)$ calculations with $SU(2) \times SU(2) $ calculations described in this appendix. 

\section{SageMath code for the character sum pairing formula for $SO(d)\times U(N)$ matrix invariants}
\label{sec:SageMathCode}

The  key formulae  \eqref{PairingSO} and \eqref{PairingO}   derived in section \ref{sec:ExactCountingAndAlgorithms} for the functions $ \cZ ( N , d, k ) $ and $ \cZ_{O} ( N , d, k )$ can be implemented in SageMath.
The source code is provided with the arXiv submission as an ancillary notebook,
{\texttt { Orthog-UNCharacterSumPairing.ipynb}}. 

The main routine is
\begin{verbatim}
	Z_mu_sum(N, d, k)
\end{verbatim}
which computes the quantity $\cZ(N,d,k)$ in equation \eqref{PairingSO} . This  counts the dimension of the space of $ U(N) \times SO(d)$ invariant polynomials of degree $k$ in the matrix variables $X^i_{ a , j }$. 

The routine evaluates the formula
\bea
\cZ(N,d,k)
= \sum_{p  \vdash k}
{ \Psi ( N , k ;  p )  \Phi(d , k ;  p )  \over \Sym ~ p }
\eea
where
\bea
\Sym p &=& \prod_i i^{p_i } p_i ! , \cr
K_k(d) &=& \sum_{S \in \cA(k,d)} \cS_S , \cr
\Phi(d , k ; p ) &=& \sum_{S\in \cA(k,d)} \chi^S_p , \cr
\Psi ( N , k;  p ) &=& \sum_{\substack{R\vdash k \\ \ell(R)\le N}} (\chi^R_p)^2 .
\eea

For each fixed \(k\), we work in a finite-dimensional vector space whose natural bases are indexed by partitions of \(k\). 
The symbols \(\mathcal S_S\), with \(S\vdash k\), denote the basis elements labelled by Young diagrams \(S\), while the power-sum basis is labelled by conjugacy classes \(p\vdash k\). 
Thus \(K_k(d)\) is the vector obtained by summing the \(\mathcal S_S\)-basis elements selected by the set \(\cA(k,d)\). 
Expanding this same vector in the power-sum basis gives the coefficients \(\Phi(d,k;p)\), equivalently the character sums appearing in (B.2).

The implementation proceeds as follows:
\begin{enumerate}
	\item Construct the allowed set $\cA(k,d)$.
	\item  Form the vector \(K_k(d)=\sum_{S\in A(k,d)}\mathcal S_S\) in the degree-\(k\) partition-labelled vector space.
	\item Extract $\Phi(d , k ; p )$ from the power--sum expansion of $K_k(d)$.
	\item  Change basis to the power-sum basis and extract \(\Phi(d,k;p)\).
	\item Evaluate the sum over partitions $p \vdash k$.
\end{enumerate}

Some sample evaluations are
\begin{verbatim}
	Z_mu_sum(8,3,10)   = 946
	Z_mu_sum(10,4,16)  = 2487395
	Z_mu_sum(10,4,18)  = 29044367
	Z_mu_sum(10,4,20)  = 349586794
	Z_mu_sum(10,4,30)  = 111959655014092
\end{verbatim}

The ancillary notebook includes comments describing the implementation of
$\cA(k,d)$ and the symmetric-function manipulations used in the
evaluation of $\Phi(d , k ; p )  $ and $\Psi ( N , k ;  p )$.

\vspace{0.5cm}

For the case of $O(d)$ invariants, the implementation is modified by restricting
the set $\cA(k,d)$ to its even sector $\cA_+(k,d)$, i.e. Young diagrams with
all row lengths even and at most $d$ rows. Equivalently, one sets
\bea
\Dim\big(V^{U(d)\to O(d)}_{\Lambda_0,T}\big)
=
\begin{cases}
	1 & T \in \cA_+(k,d) \\
	0 & \text{otherwise.}
\end{cases}
\eea

The corresponding routine is
\begin{verbatim}
	Z_mu_sum_O(N, d, k)
\end{verbatim}
which computes the $O(d)$-invariant counting function $ \cZ_O ( N , d, k )$. For odd $k$, the result
vanishes identically.

\section{Review of Black Hole formulae } \label{ReviewOfBlackHoleThermodynamics}

Here we recall the black hole formulae in $d$-dimensions, first in asymptotically flat spacetime and then we consider Schwarzschild-AdS.

\subsection{Schwarzschild Black Holes in $D$ Dimensions and Negative Specific Heat}

In $D$ spacetime dimensions, the Schwarzschild--Tangherlini metric is
\begin{equation}
   ds^2=-\left(1-\left(\frac{r_h}{r}\right)^{D-3}\right)dt^2
    + \frac{dr^2}{\left(1-\left(\frac{r_h}{r}\right)^{D-3}\right)}
    + r^2 d\Omega_{D-2}^2 \, .
\end{equation}
describes a black hole in asymptotically flat spacetime. Such black holes only exist for dimensions $D>3$.

The ADM mass, $M$ related to horizon radius $r_h$ by
\begin{equation}
  M=\frac{(D-2)\,\Omega_{D-2}}{16\pi G_D}\, r_h^{D-3} \, ,
  \label{rhMass}
\end{equation}
where $\Omega_{D-2}$, the hypersurface area of the unit $(D-2)$-sphere, is
\begin{equation}
    \Omega_{D-2}=\frac{2\pi^{(D-1)/2}}{\Gamma\!\left(\frac{D-1}{2}\right)} \, .
\end{equation}
The Bekenstein--Hawking entropy is
\begin{equation}
S=\frac{A}{4G_D}
=\frac{\Omega_{D-2}\, r_h^{D-2}}{4G_D}\quad\hbox{and}\quad M=\frac{D-2}{4\pi r_h}\frac{A}{4G_D} \, ,
\end{equation}
These imply the scaling relations
\begin{equation}
  M \propto r_h^{D-3}, \qquad S \propto r_h^{D-2}\propto M^{\frac{D-3}{D-2}}, 
	\qquad \implies M \frac{dS}{dM}=\frac{D-2}{D-3}S\, .
%	T \propto r_h^{-1} \, .
\end{equation}
The ADM mass is the natural energy for this system and the microcanonical temperature obtained from the entropy eqn (\ref{TmicrofromEntropy}) gives the microcanonical temperature as the Hawking temperature
\begin{equation}
T=\frac{D-3}{4\pi r_h}
\end{equation}
Using $M\propto T^{-(D-3)}$ so that $T\frac{dM}{dT}=-(D-3)M$ we then see that 
the specific heat is defined by
\begin{equation}
	\boxed{C = \frac{dM}{dT} =-(D-3)\frac{M}{T}} \, .
\end{equation}
Thus Schwarzschild type black holes in asymptotically flat spacetime have negative specific heat in any spacetime dimension $D \ge 4$, reflecting the fact that they become hotter as they lose energy.

\subsection{Micro-canonical $E-T$  curve for Schwarzschild--AdS black holes in $D$ dimensions}

We review properties of black holes in AdS space following  \cite{HawkPage,WittenThermal,Birmingham}. 

For a Schwarzschild black hole in asymptotically AdS$_D$ spacetime with
spherical horizon, the metric can be written
\begin{equation}
	ds^2 = -f(r)\,dt^2 + f(r)^{-1}dr^2 + r^2 d\Omega_{D-2}^2 ,
\end{equation}
with
\begin{equation}
	f(r)=1+\frac{r^2}{L^2}-{\left(\frac{r_h}{r}\right)}^{D-3} .
\end{equation}
with $L^2=-\frac{(D-1)(D-2)}{2\Lambda}$ and $\Lambda$ the cosmological constant. The black hole mass $M$ is again parameterized by $r_h$ in (\ref{rhMass}), with the horizon radius, $r_+$, satisfying $f(r_+)=0$.

\begin{wrapfigure}{r}{0.33\textwidth}
  \centering
   \includegraphics[width=0.35\textwidth]{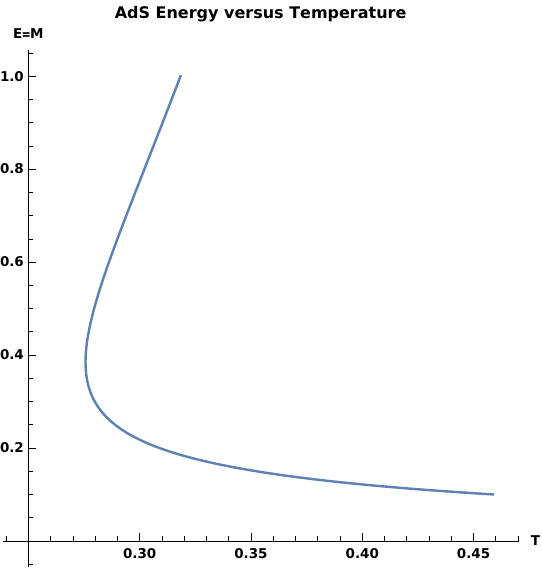} 
   \label{fig:EnergyVsTemperatureAdSFig}
\end{wrapfigure}

The Bekenstein-Hawking entropy is again given by the area law as
\begin{equation}
  S=\frac{A}{4\pi G_D}
  =\frac{\Omega_{D-2}\, r_+^{D-2}}{4G_D}
\end{equation}
and the mass $M$ is related by
\begin{equation}
  \quad\hbox{and}\quad M=\frac{D-2}{16\pi G_D}\Omega_{D-2}r_+^{D-3}(1+\frac{r_+^2}{L^2}) \, ,
\end{equation}
showing the energy $M$ is a monotonic increasing function of $r_{+}$ and $r_+$ parameterizes the entropy energy relation so the thermodynamic quantities can be expressed in terms of $r_+$. Again the obtaining the microcanonical temperature 
from the entropy via (\ref{TmicrofromEntropy}) gives the Hawking temperature as
\begin{equation}
T(r_+)=\frac{1}{4\pi r_+}\left[(D-3)+(D-1)\frac{r_+^2}{L^2}\right] .
\end{equation}
and coincides with the Hawking temperature as specified by the surface gravity
$T(r_+)=\frac{f'(r_+)}{4\pi}$.
The microcanonical energy temperature relation is then given parametrically as
$(T(r_+),E(r_+))$  which gives the caloric curve $E(T)$.
It is natural to eliminate $G_D$ by going to Planck units where $G_D=\frac{1}{8\pi m_p^{D-2}}=\frac{l_p}{8\pi}$ with $m_p$ the Planck mass and $l_p$ the Planck length. Furthermore ratios of AdS scale, $L$, arise naturally. We therefore define dimensionless scaled variables 
\begin{equation}
\bar{r}_+=\frac{r_+}{L}\, ,\qquad  \bar{E}=\frac{8\pi G_D M}{L^{D-3}}=\frac{m_p^{D-2} M}{L^{D-3}}=\frac{M/m_p}{{(L/l_p)}^{D-3}}
\end{equation}
and in these variables we have the caloric curve parameterically as
\begin{equation}
  (\bar{T},\bar{E})=\left(\frac{D-3}{4\pi \bar{r}_+}+\frac{(D-1)\bar{r}_+}{4\pi},\frac{D-2}{2}\Omega_{D-2}\bar{r}_+^{D-3}(1+\bar{r}_+^2)\right)
  \end{equation}
We plot the special case of $D=4$ in Figure \ref{fig:EnergyVsTemperatureAdSFig}

\noindent
where we see the characteristic caloric fold from a low energy negative specific heat capacity, corresponding to small black holes, which are insensitive to the AdS scale transitioning to a high temperature positive specific heat characteristic of large mass AdS black holes. The Hawking Page transition then
occurs at

\paragraph{Turning point:} Differentiating the temperature gives
\begin{equation}
  \frac{dT}{dr_+}= -\frac{D-3}{4\pi r_+^2} + \frac{D-1}{4\pi L^2}.
\end{equation}

The minimum temperature occurs at
\begin{equation}
\bar{r}_* = \sqrt{\frac{D-3}{D-1}} .
\end{equation}

The corresponding minimum temperature is
\begin{equation}
  \bar{T}_{\min}=\frac{\sqrt{(D-1)(D-3)}}{2\pi}.
\end{equation}
The energy at the turning point is
\begin{equation}
  \bar{E}_{\rm crit}=\frac{(D-2)\Omega_{D-2}}{2}\,
  \bar{r}_*^{D-3} \left(1+\bar{r}_*^2\right)=\frac{(D-2)^2}{D-1}{(\frac{D-3}{D-1})}^{(D-3)/2}.
\end{equation}

The Hawking Page transition occurs at a slightly higher temperature.
This is found by considering the free energy
\begin{equation}
  F=E-TS\quad \hbox{which gives}\quad \bar{F}=\frac{\Omega_{D-2}}{2}\bar{r}_+^{D-3}(1-\bar{r}_+^2)
  \end{equation}
and the free energy changes sign at $\bar{r}_+=1$ corresponding to
\begin{equation}
  \bar{T}_{HP}=\frac{D-2}{2\pi} \quad \bar{E}_{HP}=\frac{(D-2)\Omega_{D-2}}{2}\, .
\end{equation}
{\bf Note:} The Einstein Hilbert action $I=-\beta F$.
Hawking and Page refer to $T_{min}=T_0$ and $T_{HP}=T_1$ and the free energy here is proportional to that given by the Euclidean Einstein Hilbert action. 

\paragraph{Low Energy and High Energy regimes:}
For $T>T_{\min}$ the equation $T(r_+)=T$ has two solutions.

\begin{itemize}
	
\item Small black hole ($r_+<r_*$)
\begin{equation}
T \sim \frac{D-3}{4\pi r_+}, \qquad E \sim r_+^{D-3}.
\end{equation}
Hence
\begin{equation}
T \sim E^{-1/(D-3)} ,
\end{equation}
which corresponds to asymptotically flat Schwarzschild--Tangherlini and negative heat capacity case.
\item Large black hole ($r_+>r_*$)
\begin{equation}
T \sim \frac{D-1}{4\pi L^2} r_+, \qquad E \sim r_+^{D-1}.
\end{equation}
Hence
\begin{equation}
T \sim E^{1/(D-1)},
\end{equation}
corresponding to positive heat capacity.
\end{itemize}
These two branches meet smoothly at $(E_{\rm crit},T_{\min})$,
producing the characteristic AdS caloric fold.

\section{Large $N$ Generating Functions and Asymptotics}
\label{LargeNGeneratingFunctions}
We give here the explicit generating functions at small $d$, for the $SO(d)$ invariants in $ V_d^{ \otimes k } $. As explained in section \ref{sec:LowEnergy} , these determine the leading large $k$ asymptotics of $ U(N) \times SO(d)$ invariants of $d$ matrices for $N$ sufficiently large compared to $k$. We  include derivations of the large $k$ asymptotics of these special small $d$ cases. The techniques here complement the more general small $d$, large $k$ asymptotics in section \ref{sec:largekGend}.

\subsection{Large $N$ $SO(2)$}
\begin{eqnarray}
  \hat{Z}_2(x)&=&\int_0^{2\pi} \frac{d\theta}{2\pi}\frac{1}{1-2x\cos\theta}=\int_0^{\pi/2}\frac{d\theta}{2\pi}\frac{4}{1-4x^2+4x^2\sin^2\theta}\, ,\label{Z2SO3}
\end{eqnarray}
and shows that $\hat{Z}_2(x)$ is a function of $x^2$.
The integral in (\ref{Z2SO3}) is easily evaluated giving
\[
\boxed{\hat{Z}_2(x)=\frac{1}{\sqrt{1-4x^2}}}
\]
Expanding (\ref{Z2SO3})in $x$ gives
\[
\hat{Z}_2(x)=\sum_{n=0}^\infty\frac{\Gamma(n + \frac{1}{2})}{\sqrt{\pi} n!} (2x)^{2n}
\]
and noting that for large $n$ 
\[
\frac{\Gamma(n + \frac{1}{2})}{\sqrt{\pi} n!}=\frac{(2n)!}{2^{2n}(n!)^2}\simeq \frac{1}{\sqrt{\pi n}}
\]
we see that the exact large $N$ coefficients for $SO(2)$ are 
\[
\boxed{c_{2n}(2)=\frac{(2n)!}{(n!)^2}\sim\frac{1}{\sqrt{\pi}}\frac{2^{2n}}{\sqrt{n}}}\qquad\hbox{ for large n} 
  \]

The second integral form in (\ref{Z2SO3}) shows that there
are two equal singular contributions to the integrant, one from $x=1/2$
the second from $x=-1/2$ and that for a singularity analysis the function
is naturally treated as a function of $x^2$ rather
than $x$ or equivalently that singularities at both $x=1/2$ and $x=-1/2$
contribute to the asymptotic analysis. 

In (\ref{Z2SO3}) the singularity arises at $x=\pm\frac{1}{2}$, i.e. $x^2=\frac{1}{4}$ and $\theta=0$. Then expanding around the singular region with $1-4x^2=\epsilon$ the singular contribution to the integral is
\[
\hat{Z}_2^{sing}(x)=\int_0^a\frac{d\theta}{2\pi}\frac{4}{\epsilon+\theta^2}\sim \frac{1}{\sqrt{\epsilon}}=\frac{1}{\sqrt{1-4x^2}}
\]
and a singularity analysis reproduces the exact result.
\subsection{Large $N$ SO(3)}
For the large $N$ limit of $SO(3)$ we have
\begin{eqnarray}
  \hat{Z}_{3}(x)&=&\oint\frac{d\zeta}{2\pi i\zeta}\frac{1}{2!}\frac{(1-\zeta)(1-\zeta^{-1})}{1-x(1+\zeta+\zeta^{-1})}=\int_0^{\pi}\frac{d\theta}{2\pi}\frac{2(1-\cos\theta)}{1-x(1+2\cos\theta)}\label{Zhat3Integral}
\end{eqnarray}
The integrand has a simple pole at
$\zeta^\pm=\frac{1-x-\sqrt{(1+x)(1-3x)}}{2x}$ and $\zeta=0$. Performing the integral yields 
\[
\boxed{\hat{Z}_{3}(x)=\frac{1}{2x}\left(1-\sqrt{\frac{1-3x}{1+x}}\right)}
\]
where $\frac{1}{2x}$ comes from the pole at $\zeta=0$ while the second
contribution comes from $\zeta^-$; the pole $\zeta^+$ is outside the unit
circle and hence does not contribute.

Noting that
\[
-\sqrt{1-3x}=\sum_{n=0}^\infty\frac{(2k)!}{(2k-1)2^{2k}(k!)^2}(3x)^k
\]
is a convergent expansion for $x\le \frac{1}{3}$ in this regime one can write 
\[
\hat{Z}_3=\frac{1}{2x}+\sum_{k=0}^\infty\frac{(2k)!}{(2k-1)2^{2k}(k!)^2}\frac{(3x)^k}{2x\sqrt{1+x}}
\]
and one sees that for large $k$ the leading large $k$ asymptotics
can be obtained from noting the singularity in $x$ at $x=\frac{1}{3}$. In general the nearest singularity to the origin governs the asymptotics of the
large order coefficients and one can obtain the coefficient by evaluating
at the remaining $x$ dependence at the singularity.  Hence we see that
\begin{equation}
\boxed{c_k(3)\simeq\left. \frac{(2k)!}{(2k-1)2^{2k}(k!)^2}\frac{3^k}{2x\sqrt{1+x}}\right\vert_{x=\frac{1}{3}}\simeq \frac{3\sqrt{3}}{8\sqrt{\pi}}\frac{3^k}{k^{3/2}}}\label{Z3singAndCoeffs}
\end{equation}
Equivalently the asymptotics is governed by the non-analyticity at $x=1/3$
and with $\epsilon=1-3x$  the asymptotic coefficients are obtained by
expanding the singular contribution 
\[\hat{Z}_{3}^{sing}=-\frac{3\sqrt{3}}{4}\sqrt{\epsilon}=-\frac{3\sqrt{3}}{4}\sqrt{1-3x}
\]

One can equally extract this information from the integral form (\ref{Zhat3Integral}) by noting that in the vicinity of $\epsilon=0$ and $\theta=0$ the singularity of the integrand arises from the denominator approaching zero as $w=\epsilon+\frac{\theta^2}{3}$ when $\epsilon$ and $\theta$ go to zero. This further shows that $\theta$ scales as $\sqrt{\epsilon}$.  Expanding in $\theta$ and $\epsilon$ while retaining $w$ fixed and only keeping terms that do not go to zero with $\epsilon$ scaling yields
\[
\hat{Z}_{3}\sim \int_0^a \frac{d\theta}{2\pi}\frac{1}{2}\frac{\theta^2}{\epsilon+\frac{\theta^2}{3}}+\cdots\simeq-\frac{3\sqrt{3}}{4}\sqrt{\epsilon}+\cdots
\]
for the $\epsilon$ dependent part.
The large order behviour of the coefficients therefore is obtained from the Stirling approximation to the coefficients recovering (\ref{Z3singAndCoeffs}).
\subsection{Large $N$ $SO(4)$}
For $SO(4)$, equation (\ref{Zhat2m}) can be re-written most conveniently for the analysis here in terms of $\theta$ and $\zeta$ where $\zeta_1=\zeta$ and $\zeta_2={\rm e}^{i\theta}$ and becomes

\begin{equation}
\hat{Z}_{4}(x) =  \int_0^{2\pi}\frac{d\theta}{2\pi}\oint \frac{d\zeta}{2\pi i \zeta} \frac{(\zeta+\zeta^{-1} - 2\cos\theta)^2}{\left(1 - x(2\cos\theta+\zeta+\zeta^{-1})\right)}\; . 
\end{equation}
We first perform the contour integration over $\zeta$ noting that the expression has a double pole at $\zeta=0$ and simple poles at 
\begin{equation}
\zeta^{\pm} = \frac{1-2x\cos\theta\pm\sqrt{(1-2x\cos\theta)^2-4x^2}}{2x}
\end{equation}
with only $\zeta^-$ lies inside $|\zeta|<1$ for $0<x<\frac{1}{4}$.
The result of summing over the pole contributions is
\begin{equation}
  \hat{Z}_{4}(x) =  -\frac{1}{4x^2}+\int_0^{2\pi}\frac{d\theta}{2\pi}\frac{(1-4x\cos\theta)^2}{4 x^2 \sqrt{(1 - 2 x \cos\theta)^2 - 4 x^2}}
\end{equation}
with the first term is the result of performing the $\theta$ integral from the double pole at $\zeta=0$. In the remaining integral, the range of $\theta$
from $\pi$ to $2\pi$ is equal to that from $0$ to $\pi$ with
$x\rightarrow -x$, therefore the resulting expression is a function of $x^2$.
The integral can
be can be performed in terms of Elliptic functions with the result
\[
\boxed{\hat{Z}_{4}(x)=\frac{1}{x^2} \left(\frac{2 {\bf E}(16 x^2)-(1-16x^2){\bf K}(16 x^2)}{2\pi}-\frac{1}{4}\right)} 
\]

Expanding $\hat{Z}_4(x)$ gives 
\begin{equation}
\hat{Z}_4(x)=\sum_{n=0}^\infty\frac{((2 n)!)^2}{(n+1)^2(n!)^4}x^{2n}
\end{equation}
and the coefficients are $C_n^2$ where $C_n = \frac{1}{n+1} \binom{2n}{n} = \frac{(2n)!}{(n+1)!n!}$
are the Catalan numbers. For large $n$ we have we then have the asymptotic
result
\begin{equation}
c_{2n}(4)\sim\frac{4^{2n}}{\pi n^3}
\end{equation}

The asymptotic form can again be extracted from the integral form
noting that the singularities nearest the origin in the $x$ plane are at $x=\pm 1/4$ and at the level of the integrand arises from both $\epsilon=1-(4x)^2$
and $\theta$ near expanding the integrand near zero.
Noting again that the singularity of the integrand arises in the vicinity of $\epsilon=0$ and $\theta=0$ where the argument of the square root in the denominator approaching zero as $w=\frac{\epsilon}{2}+\frac{\theta^2}{4}$ so in the vicinity of the singularity $\theta$ scales as $\sqrt{\epsilon}$.  Scaling $\epsilon\rightarrow \rho^2\epsilon$ and $\theta\rightarrow \rho \theta$ and expanding in $\rho$ yields the singular belaviour as the coefficient of $\rho^3$ giving
\[ \hat{Z}_{4}^{sing}\sim \int_{-\delta}^\delta
\frac{2(\epsilon + \theta^2)^2}{\sqrt{2\epsilon + \theta^2}}
\sim -\frac{1}{2\pi}\epsilon^2\ln\epsilon=-\frac{1}{2\pi}(1-(4x)^2)^2\ln(1-(4x)^2)
\]

Expanding near $x=0$ gives the asymptotic form or the coefficients, as
\[
-\frac{1}{2\pi}(1-(4x)^2)^2\ln(1-(4x)^2)=\sum_k\frac{1}{\pi}\frac{4^{2n}}{n^3}x^{2n}
\]
and we recover the asymptotics 
\[
\boxed{c_{2n}(4)\sim\frac{1}{\pi}\frac{4^{2n}}{n^3}}
\]
\subsection{Large $N$ $SO(5)$}
\[
\hat Z_{5}(x)=\oint \frac{d\zeta_1}{2\pi i \, \zeta_1}
\oint \frac{d\zeta_2}{2\pi i \, \zeta_2}
\;
\frac{
\left(2 - \frac{1}{\zeta_1} - \zeta_1\right)
\left(2 - \frac{1}{\zeta_2} - \zeta_2\right)
\left(\frac{1}{\zeta_1} + \zeta_1 - \frac{1}{\zeta_2} - \zeta_2\right)^2
}{
8 \left(1 - x \left(1 + \frac{1}{\zeta_1} + \zeta_1 + \frac{1}{\zeta_2} + \zeta_2\right)\right)
}
\]
Performing the $\zeta_2$ integral by contour integration and writing $\zeta_1={\rm e}^{i\theta}$ yields
\begin{equation}
\hat{Z}_5(x)=\frac{1 + 3 x^2}{4 x^3}-\int_0^{2\pi} \frac{d\theta}{2\pi}\frac{
\left(1-3x-2x\cos\theta\right)
\left(1-x-4x\cos\theta\right)^2(1-\cos(\theta))
}{
2 x^3 \sqrt{\left(1 - x - 2x\cos\theta\right)^2-4x^2}
}\label{IngegralFormZ5}
\end{equation}
with the first factor arising from the $\zeta_2^{-3}$ pole at the origin integrated over $\theta$. The non-trivial integral due to the poles at
$$\zeta_2=\frac{1-x-2x\cos\theta \pm \sqrt{(1-x-2 x \cos(\theta))^2})-4x^2}{2 x}$$
The remaining integral can be performed noting that 
\[
\int_{0}^{2\pi}
\frac{d\theta}{2\pi}\,
\frac{1-\cos\theta}
{\sqrt{\left(1 - x - 2x\cos\theta\right)^2-4x^2}}
=
\frac{1}{\sqrt{(1+3x)(1-x)}}F(2,x)
\]
\[
\int_{0}^{2\pi}
\frac{d\theta}{2\pi}\,
\frac{(1-\cos(\theta)^2}
{\sqrt{\left(1 - x - 2x\cos\theta\right)^2-4x^2}}
=
\frac{3}{2}\frac{F(3,x)}{\sqrt{(1+3x)(1-x)}}
\]
\[
\int_{0}^{2\pi}
\frac{d\theta}{2\pi}\,
\frac{(1-\cos(\theta))^3}
{\sqrt{\left(1 - x - 2x\cos\theta\right)^2-4x^2}}
=
\frac{5}{2}\frac{F(4,x)}{\sqrt{(1+3x)(1-x)}}
\]
\[
\int_{0}^{2\pi}
\frac{d\theta}{2\pi}\,
\frac{(1-\cos(\theta)^4}
{\sqrt{\left(1 - x - 2x\cos\theta\right)^2-4x^2}}
\, d\theta
=
\frac{35}{8}\frac{F(5,x)}{2\,\sqrt{(1+3x)(1-x)}}
\]
where
\[
F(n,x)=\, F_1\!\left(
\frac{1}{2}, \frac{1}{2}, \frac{1}{2}, n,
\frac{4x}{1+3x}, \frac{4x}{1-x}
\right)
\]
giving the final expression
\[\small
\boxed{\hat{Z}_{5}(x)=\frac{1+3x^2)}{4x^3}-\frac{(1-5x)^3 F(2,x)+15(1-5 x)^2xF(3,x)+80(1-5x)x^2F(4,x)+140x^3F(5,x)}{\sqrt{1+2x-3x^2}}}
\]

The asymptotic form of the series expansion is most easily extracted from the integral (\ref{IngegralFormZ5}) by noting that the singularity nearest the origin
occurs at $x=\frac{1}{5}$ and at $\theta=0$ and arises as a branch cut of the integrand. Then expanding in $\epsilon=1-5x$ and $\theta$ one sees that the branch cut is approached as $\sqrt{5\epsilon +\theta^2}$. Taking the $\theta$ interval from $[-\delta,\delta]$ with $\delta$ a fixed infinitismal $\hat{Z}_5(x)$ becomes
\begin{equation}
 \hat {Z}_5^{sing}=-\int_{-\delta}^\delta\frac{d\theta}{2\pi}
( \frac{625\,\epsilon^3 \theta^2}{16 \sqrt{5\epsilon + \theta^2}}
+ \frac{625\,\epsilon^2 \theta^4}{16 \sqrt{5\epsilon + \theta^2}}
+ \frac{25\,\epsilon \theta^6}{2 \sqrt{5\epsilon + \theta^2}}
+ \frac{5\,\theta^8}{4 \sqrt{5\epsilon + \theta^2}})
  \end{equation}
\begin{equation}
  \hat{Z}_{5}^{sing}=-\frac{5}{8}\int_0^\delta\frac{d\theta}{2\pi}\theta^2\sqrt{5\epsilon + \theta^2}(5\epsilon+2\theta^2)^2\sim -\frac{3125}{1024 \pi}\epsilon^4 \ln\epsilon=-\frac{3125}{1024 \pi}(1-5x)^4\ln(1-5x).
\end{equation}
This contribution captutes the singularity closes to the origin in the $x$ plane, being non-analytic at $\epsilon=0$. Expanding $\hat{Z}_{5}^{sing}$ in small $x$
yields the asymptotics for large orders as
$$\boxed{c_n(5)\sim \frac{9375}{128 \pi} \frac{5^n}{n^5}\simeq 23.3137\frac{5^n}{n^5}} $$

\vfill\eject

\section{ Generalising the two summation formulae for $ \cZ ( \infty , \infty , 2n )$  for any $H \subset G$ } 
\label{sec:GenTwo}

We can  generalise the expansion $ \cZ( \infty , \infty , 2n )$ into two distinct sums over partitions. 
We replace $S_{2n}$ by a group $G$, $ \cC_p$ by any conjugacy class in $ G $ and $ S_n [ S_2]$ by any subgroup $H$ of $G$. 
\begin{comment} 
	\bea 
	\sum_{ \cC  \in  {\mathrm {Classes}  } ( G ) } \cZ^{ H ; G }_{ \cC } | \Aut_G ( \cC ) | = \sum_{\cC  \in 
		{\mathrm {Classes}  }  ( G )  } \cZ_{ H ; G }^{( \cC))}
	\eea 
	where $ \cZ^{ H ; G }_{ \cC } $ is the number of orbits of the $H$-action by conjugation on $ \cC$. 
	$  \cZ^{ H ; G }_{ \cC } | \Aut_G ( \cC ) |  $  is the sum  of orbits of $H$-action by conjugation  on $ H \cup \cC$, weighted by the positive integer $ |Aut_{ \cO } ( G  )|/ |Aut_{ \cO } ( H )|  $. 
\end{comment} 
The generalisations are as follows 
\begin{align}\label{generalisations} 
	&\Phi  (  \infty , 2n  ; p ) \rightarrow \Phi_{ ( H ; G )} ( C )  = { 1 \over |H| } \sum_{\gamma \in H } \sum_{\beta \in G }   \delta ( \gamma \beta \sigma_{\scriptscriptstyle C} \beta^{-1 } ) \cr 
	&\cZ ( \infty , \infty , 2n )  \rightarrow \cZ_{ ( H ; G )}  = \sum_{C} \Phi_{ ( H ; G )} ( C ) \cr 
	& \Phichk  (  \infty , 2n ;  q  )   \rightarrow \Phichk_{ ( H ; G ) } ( \tilde C ) = { 1 \over |H | } 
	\sum_{\gamma \in H }   \sum_{ \alpha \in \tilde C } \delta ( \gamma \alpha \gamma^{-1} \alpha^{ -1 })
\end{align}
$\Phichk_{ ( H ; G ) } ( \tilde C ) $ counts the orbits of $H$ acting by conjugation on the elements of the conjugacy class $ \tilde C \subset G$. 
With these definitions, and generalising the steps in sections   \ref{sec:twoPartExps} and  \ref{sec:IntegZInfp}   we show 
\begin{equation}\label{HGeq}  
	\boxed{~~\cZ_{ ( H ; G )}  \equiv  \sum_{C \in \clsG } \Phi_{ ( H ; G )} ( C )  =  \sum_{ \tilde C \in   \clsG }
		\Phichk_{ (H ; G)  } ( \tilde C ) ~~}
\end{equation}

First we derive the interpretation of  $ \Phi_{ ( H ; G )} ( C ) $ as a weighted counting of orbits which is a non-negative integer, following the steps in section \ref{sec:IntegZInfp}. First we recognise that $\Phi_{ ( H ; G )} ( C )$ is collecting elements in the intersection $ H \cap \cC $ and  counting them with multiplicity equal to
$ | \Aut_C ( G ) | $, which is the order of the stabiliser of elements in $ \cC\in \clsG$ under action of conjugation by $ G$. We thus have 
\bea 
\Phi_{ ( H ; G )}  ( C )  = { 1 \over |H| } \sum_{\gamma \in H } \sum_{\beta \in G }   \delta ( \gamma \beta \sigma_{\scriptscriptstyle C} \beta^{-1 } )   = { | H \cap \cC | ~ | \Aut_C ( G ) |  \over |H | }  
\eea 
Now we decompose  $  H \cap \cC  $  as a union of orbits, labelled by $s$ below,  under the action of $H$ by conjugation 
\begin{align}
	H \cap C  = \bigsqcup_{s}  ~ \cO_s ( H ; G )
\end{align}
By the orbit counting lemma 
\bea 
|  \cO_s ( H ; G ) |  =   {    |H | \over | \Aut_s ( H ) |  }
\eea
Using this decomposition 
\bea 
\Phi_{ ( H ; G )}  ( C )   && = {\Aut_C ( G ) |  \over |H | }  \sum_{s } {    |H | \over | \Aut_s ( H ) |  } \cr 
&& =  \sum_{s }   {    | \Aut_C ( G ) |    \over | \Aut_s ( H ) |  }   =  \sum_{s }   {    | \Aut_s ( G ) |    \over | \Aut_s ( H ) |  }  
\eea 
When $ H \cap \cC$ is not empty, $ \Aut_s ( H ) $ is a subgroup of $ \Aut_s ( G )   $ and this is a sum of positive integers.

Now we generalise the argument in section \ref{sec:twoPartExps} to show the equality \eqref{HGeq} 
\bea 
\cZ_{ ( H ;  G)  } = \sum_{C } \Phi_{ H ; G } ( C  ) &&  = \sum_{\cC} { ! \over |H|  } \sum_{\gamma \in H } \sum_{\beta \in G } 
\delta ( \gamma \beta \sigC \beta^{-1 } ) \cr 
&&  = \sum_{\cC} { 1 \over |H|  } \sum_{\gamma \in H } \sum_{\beta \in G }   \sum_{\sigma \in \cC } { 1 \over |C | } 
\delta ( \gamma \beta \sigC \beta^{-1 } )  \cr 
&& = \sum_{\cC} { 1 \over  |G | |H|  } \sum_{\gamma \in H } \sum_{\beta \in G }   \sum_{\sigma \in \cC }  
\delta ( \gamma \beta \sigC \beta^{-1 } )  | \Aut_C ( G ) | \cr 
&& = { 1 \over | H | ~ |G | } \sum_{ \beta \in G } \sum_{ C } \sum_{\sigma \in C  } \sum_{\gamma \in H } 
\delta ( \gamma \beta \sigma \beta^{-1} )  \sum_{ \alpha \in G }  \delta ( \sigma \alpha \sigma^{-1} \alpha^{-1 }) \cr 
&& = { 1 \over |H | ~ | G | } \sum_{\beta , \alpha \in G } \sum_{\gamma in H } 
\delta ( \beta^{-1} \gamma^{-1} \beta \alpha \beta^{-1 } \gamma^{-1} \beta \alpha  ) \cr 
&& = { 1 \over |H| } \sum_{ \alpha \in G } \delta ( \gamma^{ -1 } \alpha \gamma \alpha^{-1 })
\eea
We can now separate the $ \alpha $ sum into conjugacy classes $ \tilde C$ and recognise the definition in the last line of 
\eqref{generalisations}
\bea 
\cZ_{( H ;  G ) }  = \sum_{ \tilde C }   { 1 \over |H| } \sum_{\alpha \in \tilde \cC } \delta ( \gamma^{-1 }   \alpha \gamma \alpha^{-1 } )  = \sum_{\tilde \cC } \Phichk_{ H ; G } ( \tilde \cC ) 
\eea 
This completes the proof of \eqref{HGeq}.

\section{ An identity relating dimensions of symmetric group representations to   special functions  } 
\label{sec:AppIdDimSpec} 

We have approached the counting of $ U(N) \times SO(d) $ and $U(N) \times O(d)$ polynomial 
 invariants of multi-matrices in two ways in this paper. One approach in section \ref{sec:LowEnergy} works directly with the Molien-Weyl integration formula for the invariants, and the other in section \ref{sec:ExactCountingAndAlgorithms} approaches the counting using the combinatorics of words in $ X^{i}_{ a , j }$ approached using the representation theory of tensor product spaces. Combining these two perspectives gives some interesting identities relating the special functions in equations \eqref{SpecFuncsSO} and \eqref{SpecFuncsO} to  sums  of dimensions of irreducible 
 symmetric group representations. These sums are taken over the subsets $ \cA ( k , d )$ and $ \cA_{+ } ( k , d )$ defined in \eqref{defAkd1} and \eqref{defAkd2}.

Let \(V_d\) denote the vector representation of \(SO(d)\). The stable
generating function \(\widehat Z_d(x)\) counts the dimensions of the
invariant subspaces
\[
\left(V_d^{\otimes k}\right)^{SO(d)} .
\]
Schur--Weyl duality gives
\[
V_d^{\otimes k}
=
\bigoplus_{\substack{T\vdash k\\ \ell(T)\le d}}
V_T^{U(d)}\otimes V_T^{S_k},
\]
where \(T\) is a Young diagram with \(k\) boxes. We write
\[
d_T := \dim V_T^{S_k}
\]
for the dimension of the corresponding symmetric-group representation.
Taking \(SO(d)\)-invariants gives
\[
\dim\left(V_d^{\otimes k}\right)^{SO(d)}
=
\sum_{\substack{T\vdash k\\ \ell(T)\le d}}
d_T\,
\dim\left(V_T^{U(d)}\right)^{SO(d)} .
\]
The Cartan--Helgason selection rule gives
\[
\dim\left(V_T^{U(d)}\right)^{SO(d)}
=
\begin{cases}
	1, & T\in \cA(k,d),\\
	0, & T\notin \cA(k,d),
\end{cases}
\]
where
\[
\cA(k,d)=\cA_+ (k,d)\cup \cAm(k,d).
\]
Here
\[
\cA_+ (k,d)
=
\{\,T\vdash k:\ell(T)\le d,\ \hbox{all row lengths of }T
\hbox{ are even}\,\},
\]
and
\[
\cA_- (k,d)
=
\{\,T\vdash k:\ell(T)=d,\ \hbox{all row lengths of }T
\hbox{ are odd}\,\}.
\]
It follows that the coefficient of \(x^k\) in the \(SO(d)\) generating
function is
\begin{align}\label{IdSpCoSO} 
\boxed{ ~~
{ \rm Coefficient } ( \widehat Z_d(x) , x^k ) 
=
\sum_{T\in \cA(k,d)} d_T  ~~
} 
\end{align}
 Each allowed diagram in $ \cA(k,d)$  contributes with its Schur--Weyl
multiplicity \(d_T\).

For the \(O(d)\)-invariant tensor space, the epsilon-tensor sector is
removed. Equivalently, only the even-row sector survives:
\[
\dim\left(V_T^{U(d)}\right)^{O(d)}
=
\begin{cases}
	1, & T\in \cAp(k,d),\\
	0, & T\notin \cAp(k,d).
\end{cases}
\]
Hence the corresponding identity following from the counting of $O(d) $ invariants os 
\begin{equation}\label{IdSpCoO} 
\boxed{ ~~
{\rm Coefficient } ( \widehat Z_{O,d}(x) , x^k ) 
=
\sum_{T\in \cAp(k,d)} d_T ~~
} 
\end{equation}
This gives a direct combinatorial verification of the closed-form
generating functions for \(d=2,3,4,5\) in the \(SO(d)\) case, and for
\(d=2,3,4\) in the \(O(d)\) case. We have checked these identities \eqref{IdSpCoSO} and \eqref{IdSpCoO} for selected ranges of $k$ with Mathematica.

\section{Tables}\label{Tables}
\begin{figure}[H] %
   \centering
   \includegraphics[width=\textwidth]{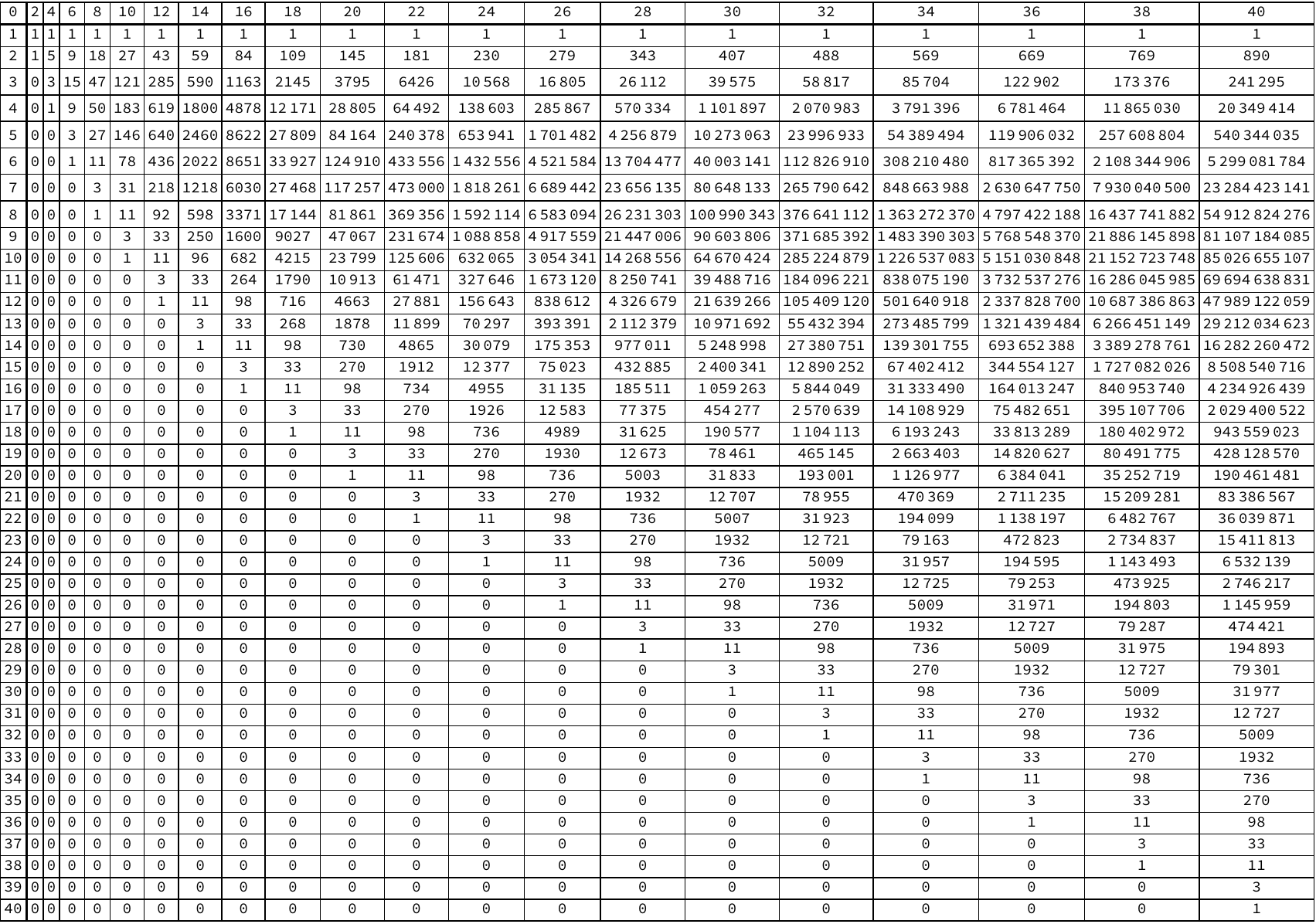} 
   \caption{Table for $SO(2)$ of $k$ vs $H$. The colums give $\Delta \cZ(H,2,k)=\cZ(H,2,k)-\cZ(H-1,2,k)$, so that $Z(N,2,k)$ is the sum of colum numbers down to $N$. The zeros at the bottom of the columns indicate that the $\cZ(N,2,k)$ has reached its stable value, i.e.
$\{2, 10, 38, 158, 602, 2382, 9142, 35492, 136936, 530404, 2053848,
     7972272, 30977742, 120576112,\\
 469915012, 1833813534, 7164469910, 28021000340, 109699469798, 429850240742\}$
   }
   \label{fig:SO2_delDegens}
\end{figure}
\vfill\eject
\begin{figure}[H] %
   \centering
   \includegraphics[width=\textwidth]{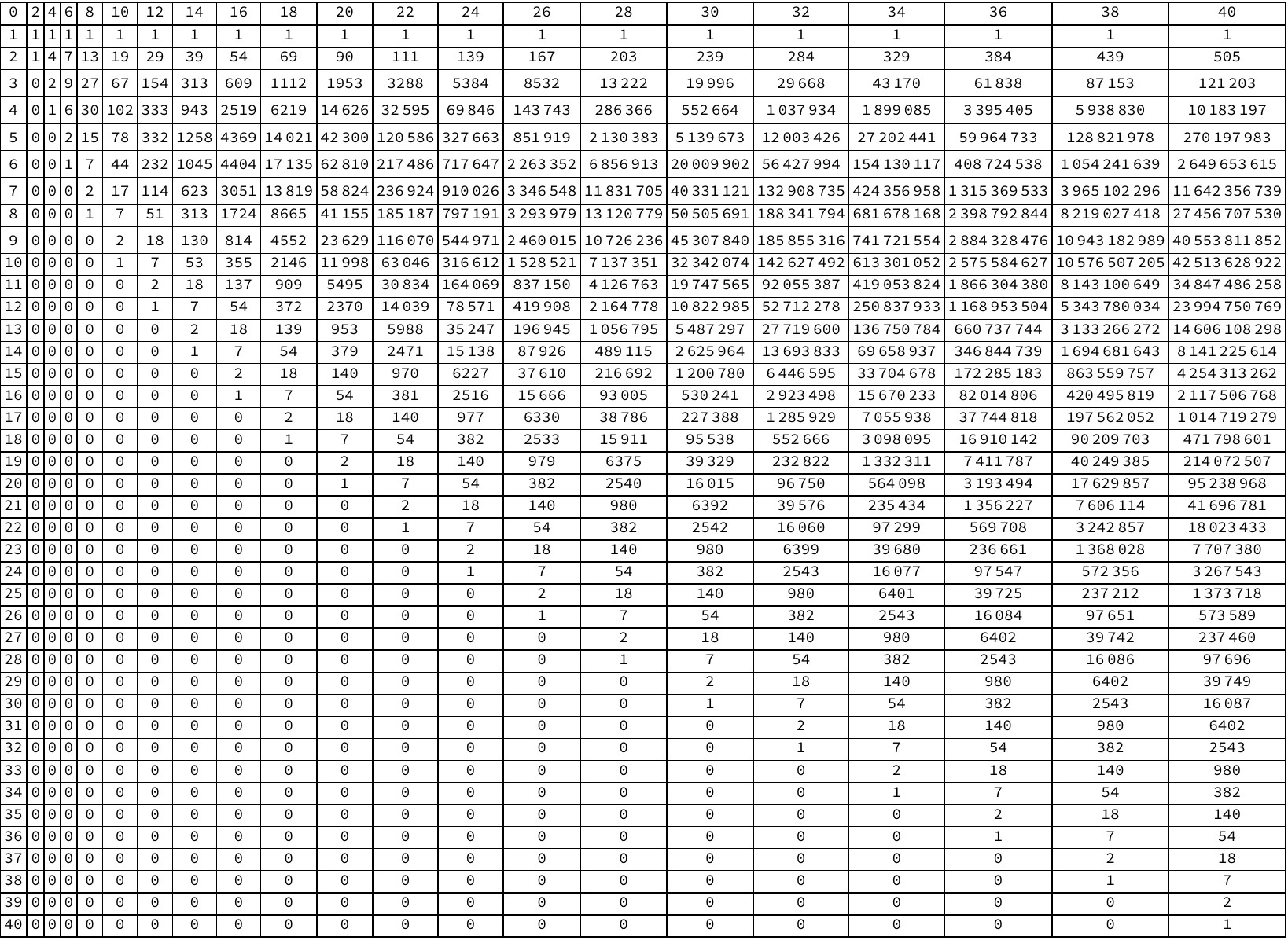} 
   \caption{Table for $O(2)$ of $k$ vs $H$. The colums give $\Delta \cZ_O(H,2,k)=\cZ_O(H,2,k)-\cZ_O(H-1,2,k)$, so that $\cZ_O(N,2,k)$ is the sum of colum numbers down to $N$. The zeros at the bottom of the columns indicate that the $\cZ_O(N,2,k)$ has reached its stable value, i.e.
     $\{2, 8, 26, 96, 338, 1274, 4746, 18119, 69241, 266805, 1030199, 3992829, 15502428, 60315503, 235012821,\\
     917018164, 3582458724,14010949457, 54850635694, 214926925836\}$
   }
   \label{fig:O2_delDegens}
\end{figure}
\vfill\eject

\begin{figure}[H] %
   \centering
   \includegraphics[width=\textwidth]{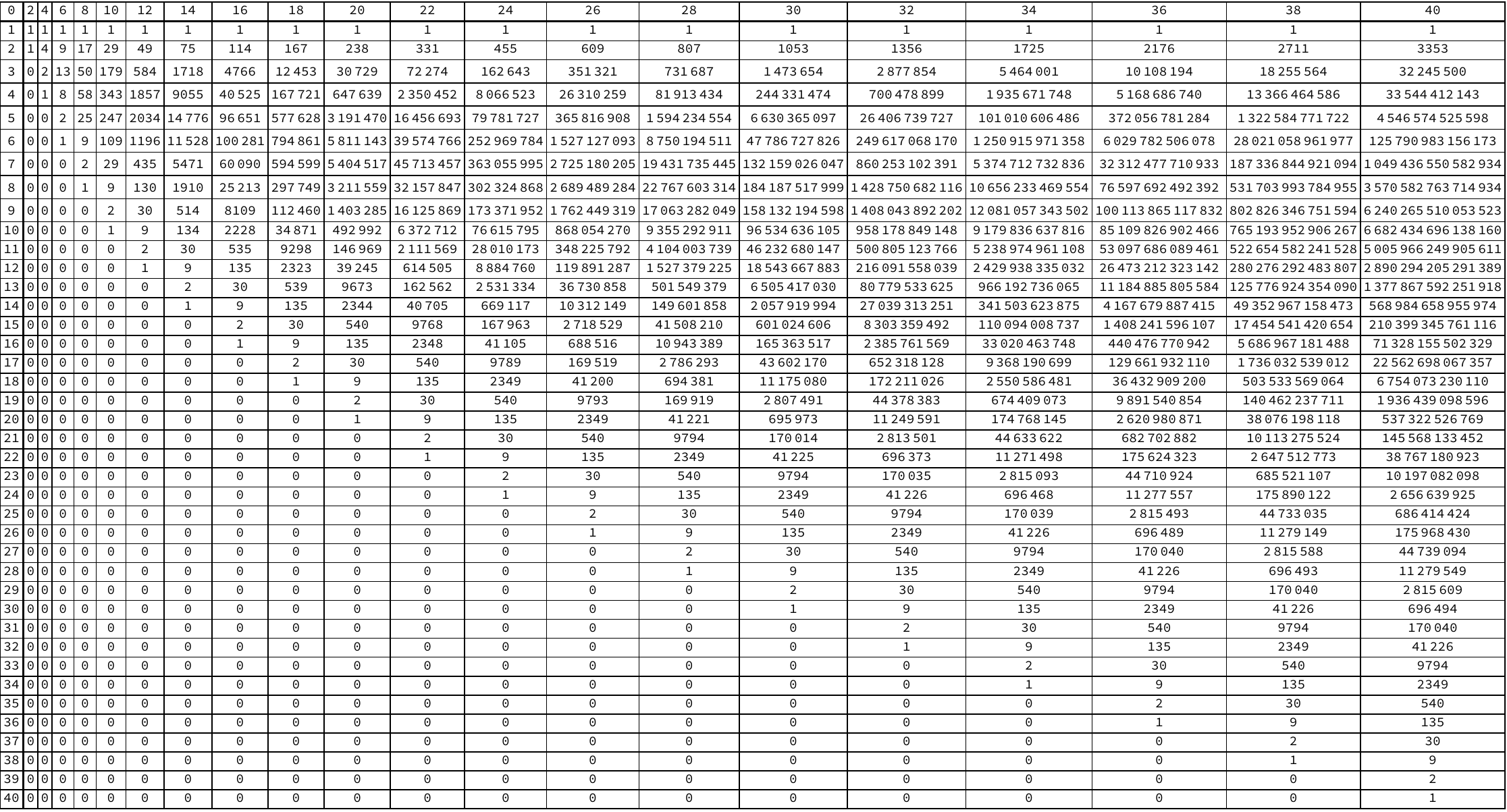} 
   \caption{Table for $SO(3)$ of  (even) $2k$ vs $H$. The colums give $\Delta \cZ(H,3,2k)=\cZ(H,3,2k)-\cZ(H-1,3,2k)$, so that $\cZ(N,3,2k)$ is the sum of colum numbers down to $N$. The zeros at the bottom of the columns indicate that the $\cZ(N,3,2k)$ has reached its stable value, i.e.
$\{2, 8, 34, 163, 949, 6328, 45224, 338690, 2604847, 20392521,
  161766576, 1296667050, 10483569978,\\
  85383679187, 699840851844, 5768242232729, 47778259622796, 397492588195576,\\
3320052228152333,27829919506597612\}$.}
   \label{fig:SO3_delDegenseven}
\end{figure}
%\vfill\eject

\begin{figure}[H] %
   \centering
   \includegraphics[width=\textwidth]{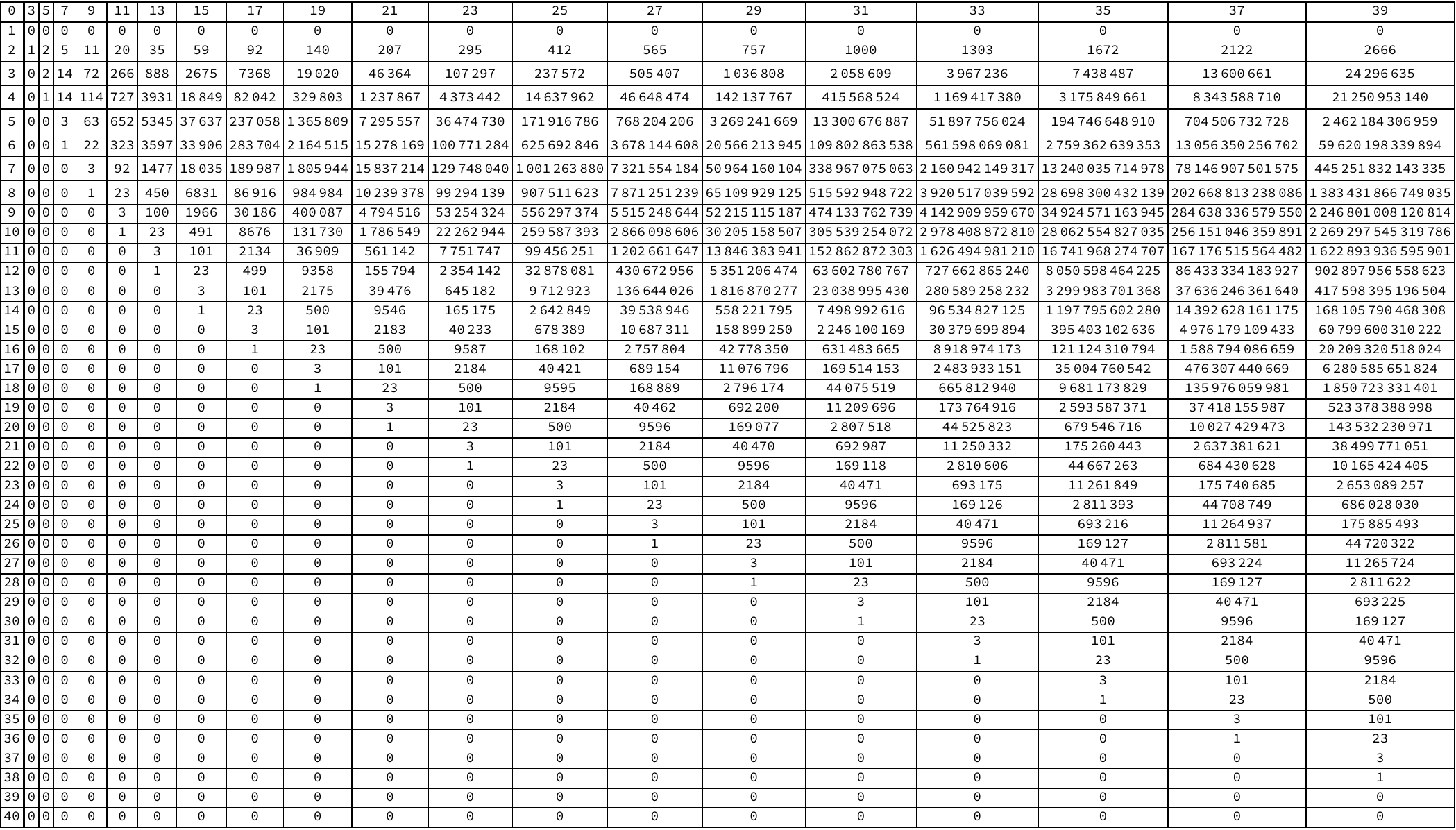} 
   \caption{Table for $SO(3)$ of (odd) $2k+1$ vs $H$. The colums give $\Delta \cZ(H,3,2k+1)=\cZ(H,3,2k+1)-\cZ(H-1,3,2k+1)$, so that $\cZ(N,3,2k+1)$ is the sum of colum numbers down to $N$. The zeros at the bottom of the columns indicate that the $\cZ(N,3,k)$ has reached its stable value, i.e.
$\{1, 5, 37, 286, 2107, 15850, 120577, 928790, 7251102, 57284590, 457255373, 3682735271, 29891529536,\\
244262141081, 2007863955974, 16591410851235, 137737822156818, 1148241301666882, 9608241369394175\}$.
   }
   \label{fig:SO3_delDegensodd}
\end{figure}
\vfill\eject
\begin{figure}[H] %
   \centering
   \includegraphics[width=\textwidth]{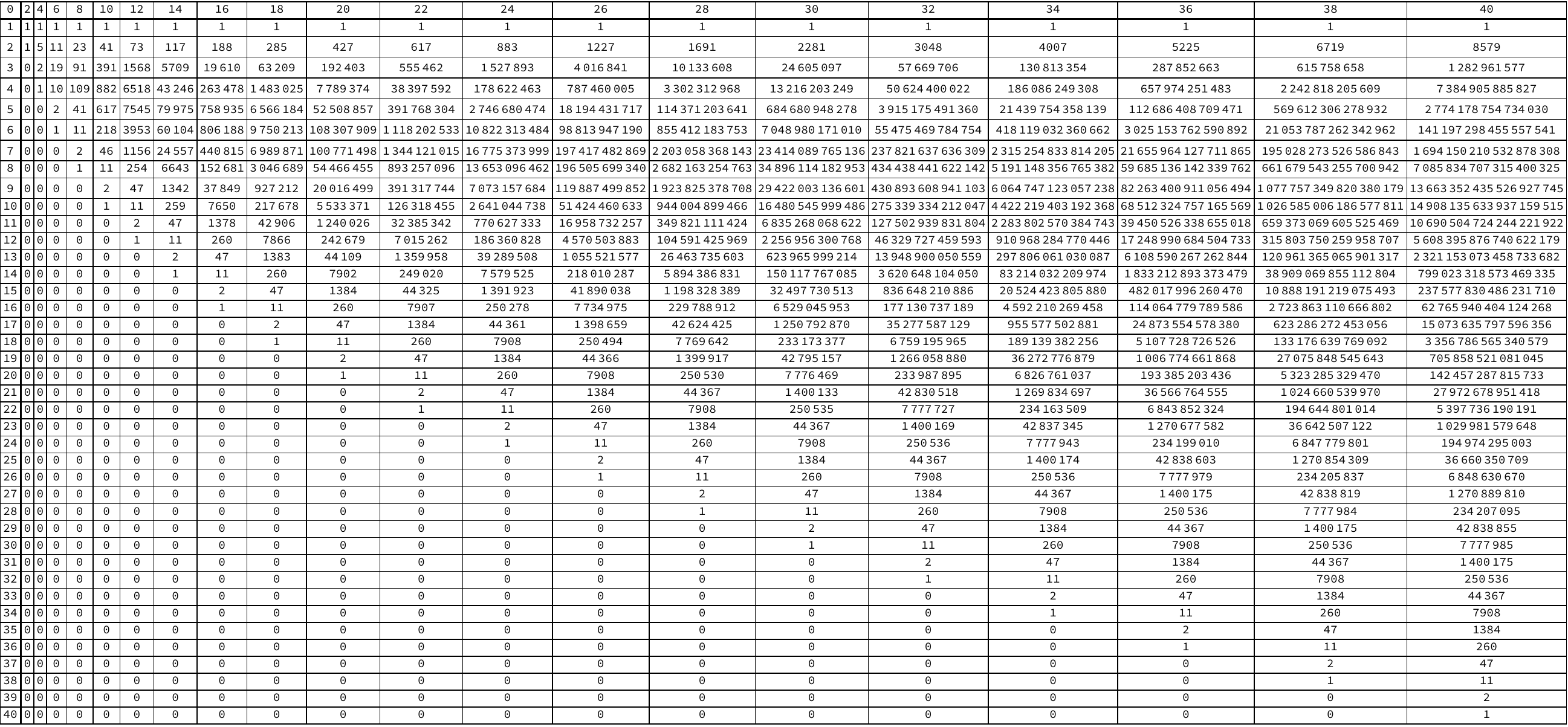} 
   \caption{Table for $SO(4)$ of $k$ vs $H$. The colums give $\Delta \cZ(H,4,k)=\cZ(H,4,k)-\cZ(H-1,4,k)$, so that $\cZ(N,4,k)$ is the sum of colum numbers down to $N$. The zeros at the bottom of the columns indicate that the $\cZ(N,4,k)$ has reached its stable value, i.e.
$\{2, 9, 44, 279, 2210, 21129, 222014, 2489094, 29096843, 351123215,
     4345002338, 54897371450,\\
     705889095824, 9214398612364, 121866526168768, 1630393933297372,
     22035211675826489, \\
300523953742514509, 4132125251307186056, 57234185993228166332\}$.}
   \label{fig:SO4_delDegens}
\end{figure}
\vfill\eject
\begin{figure}[H] %
   \centering
   \includegraphics[width=\textwidth]{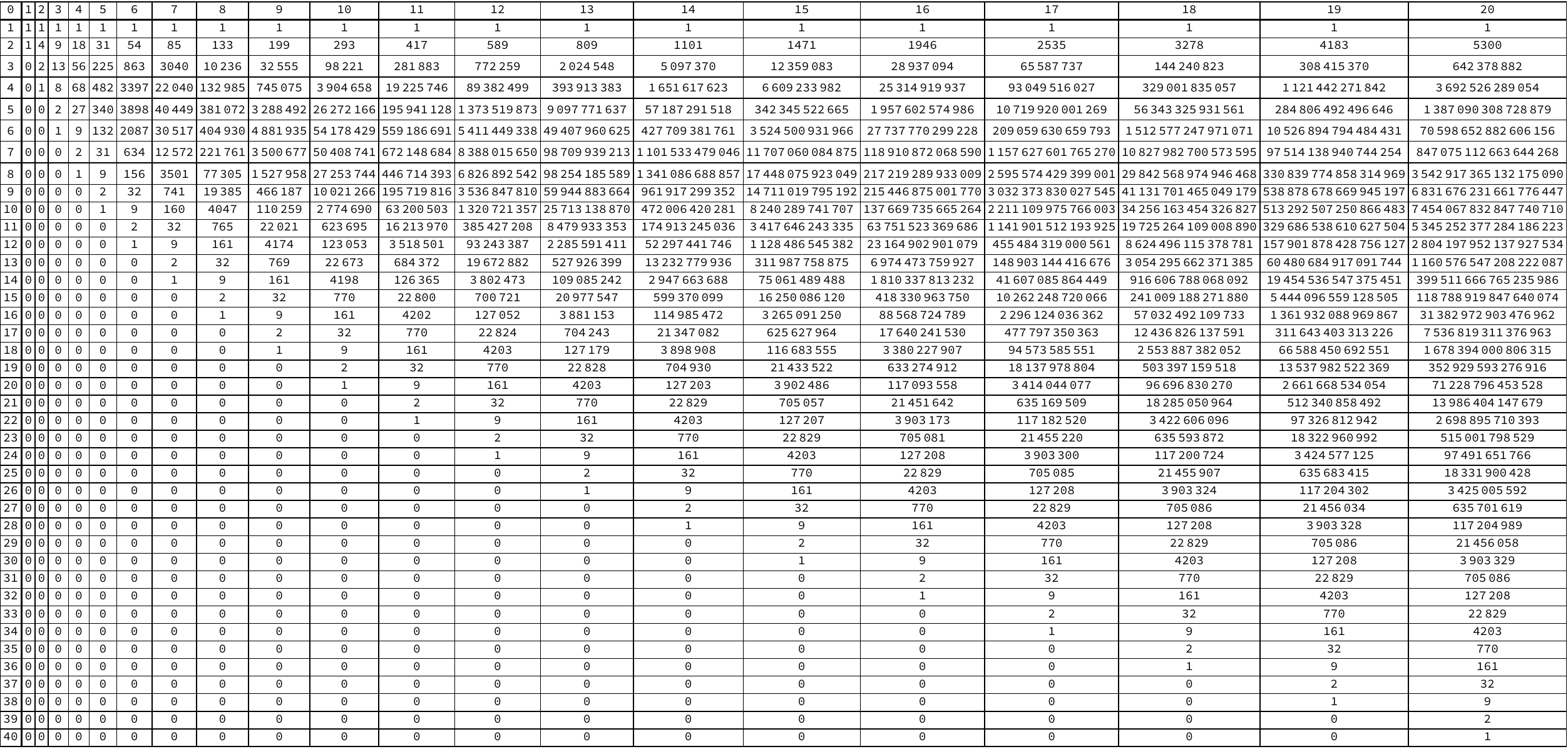} 
   \caption{Table for $O(4)$ of $k$ vs $H$. The colums
     give $\Delta \cZ_O(H,4,k)=\cZ_O(H,4,k)-\cZ_O(H-1,4,k)$, so that
     $\cZ_O(N,4,k)$ is the sum of colum numbers down
     to $N$. The zeros at the bottom of the columns indicate that
     the $\cZ_O(N,4,k)$ has reached its stable value, i.e.
$\{2, 8, 34, 182, 1254, 11134, 113150, 1252825, 14580507, 175686803,
     2172990447, 27450603643,\\
     352952072872, 4607228869017, 60933379316239,
     815197423987306, 11017607638486864,\\
     150261983964267240, 2066062653608871705, 28617093106843288058\}$
   }
   \label{fig:O4_delDegens}
\end{figure}
\vfill\eject
\begin{figure}[H] %
   \centering
   \includegraphics[width=\textwidth]{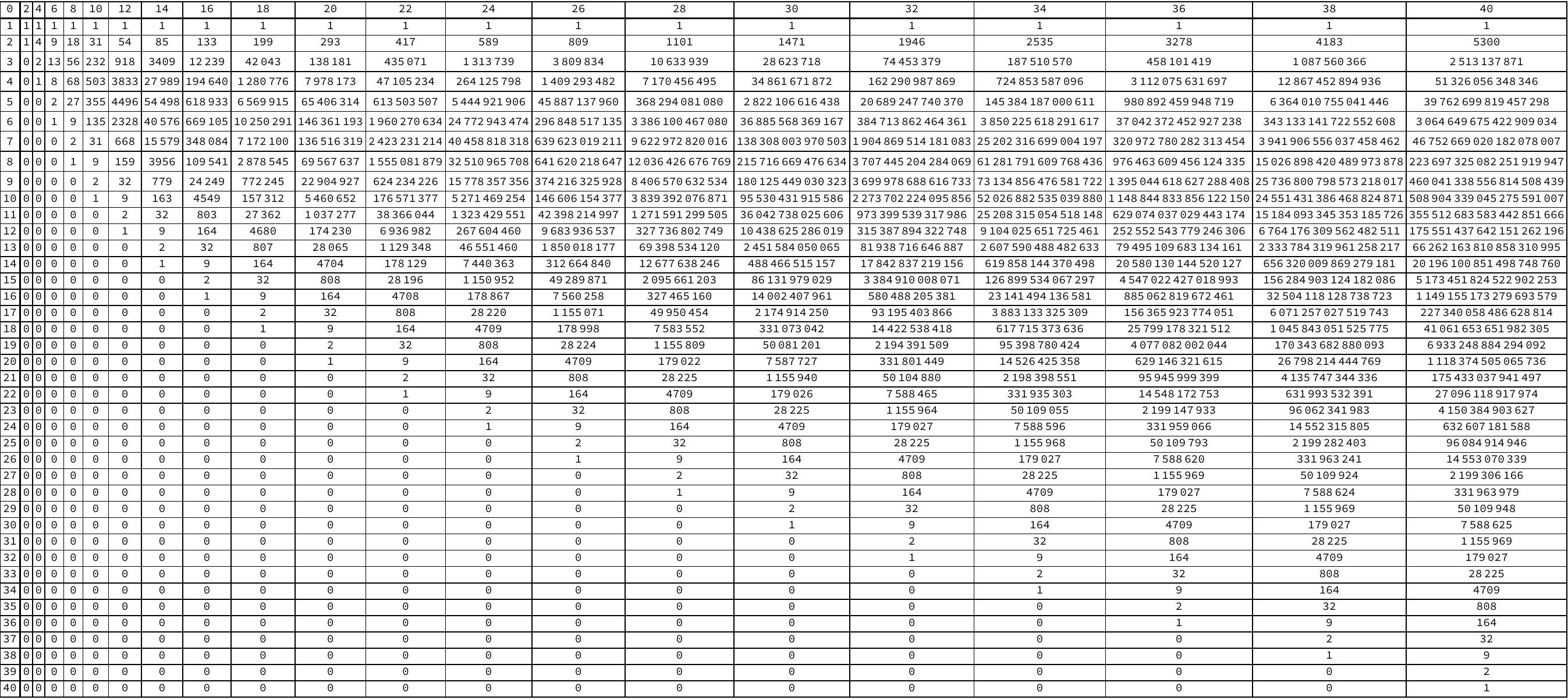} 
   \caption{Table for $SO(5)$ of  (even) $2k$ vs $H$. The colums give $\Delta \cZ(H,5,2k)=\cZ(H,5,2k)-\cZ(H-1,5,2k)$, so that $\cZ(N,5,2k)$ is the sum of colum numbers down to $N$. The zeros at the bottom of the columns indicate that the $\cZ(N,5,2k)$ has reached its stable value, i.e.
$\{2, 8, 34, 182, 1300, 12501, 147079, 1982485, 29156484, 455578982, 7447077983, 126149305741,\\
2200517530082, 39350824149656, 718947232964686, 13384205695743424, 253336625897392460,\\
4866673109966261653, 94741029311508873245, 1866621789054994963281\}$.
 }
   \label{fig:SO5_delDegenseven}
\end{figure}
%\vfill\eject

\begin{figure}[H] %
   \centering
   \includegraphics[width=\textwidth]{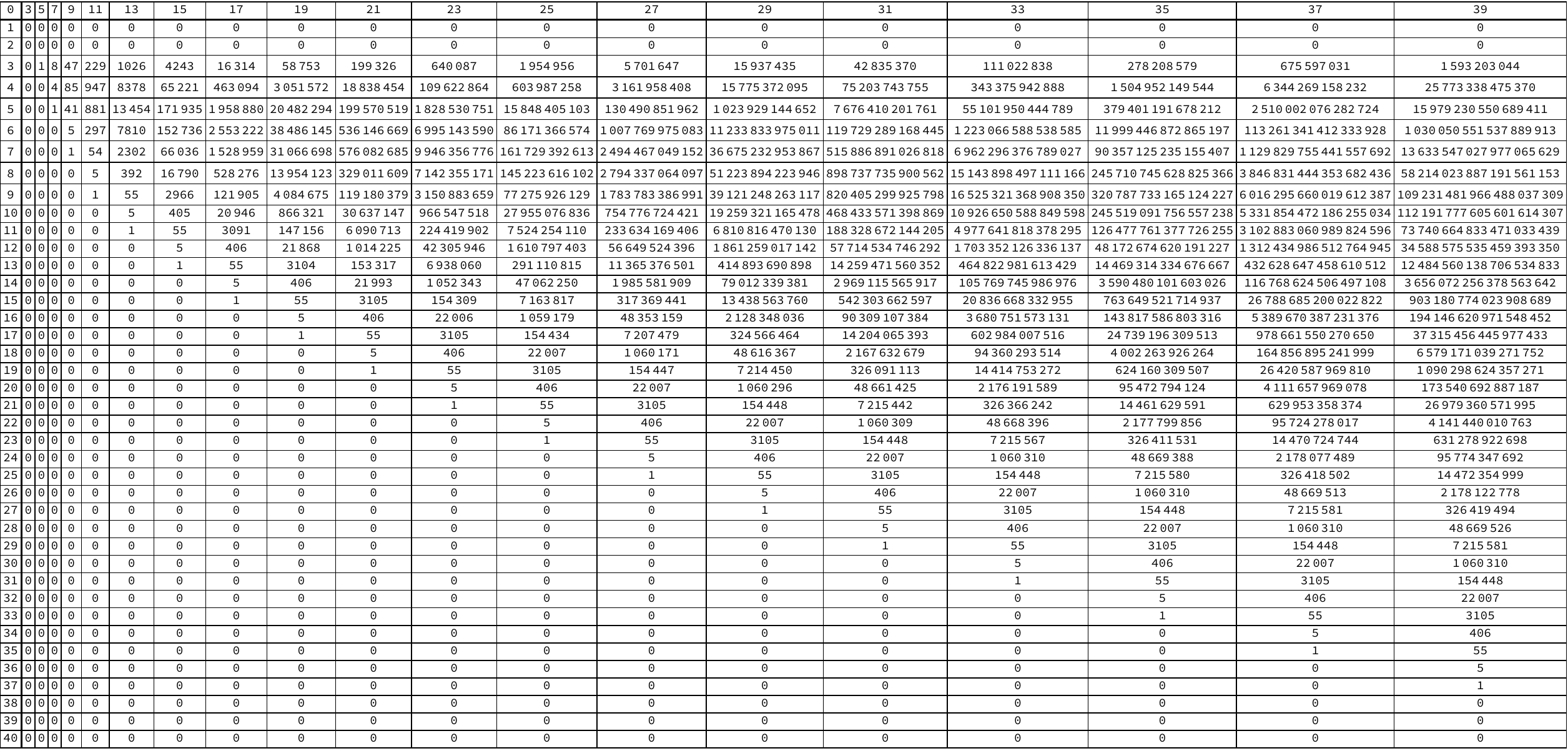} 
   \caption{Table for $SO(5)$ of (odd) $2k+1$ vs $H$. The colums give $\Delta \cZ(H,5,2k+1)=\cZ(H,5,2k+1)-\cZ(H-1,5,2k+1)$, so that $\cZ(N,5,2k+1)$ is the sum of colum numbers down to $N$. The zeros at the bottom of the columns indicate that the $\cZ(N,5,k)$ has reached its stable value, i.e.
     $\{0, 1, 13, 179, 2414, 33423, 480393, 7195154, 112223176, 1816950608, 30414976554, 524291353158, \\
     9272801534249, 167735181102552, 3094865605894513, 58113497268564597, 1108402224739584889, \\
21438657488890864917, 419929276534979888047\}$.
   }
   \label{fig:SO5_delDegensodd}
\end{figure}
\vfill\eject
\begin{figure}[H] %
   \centering
   \includegraphics[width=\textwidth]{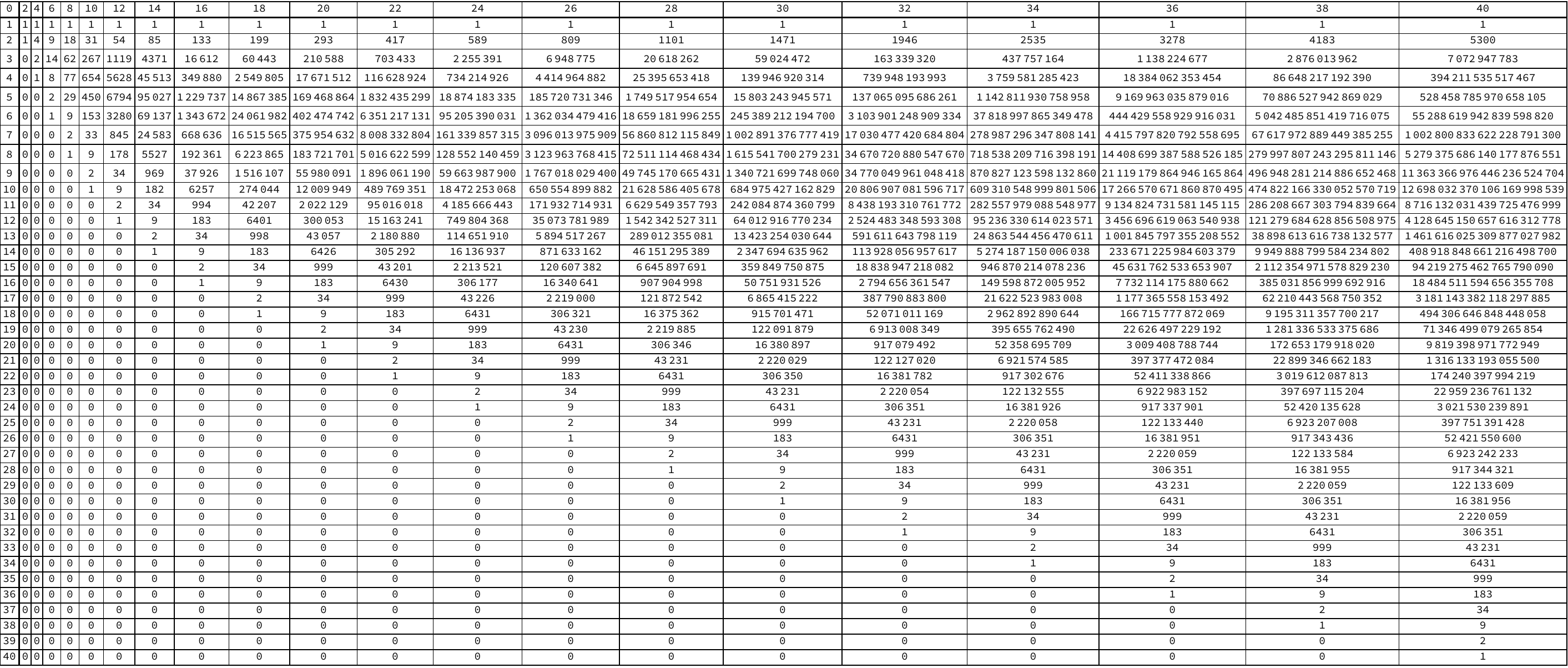} 
   \caption{Table for $SO(6)$ of $k$ vs $H$. The colums give $\Delta \cZ(H,6,k)=\cZ(H,6,k)-\cZ(H-1,6,k)$, so that $\cZ(N,6,k)$ is the sum of colum numbers down to $N$. The zeros at the bottom of the columns indicate that the $\cZ(N,6,k)$ has reached its stable value, i.e. 
$\{2, 8, 35, 199, 1601, 17945, 245441, 3846438, 66119231, 1219865266, 23824487439, 487913113256, \\
10403639970417, 229694550042371, 5227749929700846, 122210159596767816, 2925682300283728534,\\
71545638078060798440, 1783406814007389841778, 45231167315127619837762\}$.}
   \label{fig:SO6_delDegens}
\end{figure}
\vfill\eject

\begin{figure}[H] %
   \centering
   \includegraphics[width=\textwidth]{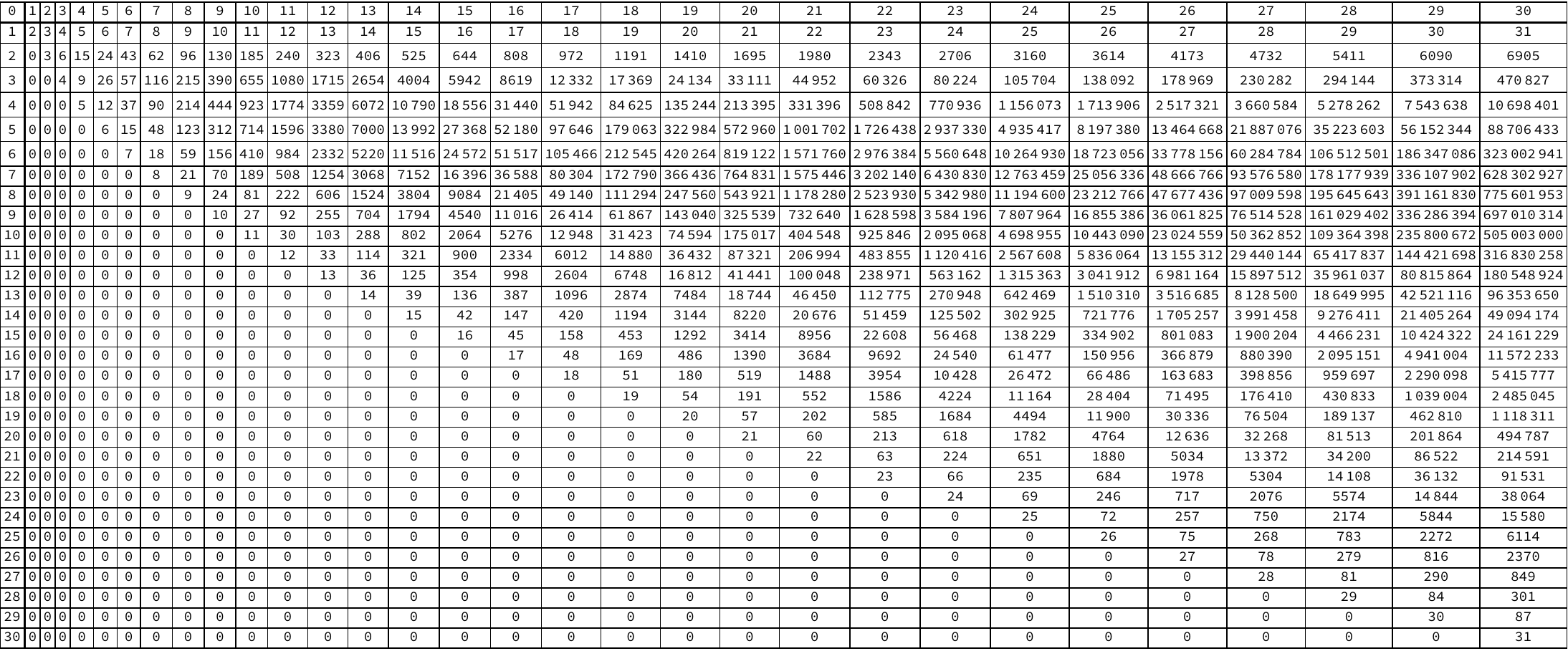} 
   \caption{Table for the two matrix model $d=2$. The colums give $\Delta Z(H,2,k)=Z(H,2,k)-Z(H-1,2,k)$, so that $Z(N,2,k)$ is the sum of colum numbers down to $N$.
     The zeros at the bottom of the columns indicate that the $Z(N,2,k)$ has reached its stable value, i.e. 
     $\{2, 6, 14, 34, 74, 166, 350, 746, 1546, 3206, 6550, 13386, 27114, 54894, 110630, 222794, 447538, 898574, \\
     1801590, 3610930\}$.}
   \label{fig:Z2_delDegens}
\end{figure}
\vfill\eject

\begin{figure}[H] %
   \centering
   \includegraphics[width=\textwidth]{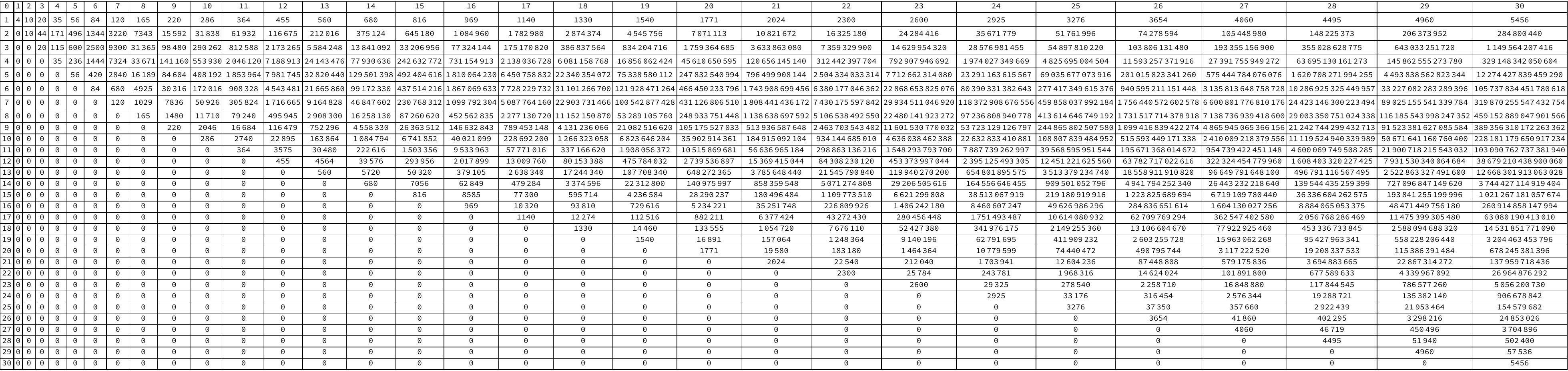} 
   \caption{Table for $d=4$ matrix model. The colums give $\Delta Z(H,4,k)=Z(H,4,k)-Z(H-1,4,k)$, so that $Z(N,4,k)$ is the sum of colum numbers down to $N$.
     The zeros at the bottom of the columns indicate that the $Z(N,4,k)$ has reached its stable value, i.e. 
     $\{4, 20, 84, 356, 1444, 5876, 23604, 94852, 379908, 1521492, 6088148, 24360548, 97451492, 389838708, \\
     1559394356, 6237711300, 24951007620, 99804576340, 399218968084, 1596878076132\}$.} The lower diagonal  entries are $n(n+1)(n+2)/6$ 
   \label{fig:Z4_delDegens}
\end{figure}
\vfill\eject

\end{appendix}

\end{document}